\documentclass[letter,11pt]{article}

\pdfoutput=1 % if your are submitting a pdflatex (i.e. if you have
\usepackage{jheppub-mod}
\usepackage{amssymb,amsmath}
\usepackage[utf8]{inputenc}
\usepackage{graphicx}
\usepackage{epstopdf,epsfig}
\usepackage[scriptsize]{subfigure}
\usepackage[export]{adjustbox}
\usepackage[T1]{fontenc}
\usepackage{xcolor}
\usepackage{marginnote}
\usepackage{mathrsfs}
\usepackage{mathalfa}
\usepackage{upgreek,textgreek}
\usepackage{comment}

\usepackage[color]{changebar}
\cbcolor{red}

\def\imagetop#1{\vtop{\null\scriptsize\hbox{#1}}}

\renewcommand{\i}{\mathrm{i}}

\renewcommand{\d}{\mathrm{d}}

\renewcommand{\Re}{\mathrm{Re}}
\renewcommand{\Im}{\mathrm{Im}}

\DeclareMathOperator\polylog{polylog}
\DeclareMathAlphabet{\mathpzc}{OT1}{pzc}{m}{it}

\newcommand{\be}{\begin{equation}}
\newcommand{\ee}{\end{equation}}
\newcommand{\ba}{\begin{equation}\begin{aligned}}
\newcommand{\ea}{\end{aligned}\end{equation}}

\newcommand{\bs}[1]{{\boldsymbol{#1}}}
\newcommand{\lap}{\bigtriangleup}
\newcommand{\const}{\mbox{const}}
\newcommand{\eq}[1]{(\ref{#1})}

\newcommand{\ins}[1]{{\mbox{\tiny #1}}}

\newcommand{\inds}[1]{{\scriptscriptstyle #1}}

\newcommand{\hh}{,\hspace{1cm}}

\newcommand{\angstrom}{\textup{\AA}}
\graphicspath{{figures/}{figures/stresses/}}

\title{\boldmath Scarf for Lifshitz}

\author[1]{A. Zelnikov,\note{Corresponding author.}}
\author{R. Krechetnikov}

\affiliation{University of Alberta, Edmonton, Canada T6G 2E1}

\emailAdd{zelnikov@ualberta.ca}
\emailAdd{krechet@ualberta.ca}

\abstract{Polarization of a vacuum as well as of dispersive and dissipative dielectric media with piece-wise and smooth inhomogeneities is studied with the goal to calculate one-loop effects and clarify the question of renormalizability of diverging electromagnetic stress-energy tensor. First, the stress tensor is computed with the Lifshitz approach to London (van der Waals) forces in the non-retarded limit, which after the substraction of the leading free space ultraviolet divergencies still retains the divergencies associated with the presence of sharp boundaries between piece-wise inhomogeneities. We call these contributions \textit{finite} because they become renormalized after a sharp interface is replaced with a dielectric permittivity $\varepsilon(\omega;\boldsymbol{x})$ changing according to a smooth function of spatial coordinates $\boldsymbol{x}$. In addition, such a smoothed out interface exhibits new \textit{subleading} ultraviolet divergencies that appear due to its internal structure. To systematically deal with the polarization of inhomogeneous media, the Hadamard expansion, based on the heat kernel method, is applied to single out both finite and subleading contributions and to unequivocally demonstrate incomplete renormalizability of the Lifshitz theory. The latter property is expected because the Lifshitz theory is an effective one due to the usage of the macroscopic dielectric permittivity, which results in the presence of a cut-off parameter reflecting an unresolved physics at smaller scales.

The above approach also allows us to reveal the nature of surface tension, which proves to be purely quantum mechanical consisting of finite cut-off independent as well as cut-off dependent contributions. The deduced theory of surface tension and its calculations for real dielectric media are favorably compared to the available experimental data. While the sharp interface limit recovers the classical boundary conditions for the electric field and uncovers the origin of the apparent local divergencies of the renormalized stresses in the sharp interface formulation previously pointed out in the literature, the problem of surface tension proves to be of a \textit{distinguished limit} type because the sharp interface formulation loses the information about the internal structure of an interface and hence cannot explain the origin of surface tension. The general theory offered here is illustrated with an exactly solvable model representing a smooth transition between two dielectric media of different dielectric permittivities, which relies upon a solution of the Schr\"{o}dinger equation with the Scarf potential.
}

\begin{document}
\maketitle
\flushbottom

\vspace{-0.5cm}

\section{Introduction}

\subsection{The circle of phenomena: quantum fluctuations of a vacuum and matter}

While the basic elements of QFT -- a vacuum, interaction between particles, and structure of simple atoms -- are reasonably well understood, real matter and its interaction with quantum fluctuations of the vacuum are less so as the accurate modelling and prediction of its properties proves to be hard \cite{Fisher:1964,Zallen:2004,Hansen:2006}. The structure and hence mechanical strength of condensed matter is determined by the forces acting between molecules -- phenomenologically, this is often described by the canonical Lennard-Jones potential $\varphi_{\mathrm{LJ}} = 4 \upsilon \left[\left(r_{\ins{m}} / r\right)^{12} - \left(r_{\ins{m}} / r\right)^{6}\right]$, which is an isotropic part of intermolecular interaction; here $\upsilon$ is the depth of the potential well and $2^{1/6} r_{\ins{m}}$ is the distance at which the particle-particle potential energy $\varphi_{\mathrm{LJ}}$ has a minimum; for brevity, we will refer to $r_{\ins{m}}$ as the intermolecular separation. The first term in $\varphi_{\mathrm{LJ}}$ is a purely heuristic way of modeling a quantum repulsion of molecules due to the Pauli exclusion principle. The last term in $\varphi_{\mathrm{LJ}}$ -- commonly known as the van der Waals interaction $\varphi_{\mathrm{vdW}} = - 4 \upsilon \left(r_{\ins{m}} / r\right)^{6}$ -- can be rigorously justified by dipole-dipole (Keesom) \footnote{Despite the fact that the energy of dipole-dipole interaction scales as $r^{-3}$, in the early twenties Keesom \cite{Keesom:1915} was the first to compute the thermal average of the force between two polar molecules and found a temperature-dependent interaction energy proportional to $r^{-6}$, which is attractive. The attractive nature of the orientation force is not difficult to understand: although the number of attractive orientations is exactly the same as the number of repulsive ones, the former are statistically favored over the latter because the Boltzmann weight $e^{-\mathcal{E}/k_{B}T}$ diminishes with increasing energy $\mathcal{E}$, while the smaller energies correspond to attractive orientations.}, dipole-induced dipole (Debye), and dispersion induced dipole-induced dipole (London) interactions.

It is the latter (dispersive) part of van der Waals forces -- present in any matter and playing a role in a host of everyday phenomena such as adhesion, surface tension, adsorption, wetting, crack propagation in solids, to name a few, and hence considered to be the most important \cite{Israelachvili:2011} -- which we will be dealing with here. These forces have been systematically accounted for in QFT, though the history of the question is not without controversies. As motivated by applications, about a century ago it was realized that the pressure $\Pi_{\mathrm{D}}$ in thin liquid films is generally different from the pressure in the macroscopic bulk liquid due to the action of van der Waals potential forces \cite{van-der-Waals:1873}. This con- or disjoining Derjaguin's pressure $\Pi_{\mathrm{D}}$ was originally calculated for pure substances and non-charged interfaces \cite{Derjaguin:1936} via pair-wise (additive) summation of intermolecular potential interactions $\varphi_{\mathrm{vdW}}^{ij} = - 4 \upsilon \left(r_{\ins{m}} / r_{ij}\right)^{6}$, i.e. between molecules $i$ and $j$:
\begin{align}\label{pressure:Derjaguin}
\Pi_{\mathrm{D}}(\ell) = - A_{\mathrm{H}} / (6 \pi \ell^{3}),
\end{align}
where $A_{\mathrm{H}}$ is the Hamaker constant specific to a given combination of substances in contact. However, a rigorous account of the subject matter started with Casimir's work \cite{Casimir:1948a,Casimir:1948b}, in whose original configuration two parallel perfectly conducting plates of an infinite extent are separated by a distance $\ell$ in the vacuum, which gives rise to a finite pressure on the plates named after Casimir
\begin{align}\label{pressure:Casimir}
\Pi_{\mathrm{C}}={\hbar c\pi^2/( 240 \ell^4)}.
\end{align}
Driven by the discrepancy between experiments \cite{Deryaguin:1954,Deryaguin:1956,Deryaguin:1960,Tabor:1968} and ``additive'' macroscopic theory \cite{Derjaguin:1934,Derjaguin:1936,Hamaker:1937}, a rigorous derivation of the pressure, with $\Pi_{\mathrm{D}}$ in the non-retarded and $\Pi_{\mathrm{C}}$ in the retarded limits, was performed by Lifshitz \cite{Lifshitz:1956} in the case of two parallel homogeneous dielectric media separated by the vacuum. His work recognized the genuine quantum nature and non-additivity of dispersion forces \cite{Farina:1999} -- unless rarefied, polarizability of a condensed matter may be vastly different from that of a sum of individual molecules contributions. Later, Dzyaloshinskii et al. (DLP) \cite{Dzyaloshinskii:1959,Dzyaloshinskii:1961} generalized the Lifshitz theory by introducing another homogeneous medium instead of the vacuum as the intervening phase.

In the Lifshitz/DLP theory, the electromagnetic (EM) fluctuations are quantized in polarizable media. Because the range of influence of van der Waals forces is much larger than the interatomic distance, $r \gg r_{\ins{m}}$ due to power-law decay $\varphi_{\mathrm{vdW}} \sim r^{-6}$ in the non-retarded limit and $\varphi_{\mathrm{vdW}} \sim r^{-7}$ in the retarded one, the calculation of the EM stress-energy tensor $T^{\mu\nu}$, comprised of the stresses $\sigma_{ij}$ and energy density $\varrho$, in macroscopic bodies can be done exclusively based on their geometry and the classical EM linear response functions such as dielectric permittivity $\varepsilon$ and magnetic susceptibility $\mu$, which are necessarily macroscopic quantities as well. In general, the Lifshitz theory of van der Waals forces mitigates between microscopic approach and the effective field theory: it is concerned with the changes in the quantized Maxwell field due to the introduction of macroscopic bodies. These changes result in forces between disjoint bodies as well as affect the energy and, naturally, the internal stresses of a single body. In formal terms, the goal of the Lifshitz theory is to construct an effective QFT describing a fluctuating quantum background field, which naturally provides local information about stresses.

In retrospect, Casimir's seminal work started the exploration of a new area of physics, where quantum fluctuations of a polarized vacuum or matter result in previously unexpected and fascinating physical effects such as quantum repulsion and levitation, quantum friction, and quantum torque, as well as the deep relationship between the vacuum energy and cosmological constant. The predicted values of the Casimir-Derjaguin force were confirmed in the earlier experiments \cite{Deryaguin:1954,Sparnaay:1958,Deryaguin:1956,Deryaguin:1960,Tabor:1968} and more recent precise measurements \cite{Lamoreaux:1997}. Nowadays the Casimir effect is considered to be important in many nano- and microelectromechanical systems \cite{Decca:2011}. Despite being of wide interest and the subject of a mature research area, there remain open questions regarding the divergence of the EM stress-energy tensor $T^{\mu\nu}$ inside the bodies and vacuum. Thus, a renaissance in studying Casimir-Derjaguin (and van der Waals, in general) forces is driven by both theoretical and applied underpinnings. The ideas, difficulties, and methods underlying the Casimir-Derjaguin effect appear in other subfields of theoretical physics. For example, the idea that the QCD vacuum can be described as a color dielectric medium \cite{Pirner:1992} with hadrons considered as bags \cite{Chodos:1974} led one to study the zero point fluctuations of the quark fields.

\subsection{The problem of divergencies: bulk and interfaces}

The Casimir-Derjaguin force, should it be in the vacuum between metal plates or in sandwiched dielectrics, arises from interactions between the fluctuating quantum fields and matter. However, traditionally, the Casimir-Derjaguin problem is posed as the response of a fluctuating quantum field to externally imposed boundary conditions (BCs), i.e. the physical interactions are replaced \textit{ab initio} by BCs. Namely, in the standard treatment of the Casimir-Derjaguin effect one investigates the EM field only and considers, for example, the metal plates as perfect conductors represented by the corresponding BCs. While BCs are a very convenient mathematical idealization in field theory, physical materials cannot constrain arbitrarily high frequency components of a fluctuating quantum field or, equivalently, wavelengths much shorter than $r_{\ins{m}}$. For example, for an imperfect conductor of characteristic skin depth $\delta_{s}$, waves of sufficiently short wavelength $\lambda \lesssim \delta_{s}$ (high frequency) penetrate a significant distance before being attenuated, and therefore do not `see' the precise position of the boundary. In general, at a very high mode frequency $\omega$ all real media become transparent and indistinguishable from the vacuum $\varepsilon,\mu \rightarrow 1$. While this property gives a hope that the divergences, resulting from high frequencies, might disappear or become weakened by dispersion, it proves to be insufficient to resolve all of them \cite{Bordag:2002} for the reason that the permittivity does not decrease with the frequency faster than $\sim \omega^{-2}$ according to Debye's relaxation model \cite{Jackson:1998}.

The presence of boundaries, without which Casimir and Derjaguin effects would not exist, leads to the distortion of the quantum state of the confined field containing no real force carriers and consisting entirely of zero-point fluctuations. Since the boundaries change the zero-point energy of the fluctuating fields, they give rise to forces between the rigid bodies or stresses on isolated surfaces. Thus, the Casimir and Derjaguin effects are inseparable from the existence of boundaries and result from a distortion of the virtual-particle `sea' occurring inevitably when there appears an inhomogeneous structure, e.g. in the form of BCs.

Since the work of Casimir it has been known that the sum over zero-point energies is highly divergent in the UV limit. Subtraction of the vacuum energy that is already present without the plates, i.e. not subject to the BCs, only removes the worst UV divergence, which is quartic in the regularization parameter in three space dimensions \cite{Milton:2016}. This divergence does not affect the total force of interaction between bodies because its contribution to the energy does not depend on the separation distance between bodies. Similar treatments can be found in the general theory of van der Waals forces by Lifshitz and DLP, where they have subtracted from the stress $T^{\mu\nu}$ a term that one would get should the medium be unbounded, uniform, and of the same local properties $\varepsilon$ and $\mu$. For this subtraction, Lifshitz explained that ``it represents the back reaction of the field produced by the body on the body itself, and is in fact compensated by similar forces on the other sides of the body,'' while DLP did so based on the assumption that the short waves do not feel the changing of $\varepsilon$ across the BCs and thus do not contribute. Casimir, Lifshitz, and DLP treatments of UV divergencies have successfully predicted the forces that agree quite well with experiments cited above, and  are widely adopted as an essential procedure (aka subtraction of the `empty space', `unbound medium', or `bare contribution').

However, as becomes obvious from the local analysis, after this subtraction there are still remnant boundary-induced divergences in stresses and energy density as has been known for a long time \cite{Brown:1969,Deutsch:1979,Philbin:2010,Xiong:2013,Simpson:2013}. One often considers this remaining effect as determined by the part of the zero-point energy \cite{Bordag:1992}, which is infrared (IR) and hence non-local being associated with the BCs. This is in contrast to the local nature of UV divergencies, the renormalization of which assumes that all local divergent terms can be combined with the bare coupling constants of the field theory to produce finite (renormalized) physically observable local characteristics of the system. In the case of a pure Maxwell field, i.e. not affected by the real media polarization, the stresses at the boundary happen to be finite \cite{Candelas:1982}, only because of the special symmetry of the Maxwell field in the vacuum in the presence of an infinite plane boundary, which allows for the diverging contributions from electric and magnetic fields to cancel each other locally. Dispersive media break this symmetry resulting in the stresses diverging at the sharp boundaries. Altogether, this generic unboundedness of the UV-renormalized stress-energy tensor $T^{\mu\nu}$ as the boundary is approached arises from the high-frequency and short wavelength modes. In fact, $T^{\mu\nu}$ would attain extreme values, which should depend on a molecular structure of the matter, and may seem to imply diverging effects \cite{Actor:1995} such as surface tension.

There have been many attempts using various methods \cite{Fulling:2018} -- from point splitting \cite{Parashar:2018}, heat kernel expansion \cite{Vassilevich:2003}, proper time and wavenumber cut-off \cite{Bordag:1998,Baacke:1985b}, zeta-function \cite{Blau:1988}, dimensional \cite{tHooft:1972tcz} to other approaches \cite{Bao:2016,Li:2019} -- to achieve local regularization near boundaries. As noted by Fulling \cite{Fulling:2010zz}, zeta-function and dimensional regularization methods hide the divergencies in an \textit{ad hoc} way and yield global energies that may be inconsistent with the local $T^{\mu\nu}$. While the cut-off methods based on limiting the range of wavenumbers from above due to the microstructure of materials, the cut-off dependent values of stresses and energy density render the theory dissatisfying as well as leave some ambiguities in the interpretation of surface tension and energy. It was also shown \cite{Fulling:2010zz,Fulling:2012} that some versions of a finite UV cut-off near a reflecting boundary (the Dirichlet boundary condition on the ``conducting'' boundaries) may lead to violation of a stress-energy tensor $T^{\mu\nu}$ conservation. There have been other formal solutions offered to cure the problem of boundary divergencies, e.g. to cancel them by introducing \textit{ad hoc} surface dependent counterterms \cite{Symanzik:1981,Actor:1995,Vassilevich:2009uf}.

Another approach put in practice is to replace sharp boundaries by steeply rising potential barriers \cite{Milton:2011b,Bouas:2012,Murray:2016,Milton:2016,Fulling:2018}, e.g. by modelling a wall with a potential such as the Dirac $\delta$-function \cite{Graham:2002fi,Graham:2002xq,Graham:2002fw,Graham:2003ib,Milton:2004} or a smooth power-law function \cite{Milton:2011b,Fulling:2012,Milton:2016}; in the case of a metal plate, the BCs with respect to the behavior of the electrons are idealized with a potential well (Sommerfeld pot model) represented by the $\delta$-function \cite{Bordag:1992}.
Graham et al. \cite{Graham:2002} constructed a model of an inhomogeneous medium by coupling the fluctuating field to a smooth background potential that implements the boundary condition in a certain limit. The same authors \cite{Graham:2003} considered the Casimir problem as the limit of a conventional quantum fluctuating field $\phi(\boldsymbol{x})$ coupled to a smooth non-dynamical background field $\sigma(\boldsymbol{x})$ representing the material in the corresponding Lagrangian $\mathcal{L} \sim \sigma(\boldsymbol{x}) \phi^{2}(\boldsymbol{x})$; in their view, this method of renormalization in a continuum QFT without boundaries provides a physically reasonable way to remove divergences. However, in the Dirichlet limit when the external potential becomes sharply peaked, all modes of the fluctuating field vanish as it should be in the classical Casimir problem statement leading to divergent energy density at the sharp interface, i.e. in this limit the divergence cannot be removed by renormalization. Moreover, the Lagrangian for an effective theory of real media may not be written explicitly, which also renders the theory non-renormalizable from a practical point of view.

A more natural way to avoid the above mentioned unphysical predictions seemed to smooth out the interface \cite{Philbin:2010,Simpson:2013} with some spatially-dependent permittivity $\varepsilon(\boldsymbol{x})$ and permeability $\mu(\boldsymbol{x})$. It is intuitively clear that in inhomogeneous media the Casimir-Derjaguin effect can act inside the material with the resulting internal EM stresses being particularly strong near discontinuities in the refractive-index profile \cite{Griniasty:2017b}. In solids, this stress is negligible in comparison with the interatomic forces, but in fluids it may build up sufficient pressure forcing them to move until an equilibrium is reached.

\subsection{Paper outline}

Regardless of the method for removing the boundary-induced divergencies remaining after the Lifshitz regularization, the lingering question is if they reflect physically significant cut-off dependence, e.g. related to tension of the interface (aka surface tension) \cite{Hoye:2017}, or are merely unphysical artifacts of the calculation method. The goal of the present study is to address the question of the space structure (should it be a vacuum or matter) distortion exhibited in the stress-energy tensor $T^{\mu\nu}$, which becomes unbounded as the distance to the sharp boundary in the models governed by BCs vanishes. This is the key subject of our study aiming to address the associated boundary-induced divergencies as well as their relation to the usual UV bulk divergencies. The accompanying question is on the nature of surface tension, which may not be properly accounted for by the sharp interface models as it strongly depends on the details of the inter-molecular interactions near the interface \cite{Graham:2002fi,Graham:2002xq,Graham:2002fw,Graham:2003ib}. To resolve all these quandaries, we revisit the idea of smoothing out the interface as it corresponds to real physics: microscopically, the transition between phase densities and dielectric permittivities is never truly discontinuous. However, it is known that even for a smoothed out interface there are UV divergencies of the stress tensor, which remain after renormalizations and have been previously deemed unphysical \cite{Philbin:2010,Simpson:2013}, thus requiring one to revisit the issue with proper renormalization and interpretation.

To make the theory specific we will focus here on the non-retarded limit of van der Waals forces corresponding to Derjaguin's disjoining pressure case \cite{Derjaguin:1934,Hamaker:1937} when the materials involved are dielectrics, as formulated in \S \ref{sec:Lifshitz} in the course of discussing the troubles with the Lifshitz theory. This electrostatic limit enables a significant simplification of the algebra hence making the analysis transparent and highlighting not only the nature of divergences, but also the procedure for their potential renormalization. Physically, this Derjaguin limit also corresponds to the leading order effect $\sim \ell^{-3}$ compared to that of Casimir $\sim \ell^{-4}$, when it concerns the attractive part of van der Waals forces $\sim r^{-6}$ and surface tension calculations.

To deal with the boundary-induced divergencies, we consider an interface between two media with dielectric constant $\varepsilon(\boldsymbol{x})$ smoothed out over the width $w$ in one direction $z$ and apply the standard QFT methods of dealing with UV divergencies applicable in the bulk. Hence, in our approach we do not need to make any extra assumptions as to how to renormalize the interface itself: renormalization procedure in the bulk uniquely defines the way of dealing with UV divergencies in the smoothed out interface. As a result, the procedure is unambiguous, because the interface is just a smooth inhomogeneity of characteristic width ${w}$ in the bulk. A particular choice of the interface inhomogeneity $\varepsilon(z)$ profile leading to the stationary Schr\"{o}dinger equation with the Scarf potential $V(z)$ enables us to construct the pertinent propagator (Green's function) in an analytic form (\S \ref{subsec:Scarf}) and thus compute all stresses explicitly. Given the exact representation of the propagator, in line with the Wald axiomatic approach \cite{Hack:2012,Hollands:2005} it proves to be natural and most convenient to use proper time regularization with the Hadamard expansion in \S \ref{subsec:proper-time-regularization} and the point-splitting method of regularization \S \ref{subsec:point-splitting}, which is constructed here with the help of the heat kernel method \cite{DeWitt:1965,Vassilevich:2003}. The resulting qualitative picture, though, is the same for all smoothed out interfaces and hence our choice of the Scarf potential $V(z)$ is just a matter of convenience. The case of a sharp interface is recovered in the limit when its width $w$ vanishes: the stresses \eqref{sigmatau1}, which are finite in the case of a smoothed out interface, grow in this limit and account for the divergencies \eqref{sigmayy} that appear in the computations using the conventional sharp BCs approach (\S \ref{sec:Lifshitz}). However, in addition, there appear subleading divergent stresses \eqref{sigma-div:subleading}, which are absent in the solution of the sharp interface problem and account for the intrinsic interfacial structure. Finally, the laid out analysis also reveals the quantum mechanical nature of surface tension in a rigorous manner (\S \ref{sec:ST}) with the concise expression (\ref{gamma1},\ref{eqn:ST-total}) comparing favorably with the available experimental data (\S~\ref{subsec:ST-calculation}).

\section{The trouble with the Lifshitz theory} \label{sec:Lifshitz}

The underlying idea of the Lifshitz theory is similar to that of Casimir: instead of the vacuum gap of width $\ell$ between two metal plates, the Maxwell field fluctuations are quantized in the vacuum gap between two polarizable dielectric media \cite{Lifshitz:1956}. The corresponding energy of EM zero-point fluctuations is also formally divergent, which is not observable, but any variation of this energy results in an actual force between dielectrics. As in the case of Casimir's study, subtraction of the vacuum energy $\mathcal{E}^{\ins{vac}}(\infty)$ when the dielectrics are absent cancels the contribution of all the modes outside the gap and results in a finite effect, $\mathcal{E}^{\ins{fin}}(\ell) = \mathcal{E}^{\ins{vac}}(\ell) - \mathcal{E}^{\ins{vac}}(\infty)$. Because of the energy conservation of the system, the Maxwell field energy can be interpreted as the result of work against the Casimir-Derjaguin pressure between the plates when we change the distance $\ell$ between them. We will focus on this case in the present section as well, as it suits the purpose of demonstrating the key trouble with the Lifshitz theory.

\subsection{Non-retarded limit of the Lifshitz theory} \label{subsec:Derjaguin-limit}

The EM stress-energy tensor of the Maxwell field reads \cite{Misner:1973}
\ba
T^{\mu\nu}=F^{\mu\alpha}\,D^{\nu}{}_{\alpha}-\frac{1}{4}\,g^{\mu\nu}\,F_{\alpha\beta}\,D^{\alpha\beta},
\ea
where $F^{\mu\nu}$ and $D^{\mu\nu}$ are the Maxwell field and the displacement tensors, correspondingly, and $g^{\mu\nu}$ the Minkowski metric tensor with signature $(-+++)$. Throughout the text we will adopt the Planck units, in which $\hbar=c=k_{\ins{B}} \equiv 1$, and restore SI units whenever computations are performed. In the equilibrium case there are no fluxes and the stress-energy tensor takes the form \cite{Misner:1973}
\ba
\label{tensor:stress-energy}
T^{\mu\nu}=\begin{bmatrix}
    \varrho &0 \\
    0    & -\sigma^{ij}
\end{bmatrix}\hh i,j=(x,y,z), \ \mu,\nu=(t,x,y,z),
\ea
where $\varrho=T_{00}$ is the EM energy density and $\sigma_{ij}$ is the EM stress tensor with the standard convention on its signs \cite{Lifshitz:2012}. In dispersive media the Lifshitz theory is formally applicable to arbitrary wavelengths and frequencies. Both the electric and magnetic fields written here in the frequency domain\footnote{because in the time-domain the relation between the displacement vector $\bs{D}$ and the electric field vector $\bs{E}$ is nonlocal, $\bs{D}(t,\bs{x}) = \bs{E}(t,\bs{x}) + \int_{0}^{\infty}{f(\tau) \bs{E}(t-\tau,\bs{x}) \, \mathrm{d}\tau}$, while in the frequency-domain it becomes $\widehat{\bs{D}}(\omega;\bs{x}) = \varepsilon(\omega) \widehat{\bs{E}}(\omega;\bs{x}), \ \varepsilon(\omega) = 1 + \int_{0}^{\infty}{f(\tau) \, e^{\i \omega \tau} \, \d \tau}$.}
\ba
\widehat{D}_{0j}(\omega;\bs{x})=\varepsilon(\omega;\bs{x}) \widehat{F}_{0j}(\omega;\bs{x}), \
\widehat{H}_{ij}(\omega;\bs{x})={1\over \mu(\omega;\bs{x})}\widehat{F}_{ij}(\omega;\bs{x})
\ea
contribute to the stresses. If the medium is dissipative, $\varepsilon(\omega;\bs{x})$ and $\mu(\omega;\bs{x})$ are necessarily complex with the imaginary parts being always positive, as they account for energy dissipation (absorption) of the EM waves propagating in the medium \cite{Dzyaloshinskii:1961}. Dispersion forces are due to the molecules polarizability, which in turn is related to the frequency dependent refractive index and thus dispersion. This is an important point because absorption frequency $\omega_{a}$ of the real media provides a natural length-scale $\lambda_{a}$ where Casimir (retarded) effect transitions to Derjaguin (non-retarded) one. Physically, the latter limit \cite{Barash:1975} corresponds to the case when the characteristic distance $\ell$ involved (e.g. the size of the gap separating materials or the length-scale on which the interaction is considered) is much larger than the interatomic distance $r_{\ins{m}}$, i.e. $\ell \gg r_{\ins{m}}$, while much smaller than the wavelength $\lambda_{a}$ characteristic of these atoms absorption, $\ell \ll \lambda_{a}$. The latter condition follows from the non-retarded limit, since the time $\Delta t = \ell$ over which EM field propagates on the distance $\ell$ should be much smaller than the period of molecules vibrations, so that the EM interaction is basically electrostatic.

Thus, we are interested in the forces on the (short) distances where retarded effects are no longer important, so that the Maxwell equations $\widehat{D}^{\mu\nu}{}_{;\nu}= 0$, which are local in the frequency domain, in the absence of free charges reduce to electrostatics and hence the components of the vector potential reduce to $A_{0} = - \phi$ and $A_{i}=0$. This interaction is described by the non-retarded limit of the Lifshitz theory and in the Coulomb gauge the corresponding Green's function for the mode with frequency $\omega$ reduces to
\ba\label{GG}
\widehat{G}(\omega;\bs{x},\bs{x'})\equiv \widehat{G}_{00}(\omega;\bs{x},\bs{x'})
\ea
and satisfies the equation
\ba\label{eqhatG}
\nabla^i \big[\varepsilon(\omega;\bs{x}) \nabla_i\widehat{G}(\zeta;\bs{x},\bs{x'})\big]=\delta(\bs{x}-\bs{x'}).
\ea
Note that in this equation only spatial covariant derivatives appear and there is no potential term, which is the consequence of the Maxwell equations in homogeneous dielectric media. It reflects the fact that interaction of electric dipoles only is important in the non-retarded limit, in spite of the fact that the quantum fluctuations of Maxwell field are time-dependent. The inverse Fourier transform of the solution to \eqref{eqhatG} corresponds to the Feynman Green's function (propagator) $G^F(t,\bs{x};t',\bs{x}')$ which is an expectation value of time ordered product of field operators -- in our case finite temperature quantum mean value
\footnote{For signs of various Green's functions here and later in the text we refer reader to Fursaev and Vassilevich \cite{Fursaev:2011} for our choice of the metric signature. For other choices of the metric, one may consult \cite{DeWitt:1965,Itzykson:1980}.}:
\begin{align}\label{Green-function:Feynman}
G^F(t,\bs{x};t',\bs{x}')\equiv G_{00}^F(t,\bs{x};t',\bs{x}')= \i\langle \mathrm{T}\,\mathrm{A}_0(t,\bs{x})\mathrm{A}_0(t',\bs{x}') \rangle .
\end{align}
Here $\mathrm{A}_0$ and upright equivalents of classical quantities elsewhere will be understood as operators, the angle brackets $\langle\ldots\rangle$ denote averaging w.r.t. the ground state of the system and symbol $\mathrm{T}$ the chronological product: the operators following it are to be arranged from right to left in the order of increasing times. Fundamentally, the constructed Green's function is a linear response to an external source introduced in the Maxwell equations, which was originally done by Lifshitz \cite{Lifshitz:1956} and later by Schwinger et al. \cite{Schwinger:1978}, who repeated Lifshitz calculations by adding such an external source to Amp\'{e}re's circuital law. According to the linear response theory and fluctuation-dissipation theorem \cite{Rytov:1953}, the Green's function -- the correlation function, determining the average value of the product of components of the quantum field at two different points in space -- depends on the imaginary part of the dielectric permittivity $\Im[\varepsilon]$. As a consequence of causality, i.e. because $\bs{D}(t,\mathbf{x})$ at any instant cannot be affected by the values of $\bs{E}(t,\mathbf{x})$ at future times, $\varepsilon$ is an analytic function in the upper half-plane of the complex frequency $\omega$-plane, which implies that the real $\Re[\varepsilon]$ and imaginary $\Im[\varepsilon]$ parts of the dielectric permittivity $\varepsilon$ are not independent, but related via the Sokhotski-Plemelj theorem \cite{Gakhov:1966}, known in physics as the Kramers-Kronig formula \cite{Lifshitz:1980}:
\ba\label{KK-relation}
\Re[\varepsilon(\omega;\bs{x})]=1+{1\over \pi}\mathcal{P}\int\limits_{0}^{\infty}\d\omega'\, {\Im[\varepsilon(\omega';\bs{x})]\over\omega^{\prime}-\omega}
\hh
\bs{x}=(x,y,z),
\ea
where we took into account the oddness of $\Im[\varepsilon]$ w.r.t. $\omega$.

The Green's function of a macroscopic system (as ours) at finite temperatures $G^{\beta}$ differs from that at zero temperature $G^F$ only in that the vacuum mean value of the product of fields is replaced by averaging over the Gibbs distribution of states at temperature $T$, which in QFT for convenience is accounted for by the inverse temperature parameter defined as $\beta = 1/T$. In the coordinate space the thermal (finite temperature) Green's function $G^{\beta}(t,\bs{x};t',\bs{x}')$ can be obtained from the Feynman propagator $G^F(t,\bs{x};t',\bs{x}')$ \eqref{Green-function:Feynman} using the Wick rotation\footnote{The sign convention \cite{Wald:1979,Fulling:1987} used here is dictated by the need to regularize the Feynman path integral via an analytic continuation from real $t$ to complex $t$ with negative imaginary part. However, the opposite sign convention is often used \cite{Lifshitz:1980}, which also transforms the Lorentzian into the Euclidean metric.} -- substitution of $t = - \i t_\ins{E}$ in the Lorenzian Green's function to produce a Euclidean one\footnote{The terminology comes from the fact that the Minkowski metric becomes the Euclidean one under the Wick rotation.} $G^{\beta}(t_\ins{E},\bs{x},\bs{x'})=-\i G^F(\i t_\ins{E},\bs{x},\bs{x'})$, which is periodic in the Euclidean compactified time $t_\ins{E}$ with the period $\beta$. Here we took into account the homogeneity of the Green's function in time $t_\ins{E}$ and inhomogeneity in space in view of the presence of interfaces or boundaries. Due to the periodicity in Euclidean time, one can decompose $G^{\beta}(t_\ins{E},\bs{x},\bs{x'})$ in a Fourier time series:
\be
\label{FT:Green-thermal}
G^{\beta}(t_\ins{E},\bs{x},\bs{x'})={1\over\beta}\sum_{n=-\infty}^{\infty}\widehat{G}(\zeta_n;\bs{x},\bs{x'}) \, e^{\i\zeta_nt_\ins{E}},
\ee
where $\widehat{G}(\zeta_n;\bs{x},\bs{x'}) = \int_{0}^{\beta}{G^{\beta}(t_\ins{E},\bs{x},\bs{x'}) \, e^{- \i\zeta_nt_\ins{E}} \d t_{\ins{E}}}$ are the Fourier components -- the solutions of \eqref{eqhatG} -- and $\zeta_n={2\pi n / \beta}$ the Matsubara frequencies. Stress tensor of EM field and other physical quantities can be computed from the derivatives of the Green's function \eqref{FT:Green-thermal} in the limit of coincident points $t_\ins{E}=0$, $\bs{x}=\bs{x'}$. Note that although formally the resulting expressions are proportional to the temperature $T$ as \eqref{FT:Green-thermal}, they do not vanish when $T \rightarrow 0$, because in this limit the distance between Matsubara frequencies $\zeta_n$ shrinks and therefore the summation can be replaced with integration $T\sum_n (\dots)\to \int \frac{\d\omega}{2\pi} (\dots)$ \cite{Lifshitz:1980}, which generally produces a non-vanishing and temperature independent quantity proportional to the Planck constant $\hbar$ (in SI units).

Next, note that because of the Wick rotation, the frequency $\omega$ in the dielectric permittivity becomes purely imaginary -- with the help of \eqref{KK-relation} we can express it as
\ba\label{epsilon-Wick}
\varepsilon(\i\zeta;\bs{x})=1+{2\over \pi} \int\limits_{0}^{\infty}\d\omega'\, {\omega'\,\Im[\varepsilon(\omega';\bs{x})]\over\omega^{\prime 2}+\zeta^2},
\ea
i.e. indeed the results throughout the paper will be expressed in terms of $\Im[\varepsilon(\omega;\bs{x})]$ in accordance to the fluctuation-dissipation theorem. Due to \eqref{epsilon-Wick}, the Fourier coefficients in \eqref{FT:Green-thermal} are symmetric, $\widehat{G}(\zeta_{-n};\bs{x},\bs{x'}) = \widehat{G}(\zeta_n;\bs{x},\bs{x'})$. In the context of the above transition from zero to a finite temperature Green's function, it must be mentioned that the Lifshitz theory is valid when fluctuations are predominantly quantum \cite{Barash:1975,Barash:1984}, i.e. it must be $k_{\ins{B}} T \ll \hbar \omega$, where $\omega$ are the frequencies in the neighborhood of absorption $\omega_{a}$, since according to the fluctuation-dissipation theorem the main contribution to the interaction (and hence stresses) comes from this range of frequencies. For example, for water $\omega_{a} \sim 2 \pi \cdot 10^{15} \, \mathrm{rad \, s^{-1}}$ \cite{Israelachvili:2011} and thus we get the upper bound temperature on the order of $T \sim 10^{3} \, \mathrm{K}$. Since the frequency $\omega$ in the above inequality $k_{\ins{B}} T \ll \hbar \omega$ is related to some characteristic time scale $\tau = 2\pi/\omega$, the same inequality must be satisfied when applied to the time scale of the inhomogeneity, which in our case is related to the interface width $\tau = w/c$.

In classical electrodynamics the total Helmholtz stress tensor in isotropic dielectric media is a sum of the isotropic mechanical (elastic) stress tensor of the medium $\sigma_{ij}^\ins{(m)} = - p^\ins{(m)} \delta_{ij}$\footnote{We do not use notation for Kroneker delta-function different from Diract delta-function because they are, in essence, the same when viewed as arising from normalization of orthogonal modes with the only difference that discrete indices are replaced with continuous ones -- both can be viewed together in the DeWitt notation.} and the EM part $\sigma_{ij}$ \cite{Stratton:2007}:
\ba\label{sij}
\sigma_{ij}^\ins{(tot)}=\sigma_{ij}^\ins{(m)}+\sigma_{ij}, \
\sigma_{ij}=\varepsilon E_i E_j-{1\over 2}\delta_{ij}\Big(\varepsilon-\rho {\partial\varepsilon\over\partial\rho}\Big) E_k E^k,
\ea
where $\rho$ is the medium mass density. The eletrostriction effect, which is described by the $\rho\partial_\rho\varepsilon$ term in brackets, is of the same order as other terms, as can be seen from the Clausius-Mossotti formula \cite{Debye:1929,Stratton:2007} $\rho \, {\partial\varepsilon / \partial\rho} = \left(\varepsilon - 1\right)\left(\varepsilon + 3\right)/3$ leading to $\varepsilon - \rho \, {\partial\varepsilon / \partial\rho} = - \left(\varepsilon^{2} - 2 \varepsilon - 2\right)/3$. The stress tensor $\sigma_{ij}$ in quantum electrodynamics can be obtained from the classical one \eq{sij} by substitution
\be
E_i(t,\bs{x}) E_j(t',\bs{x}') \to \langle \mathrm{E}_i(t,\bs{x}) \mathrm{E}_j(t',\bs{x}')\rangle,
\ee
where $\langle \mathrm{E}_i \mathrm{E}_j\rangle$ is a quantum average of the product of field operators\footnote{As noted by Lifshitz and Pitaevskii \cite{Lifshitz:1980}, the resulting expression coincides with the classical Maxwell stress for a constant electric field -- this, however, does not imply the existence of a general expression for the stress tensor in an arbitrary variable EM field and in dispersive and dissipative media.}; the resulting expression is valid for the fluctuating EM field, which is in a thermodynamic equilibrium with the medium. Since we are working in the frequency space as dictated by the form of the Maxwell equation \eqref{eqhatG}, we need to compute the quantum average $\langle \mathrm{E}_i(\zeta,\bs{x}) \mathrm{E}_j(\zeta,\bs{x})\rangle$ for every mode $\zeta_n$. In terms of the thermal Green's function its regularized version (achieved by point-splitting) takes the form
\be
\langle \mathrm{E}_i(\zeta,\bs{x}) \mathrm{E}_j(\zeta,\bs{x}')\rangle =\nabla_{i}\nabla_{j'}\widehat{G}(\zeta;\bs{x},\bs{x'}),
\ee
where the Green's function has a meaning of a quantum average
\ba
\label{G-hat}
\widehat{G}(\zeta;\bs{x},\bs{x'})=\langle \mathrm{A}_0(\zeta;\bs{x}) \mathrm{A}_0(\zeta;\bs{x}')\rangle.
\ea

The quantum mean values are typically prone to UV divergencies. For our purposes it is convenient to use a point-splitting regularization as it helps us to regularize the quantities we are interested in and, at the same time, we can take an advantage of knowledge of the exact solution for the Green's function. In this approach the quantum mean value of the stress-energy tensor \eqref{tensor:stress-energy} is expressed in terms of the derivatives of the finite temperature Green's function \eqref{FT:Green-thermal}. Namely, for every component $\widehat{\sigma}_{ij}(\zeta_n;\bs{x})$ of the Fourier expansion of the stress in the complex frequency $\zeta_n$ modes, similar to \eqref{FT:Green-thermal}, we write the regularized version of \eq{sij}:
\begin{subequations}
\label{hatsigmaij}
\begin{align}
\label{dfn:limit}
\widehat{\sigma}_{ij}(\zeta;\bs{x})
&= \left[\widehat{\sigma}_{ij}(\zeta;\bs{x},\bs{x}')\right] \equiv \lim_{\bs{x}'\rightarrow\bs{x}} g_{j}{}^{j'} \widehat{\sigma}_{ij'}(\zeta;\bs{x},\bs{x}'), \\
\widehat{\sigma}_{ij'}(\zeta;\bs{x},\bs{x}')&={\sqrt{\varepsilon(\i\zeta;\bs{x})\varepsilon(\i\zeta;\bs{x}')}}
\left(\nabla_{i}\nabla_{j'}-{1\over 2}g_{ij'}g^{kk'}\, \nabla_{k}\nabla_{k'}\right)\widehat{G}(\zeta;\bs{x},\bs{x'}) \nonumber \\
&+{1\over 2}g_{ij'}\sqrt{\rho(\bs{x}){\partial\varepsilon(\i\zeta;\bs{x})\over\partial\rho}
\rho(\bs{x'}){\partial\varepsilon(\i\zeta;\bs{x'})\over\partial\rho}} \,
g^{kk'}\, \nabla_{k}\nabla_{k'}\widehat{G}(\zeta;\bs{x},\bs{x'}).
\end{align}
\end{subequations}
Here $g_{ij'}=g_{ij'}(\bs{x},\bs{x'})$ is the spatial parallel transport operator, which becomes trivial and reduces to unity $g_{ij'}(\bs{x},\bs{x'})=\delta_{ij'}$ since we work in the Cartesian coordinates $x,y,z$. Because we use a point-splitting regularization in the $z$-direction orthogonal to the interface, the Euclidean time can be put to zero, $t_\ins{E}=0$, in \eqref{FT:Green-thermal} at the very beginning. Note that the choice of point-splitting in the $z$-direction is just a matter of convenience. Because the treatment of the static system at finite temperature is equivalent to the analysis in the Euclidean signature, other choices will lead to the same results for the renormalized finite physical observables.

\subsection{Stresses between parallel plates} \label{subsec:Lifshitz-divergence}

\begin{figure}[!htb]%
    \centering
    \includegraphics[width=0.55\textwidth]{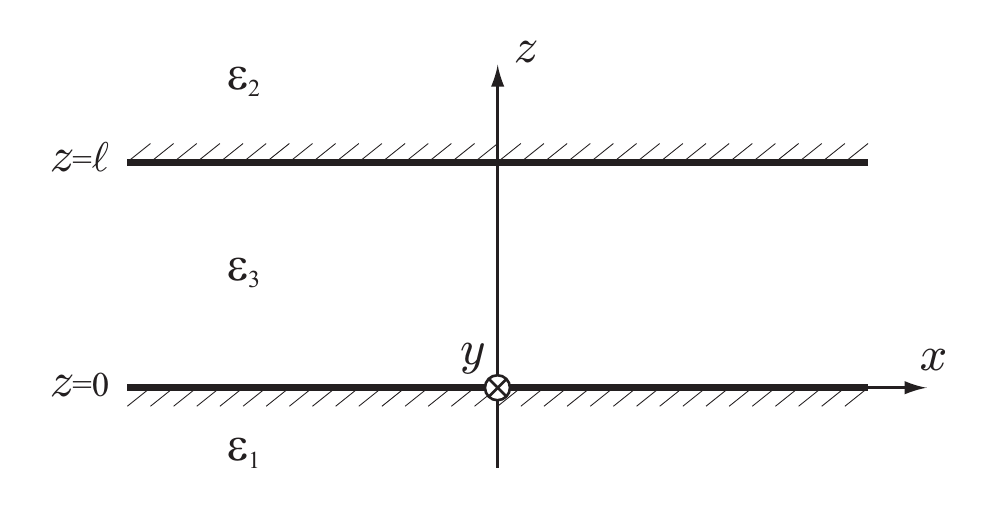}\\[-15pt]
    \caption{The system consists of three parallel layers of dielectrics characterized by the corresponding $\varepsilon(\i\zeta)$. The distance between two interfaces is $\ell$.}
    \label{fig:Lifshitz}
\end{figure}

As an illustration we now briefly derive local stresses for a system consisting of the vacuum gap, $\varepsilon_{3}=1$, between two layers of homogeneous dielectrics, cf. figure~\ref{fig:Lifshitz}. Because $\varepsilon_{3}=1$, there are no electrostriction effects in that layer. This example is a good test of our approach and demonstrates the origin of divergent stresses near the sharp interface.
Because in the $x$ and $y$-directions the problem is homogeneous, it is convenient to expand the Green's function in the Fourier modes
\be\label{hatG}
\widehat{G}(\zeta;\bs{x},\bs{x'})
=\int\limits_{-\infty}^{\infty} {\d^{2} \bs{q}\over (2\pi)^2} \,e^{\i \bs{q} \cdot \bs{x}}
\,\widetilde{G}(\zeta,\bs{q};z,z'),
\ee
where $\bs{q}=(q_x,q_y)$. In every layer with constant $\varepsilon=\varepsilon_{1},\varepsilon_{2},\varepsilon_{3}$, the corresponding $\widetilde{G}$ for every mode satisfies
\ba\label{eqn:Green-Lifshitz}
\varepsilon(\partial^2_{zz}-q^2)\widetilde{G}(\zeta,\bs{q};z,z')
=\delta(z-z')\hh q=|\bs{q}|=\sqrt{q_x^2+q_y^2}.
\ea
On the interfaces between the layers the Green's function has to satisfy the BCs which follow from the Maxwell equations, i.e. the continuity across the interface of the tangential component of the electric field strength $\bs{E}$ and of the normal component of the electric displacement field $\bs{D}$:
\begin{subequations}\label{BCs:sharp}
\begin{align}
\widetilde{G}^{(3)}(z,z')\big|_{z=0} &=\widetilde{G}^{(2)}(z,z')\big|_{z=0}, &
\widetilde{G}^{(2)}(z,z')\big|_{z=\ell} &=\widetilde{G}^{(3)}(z,z')\big|_{z=\ell}, \\
\varepsilon_{3}\partial_z\widetilde{G}^{(3)}(z,z')\big|_{z=0} &=\varepsilon_{1}\partial_z\widetilde{G}^{(1)}(z,z')\big|_{z=0}, &
\varepsilon_{2}\partial_z\widetilde{G}^{(2)}(z,z')\big|_{z=\ell} &=\varepsilon_{3}\partial_z\widetilde{G}^{(3)}(z,z')\big|_{z=\ell},
\end{align}
\end{subequations}
where for conciseness we omitted dependence on $\zeta$ and $q$. The solution to (\ref{eqn:Green-Lifshitz},\ref{BCs:sharp}) in layer $3$ reads:
\ba\label{G3}
\widetilde{G}(z,z')&=-{1\over 2q W{\varepsilon_3}}\Big[
-e^{-q(z+z'-2\ell)}{
(\varepsilon_1-\varepsilon_3)(\varepsilon_2+\varepsilon_3)}\\
&+
(\varepsilon_1-\varepsilon_3)(\varepsilon_2-\varepsilon_3)e^{q(z-z')}
+(\varepsilon_1+\varepsilon_3)(\varepsilon_2-\varepsilon_3)\big(e^{-q(z+z')}-e^{q(z+z')}\big)
\Big],
\ea
where
\ba
W=
(\varepsilon_1+\varepsilon_3)(\varepsilon_2+\varepsilon_3)e^{2q\ell}
-(\varepsilon_1-\varepsilon_3)(\varepsilon_2-\varepsilon_3).
\ea
The Green's function \eqref{G3} can be rewritten as a sum of UV divergent and finite parts:
\ba\label{G3:sum}
\widetilde{G}(z,z')=\widetilde{G}^{\ins{(div)}}(z,z')
+\widetilde{G}^{\ins{(fin)}}(z,z'),
\ea
where the UV divergent part is
\ba
\widetilde{G}^{\ins{(div)}}(z,z')&=-{1\over 2q{\varepsilon_3}}e^{-q|z-z'|}
\ea
and the finite part, e.g. in layer $3$ for $0<z,z'<\ell$, follows from (\ref{G3},\ref{G3:sum}).

Then the renormalized components of the EM stress tensor $\sigma^\ins{(fin)}_{ij}(z)$ in medium $3$ can be computed in the limit of coincident points:
\begin{subequations}
\begin{align}
\sigma^\ins{(fin)}_{xx}(z)&=-{1\over2\beta}\sum_{n=-\infty}^{\infty}{{\varepsilon}(\i\zeta_n)}\int {\d q_x \d q_y \over (2\pi)^2}
\Big\{[\partial_{z}\partial_{z'}+q_y^2-q_x^2]\widetilde{G}^{\ins{(fin)}}(\zeta_n,q;z,z')\Big\}\Big|_{z=z'}, \\
\sigma^\ins{(fin)}_{yy}(z)&=-{1\over2\beta}\sum_{n=-\infty}^{\infty}{{\varepsilon}(\i\zeta_n)}\int {\d q_x \d q_y \over (2\pi)^2}
\Big\{[\partial_{z}\partial_{z'}-q_y^2+q_x^2]\widetilde{G}^{\ins{(fin)}}(\zeta_n,q;z,z')\Big\}\Big|_{z=z'}, \\
\sigma^\ins{(fin)}_{zz}(z)&={1\over2\beta}\sum_{n=-\infty}^{\infty}{{\varepsilon}(\i\zeta_n)}\int {\d q_x \d q_y \over (2\pi)^2}
\Big\{[\partial_{z}\partial_{z'}-q^2]\widetilde{G}^{\ins{(fin)}}(\zeta_n,q;z,z')\Big\}\Big|_{z=z'}.
\end{align}
\end{subequations}
Note that from the general Kramers-Kronig relation \eq{KK-relation} (see also the phenomenological one \eq{formula:Debye} below), it follows that when the real part of the dielectric permittivity $\Re[\varepsilon(\i \zeta_{n};\bs{x})]$, denoted from this point on as simply $\varepsilon$, is evaluated at the imaginary Matsubara frequencies, it stays real. When $x=x'$ and $y=y'$, the integration over $q_x,q_y$  reduces to
\begin{subequations}
\begin{align}
\sigma^\ins{(fin)}_{xx}(z)&=\sigma_{yy}(z)=-{1\over 4\pi\beta}\sum_{n=-\infty}^{\infty}{{\varepsilon}(\i\zeta_n)}\int_0^{\infty} \d q\,q\,
\Big\{\partial_{z}\partial_{z'}\widetilde{G}^{\ins{(fin)}}(\zeta_n,q;z,z')\Big\}\Big|_{z=z'}, \\
\sigma^\ins{(fin)}_{zz}(z)&={1\over 4\pi\beta}\sum_{n=-\infty}^{\infty}{{\varepsilon}(\i\zeta_n)}\int_0^{\infty} \d q\,q\,
\Big\{[\partial_{z}\partial_{z'}-q^2]\widetilde{G}^{\ins{(fin)}}(\zeta_n,q;z,z')\Big\}\Big|_{z=z'}.
\end{align}
\end{subequations}
Substituting here the derived Green's function \eq{G3} we recover the non-retarded limit of the Lifshitz theory \cite{Lifshitz:1956}:
\begin{subequations}
\label{sigmas:Lifshitz}
\begin{align}
\label{sigmazz}
\sigma^\ins{(fin)}_{zz}(z) & ={1\over 8\pi\beta}
\sum_{n=-\infty}^{\infty}\int_0^{\infty} \d q\,q^2{4({\varepsilon_1}-{\varepsilon_3})
({\varepsilon_2}-{\varepsilon_3})
\over({\varepsilon_1}+{\varepsilon_3})({\varepsilon_2}+{\varepsilon_3})e^{2q\ell}
-({\varepsilon_1}-{\varepsilon_3})({\varepsilon_2}-{\varepsilon_3})
}, \\
\sigma^\ins{(fin)}_{xx}(z)&=\sigma_{yy}(z)=-{1\over 8\pi\beta}\sum_{n=-\infty}^{\infty}\int_0^{\infty} \d q\,q^2 \nonumber \\
\label{sigmayy}
&\times {2({\varepsilon_1}-{\varepsilon_3})({\varepsilon_2}-{\varepsilon_3})
+({\varepsilon_1}+{\varepsilon_3})({\varepsilon_2}-{\varepsilon_3})e^{2qz}
+({\varepsilon_1}-{\varepsilon_3})({\varepsilon_2}+{\varepsilon_3})e^{2q(\ell-z)}
\over({\varepsilon_1}+{\varepsilon_3})({\varepsilon_2}+{\varepsilon_3})e^{2q\ell}
-({\varepsilon_1}-{\varepsilon_3})({\varepsilon_2}-{\varepsilon_3})
},
\end{align}
\end{subequations}
where $0 < z < \ell$. Note that, should one think in terms of real photons, the integrals over frequency in \eqref{sigmas:Lifshitz} would have to be limited by the (real) absorption frequency $\omega_{a}$. Instead, we sum up over all (imaginary) Matsubara frequencies $\zeta_{n}$ because this corresponds to summing up w.r.t. all modes, not necessarily real photons (cf. discussion in \S~\ref{subsec:proper-time-regularization}), in the response $\varepsilon(\omega)$ of the medium to the $\delta$-function decomposed in the time Fourier series containing harmonics of all frequencies. Obviously, the pressure in the $z$-direction $\sigma^\ins{(fin)}_{zz}(z)|_{0<z<\ell}$ is finite, i.e. independent of the UV cut-off, and does not depend on $z$; with the change of variable $2 q \ell = p$ one can see that $\sigma^\ins{(fin)}_{zz} \sim \ell^{-3}$. The EM pressure in the gap (or a dielectric film \cite{Dzyaloshinskii:1961}) corresponds to the excess mechanical pressure compared to the pressure outside the gap (film). When $\varepsilon_1>\varepsilon_3$ and $\varepsilon_2>\varepsilon_3$, it leads to attraction between plates. In the case when $\varepsilon_1>\varepsilon_3$ and $\varepsilon_2<\varepsilon_3$ it produces a repulsion force. When the gap (film) is thick, $\ell \rightarrow \infty$, this excess pressure vanishes. As for the tangent components of the pressure $\sigma^\ins{(fin)}_{xx}(z) =\sigma^\ins{(fin)}_{yy}(z)$, regardless how thick the gap is, they diverge as $z$ approaches interfaces at $z=0$ and $z=\ell$ because the integral in \eqref{sigmayy} over the momentum $q$ becomes divergent on the interfaces.

Intuitively, the divergence of the along-the-interface stresses \eqref{sigmayy} can be appreciated on the example of the Casimir problem from the \textit{leakage} of a point continuous spectrum of the along-the-interface wavenumber components $\bs{q}$ of the total wavenumber $\bs{k}=(\bs{q},k_{z})$ when the free space energy $\mathcal{E}^{\ins{vac}}(\infty) = \frac{1}{(2\pi)^{3}} \int{\d^{3}\bs{k} |\bs{k}|} = \frac{1}{(2\pi)^{3}} \int{\d^{2}\bs{q} \int{\d k_{z} \sqrt{q^{2} + k_{z}^{2}}}}$ is subtracted from that of the finite width $\ell$ cavity $\mathcal{E}^{\ins{vac}}(\ell) = \frac{1}{2} \sum_{n,q}{\omega_{n,q}}$. Here $\omega_{n,q} = \sqrt{q^{2} + (n\pi/\ell)^{2}}$ and the summation is performed over all integers $n$, due to discreteness of the numeration of the point spectrum in the $z$-direction, and over all continuous wavenumbers $q$ in the $x$- and $y$-directions. The leakage occurs because of the square root in the dispersion relation $\omega \sim |\bs{k}|$, which, despite being linear, mixes the discrete $k_{z}$ and continuous wavenumbers $\bs{q}$ in the difference $\mathcal{E}^{\ins{vac}}(\ell)-\mathcal{E}^{\ins{vac}}(\infty)$. This decomposition of the wavenumbers $\bs{k}=(\bs{q},k_{z})$ and the dispersion relation $\omega \sim |\bs{k}|$ itself are analogous to the Klein-Gordon model
\begin{align}
\label{eqn:Klein-Gordon}
\phi_{tt} - \phi_{zz} + m^{2} \phi = 0,
\end{align}
the free particle solution $\phi \sim e^{\i (k_{z} z - \omega t)}$ of which yields the dispersion relation of our type, $\omega = \sqrt{k_{z}^{2} + m^{2}}$, i.e. our wavenumbers $q$ play the role of a mass in the Klein-Gordon equation. If we think of equation \eqref{eqn:Klein-Gordon} as governing oscillations of a string, then the second term is a restoring tension force from neighboring points as in the standard wave equation, while the last term is a harmonic oscillator restoring force, which naturally grows with mass $m$, or equivalently with the wavenumber $\bs{q}$ in our case. As explained above, this term can be unbounded due to sharp interface approximation when neglecting the microscopic structure of matter and leaks into the Casimir effect after the free space contribution $\mathcal{E}^{\ins{vac}}(\infty)$ is subtracted. The same Klein-Gordon model \eqref{eqn:Klein-Gordon} also explains the localization of the divergence near the interface: indeed, in the steady case we find $\phi(z;q) \sim e^{-m z}$ behaving as surface EM waves (plasmons) \cite{Barash:1975}.

\section{Trouble resolution: Scarf for Lifshitz}

\subsection{Current understanding of the role of divergencies}

As we saw in \S \ref{subsec:Lifshitz-divergence}, in the Lifshitz approach \cite{Lifshitz:1956,Dzyaloshinskii:1959,Dzyaloshinskii:1961}
as well as in the original Casimir work \cite{Casimir:1948b}, the stress tensor contains a UV divergent part. This property is reflected in the fact that the stress tensor with separated points $\widehat{\sigma}_{ij}(\zeta;\bs{x},\bs{x}')$ diverges in the limit $\bs{x}\to\bs{x'}$. In accordance to a conventional QFT approach it should be written as a sum of a UV divergent part and a regular one
\ba\label{stress:split:divfin}
\widehat{\sigma}_{ij}(\zeta;\bs{x},\bs{x}')=\widehat{\sigma}^\ins{(div)}_{ij}(\zeta;\bs{x},\bs{x}')
+\widehat{\sigma}^\ins{(fin)}_{ij}(\zeta;\bs{x},\bs{x}')
\ea
with similar decomposition applied to all other quantities such as the energy density $\widehat{\varrho}(\zeta;\bs{x},\bs{x'})$ and the Green's function $\widehat{G}(\zeta;\bs{x},\bs{x'})$. The UV divergent contributions are present even in homogeneous media without boundaries. However, due to equilibrium of the media these formally diverging stresses must be balanced by mechanical ones. The observable regular stresses in a piece-wise homogeneous system, that is described by constant dielectric permittivities $\varepsilon(\i\zeta)$, are typically achieved through subtracting from the total Green's function an auxiliary one \cite{Lifshitz:1956,Dzyaloshinskii:1959,Dzyaloshinskii:1961}
\ba
\widehat{G}^\ins{(div)}(\zeta;\bs{x},\bs{x}') = -{1\over 4\pi \varepsilon(\i\zeta) |\bs{x}-\bs{x}'|}
\ea
associated with an infinite homogeneous medium having the constant dielectric permittivity $\varepsilon(\i\zeta)$ equal to that of the corresponding phase. The resulting finite stresses $\widehat{\sigma}^\ins{(fin)}_{ij}(\zeta;\bs{x},\bs{x}')$ are then obtained by acting on  $\widehat{G}^\ins{(fin)}(\zeta;\bs{x},\bs{x}')=\widehat{G}(\zeta;\bs{x},\bs{x'})-\widehat{G}^\ins{(div)}(\zeta;\bs{x},\bs{x}')$ by the same differential operator as in \eq{hatsigmaij}. Note that separation of the divergent part of $\widehat{G}$ is based not on the divergence of $\widehat{G}$ itself, but the part of it that leads to diverging stresses $\widehat{\sigma}^\ins{(div)}_{ij}$ in \eqref{stress:split:divfin}.
Thus, after this regularization procedure, the obtained finite pressure tensor component \eqref{sigmazz} orthogonal to the interface \eq{sigmazz} exactly reproduces the Lifshitz result. Evidently, this Lifshitz subtraction does not cure all divergencies in inhomogeneous media: the problem is that the local tangent pressure tensor components \eqref{sigmayy} diverge on the interface. The energy density $\varrho(z)$ is also often deemed divergent near the sharp interface \cite{Candelas:1982,Deutsch:1979,Milton:2011a}. This observation is not new as discussed in the Introduction: in the presence of boundaries the regularized Casimir-Derjaguin type stresses of a scalar and other fields are known \cite{Brown:1969,Deutsch:1979,Candelas:1982,Philbin:2010,Xiong:2013,Simpson:2013} to be still divergent. Subtraction of the zero-point energy of each field mode renders the local energy density finite at any non-zero distance to the boundary, but leaves a non-integrable singularity at the boundaries.

As pointed out in the Introduction, the diverging stresses resulting from the EM fluctuations in polarizable media have been given much attention in the literature with the key approaches being the replacement of sharp interfaces by steeply rising potential barriers
\cite{Milton:2011b,Bouas:2012,Murray:2016,Milton:2016,Fulling:2018} or by dielectric constants $\varepsilon(\omega;\bs{x})$ smoothly changing over the length-scale ${w}$, i.e. the consideration of inhomogeneous dielectric media. In the latter formulation of the problem, the UV renormalization prescription is the same at every point in the bulk, and one does not need any extra assumptions as to how to regularize the theory on the interface itself. In the limit when the characteristic width of the interface ${w}$ goes to zero, one should be able to reproduce the sharp interface recovering the classical BCs \eqref{BCs:sharp}, cf. Appendix \ref{apx:Schwartz}. When trying to calculate Casimir-Derjaguin forces in the DLP configuration with the intervening medium being inhomogeneous, the authors of Refs. \cite{Philbin:2010,Xiong:2013} ruled out the feasibility of the Lifshitz regularization and introduced another one, which resulted in divergences on the boundaries with the homogeneous media, an outcome they considered to fall ``outside the current understanding of the Casimir effect.'' Another attempt to regularize the inhomogeneous medium was carried out by Simpson et al. \cite{Simpson:2013} using a modified Lifshitz regularization based on a piecewise homogeneity approximation. They concluded that their piecewise method is not likely to give the correct solution.

Also, Philbin et al. \cite{Philbin:2010} studied the Casimir/Lifshitz `self-force' in an inhomogeneous dielectric, using a simple model for the dielectric permittivity. As far as they have been able to ascertain, this is the first analysis of the Casimir effect for inhomogeneous media, as opposed to piece-wise homogeneous media. Although the standard Lifshitz regularization prescription was formulated with the general case of inhomogeneous dielectrics in mind, an attempt to extract a finite Casimir force per unit volume from the diverging stress failed. These authors also tried a new regularization method, which aims to remove the contribution to the Casimir force arising from the inhomogeneity over short length-scales where the use of macroscopic electromagnetism is unphysical. The new regularization gives a finite Casimir stress inside the inhomogeneous medium in the example considered, but the stress and force per unit volume increase without limit at the boundaries joining the inhomogeneous dielectric to homogeneous regions. Thus, even with their new regularization the divergency problem still persists.

In summary, the current status quo in the literature is the lack of a universal approach to deal with boundary-induced divergencies, regardless of the boundary being sharp or smoothed out. The divergences that occur in the local energy-momentum tensor near surfaces are different from UV divergences (and from the divergences in the total energy). The UV-regularized, i.e. after application of the Lifshitz regularization, energy of interaction between distinct rigid bodies of whatever type is finite, as it is related to observable forces and torques between the bodies, which can be unambiguously calculated. The remaining, after the regularization, divergent local stresses and energy density near surfaces are the result of the sharpness of interfaces -- the idealized BCs, such as the ones of a `perfect conductor', constrain all high frequencies and short wavelengths -- and neither affect the total energy nor lead to additional net forces between interacting bodies. However, as we will show in \S \ref{sec:ST}, the divergent local stresses are closely related to the surface tension of the interface. In what follows we will address the problem of how to extract finite terms from the energy-stress tensor, whose physical meanings are unambiguous.

\subsection{Proper time regularization} \label{subsec:proper-time-regularization}

The problem we will focus on is the polarization of a dielectric medium or, equivalently, the vacuum polarization in a background field given by an electric permittivity $\varepsilon(\omega;\bs{x})$. The calculation of the quantum stress-tensor should be based on some kind of regularization scheme, which allows one to separate the effect of the low-energy modes of the vacuum or polarizable media from the diverging contribution of the high-energy degrees of freedom (UV divergencies). This situation is typical in condensed matter physics, e.g. in the Debye theory of specific heat in solids, where the UV characteristic frequencies are typically on the order of the Debye frequency dictated by the discreteness of the media, i.e. the fact that the minimum wavelength of a phonon is twice the interatomic distance $r_{\ins{m}}$ leading to the definition $\omega_{\mathrm{D}} = c_{s}/r_{\ins{m}}$, where $c_{s}$ is the speed of sound. While in the Lifshitz theory we deal with the Maxwell field instead, in analogy to the Debye theory it may seem that the shortest wavelength is dictated by the intermolecular distance $r_{\ins{m}}$ and hence the associated Debye-like frequency sets the upper limit $\omega_{\mathrm{max}} = c/r_{\ins{m}}$, where $c$ is the speed of light (restoring the SI units for clarity here). All other characteristic frequencies -- the one dictated by the inhomogeneity (interface width) $\omega_{\mathrm{inh}} = c/w$ and the absorption frequency $\omega_{a}$ -- are significantly below $\omega_{\mathrm{max}} = c/r_{\ins{m}}$:
\begin{align}
\omega_{\mathrm{max}} = c/r_{\ins{m}} \gg \omega_{a} \gg \omega_{\mathrm{inh}}.
\end{align}
The absorption frequency is associated with the length-scale $c/\omega_{a}$, which corresponds to the transition between the non-retarded (Derjaguin) limit we consider and the retarded (Casimir) one (cf. \S \ref{subsec:Derjaguin-limit}), e.g. for water it is $\sim 0.1 \, \mathrm{\mu m}$. One can think of the non-retarded limit of the Lifshitz theory as defined by longitudinal quasi-electrostatic modes (also known as virtual photons in the physics folklore \cite{Dzyaloshinskii:1961}), while the retarded one as related to transverse electromagnetic modes (photons). Note that at zero temperature there are time dependent quantum fluctuations of all kinds, but there are no real propagating photons. Physically, we sum up the molecular interactions to compute stresses at a given point $\bs{x}$ over the distances $\ell \lesssim c/\omega_{a}$ in the non-retarded theory, which is justified by the fact that its contributions $\sim \ell^{-3}$ are dominant over the retarded $\sim \ell^{-4}$ ones. When the length-scale of the inhomogeneity $w$ is much shorter than the wavelength $\lambda_{a}$ corresponding to the absorption frequency $\omega_{a}$, retarded effects are smaller by a factor of $\sim \left(w \, \omega_a/c\right)^{4}$ than the non-retarded ones \cite{Lifshitz:1980}.

It is commonly accepted that the Casimir vacuum pressure does not depend on the UV cut-off. However, quantum fluctuations can lead to observable effects that actually depend on ``trans-Planckian'' physics \cite{Volovik:2001qu}, i.e. on the frequencies $\omega \in [\omega_{\mathrm{max}},\infty)$ -- this will prove to be the case for surface tension phenomena (\S \ref{sec:ST}). As we will see, there are two crucial differences from the Debye theory. First, the Lifshitz theory will prove to be not closed (\S \ref{subsec:ST-calculation}), since there is no a priori obvious cut-off $\omega_{\mathrm{max}}$. Second, while above we reason in terms of frequencies as reflected in the fact that we sum over all of them \eqref{FT:Green-thermal} as a response of dielectric permittivity to modes of different frequencies, the energy of interaction of fluctuating EM dipoles decays fast with frequency \cite{Berestetskii:1982}, i.e. the medium becomes transparent at high frequencies and short wavelengths \eqref{formula:Debye}. Hence one would expect weak dependence on the cut-off parameter, in contradistinction to the actual result, e.g. for surface tension \eqref{eqn:ST-total}, which demonstrates dependence on ``trans-Planckian'' physics.

In order to understand the role of UV divergent terms in the stress tensor in the present section we start with Schwinger's proper time regularization of the effective action \cite{Schwinger:1951,Christensen:2019}, because it respects the symmetries of the system and keeps track of the divergent parts. Later on (\S \ref{subsec:point-splitting}) we will also use the point-splitting regularization, which, for purely technical reasons, happens to be more convenient for computations of the finite part of the stress tensor. There is no disagreement between different regularization schemes about finite contributions to the quantum stress tensor, which do not depend on regularization parameters. As for the divergent parts, they are dealt with differently in different regularization schemes. Usually they are removed by the introduction of corresponding counterterms in the bare Lagrangian of the system. Other methods simply omit divergent terms: for example, the zeta-function regularization automatically discards all divergencies, while the dimensional regularization keeps track of logarithmic divergencies and discards power law divergencies.

In our case of a real condensed matter system we have to be vigilant and keep track of all divergent contributions. These UV divergent terms formally appear because the Lifshitz theory is effective, rather than microscopic, and hence deals only with low-energy phenomena of van der Waals forces by accounting for EM interactions of polarizable molecules only. Divergent terms depend on the microscopic structure of the media and can lead to physically observable effects. In particular, as we will show in \S \ref{sec:ST}, the main contribution to surface tension is explained by a specific divergent term which depends on both the Debye-like cut-off parameter and the width ${w}$ of the interface.

\subsubsection{Calculation of stresses}

Schwinger's proper time regularization \cite{Schwinger:1951,Christensen:2019} keeps information about all divergent terms and does not require averaging over all directions, as in the point-splitting method; therefore it is the best choice for our analysis of divergent contributions. Using Schwinger's approach the stress tensor of the system can be computed from the classical action and then quantization. While in our problem the metric is flat $g_{ij}=\delta_{ij}$, $g_{tt}=-1$, and $\sqrt{g(\bs{x})}=1$, we temporarily keep it arbitrary which will be handy in further analysis of divergencies via the calculations based on variation with respect to the metric, in particular, because the resulting Hilbert stress-energy tensor
\ba
T^{\mu\nu}={2\over\sqrt{|g|}}{\updelta \mathcal{S}\over \updelta g_{\mu\nu}} \hh T_{\mu\nu}=g_{\mu\alpha}g_{\nu\beta}T^{\alpha\beta}
\ea
is symmetric as opposed to the canonical stress-energy tensor requiring the Belinfante-Rosenfeld modification to make it symmetric. Note that in this variation 4D metric $g_{\mu\nu}(t,\bs{x})$ depends on all coordinates, and only after the variation we can put $g_{\mu\nu}=\mathrm{diag}(-1,1,1,1)$. As per \eqref{tensor:stress-energy}, the spatial components of the stress-energy tensor define the stress tensor
\ba
\sigma_{ij}=-T_{ij},
\ea
which is used in the theory of continuous media \cite{Landau:1980,Thorne:2017,Misner:1973}.

Given the standard logic of application of the Maxwell equations to general isotropic dielectric media with dispersion and dissipation (\S \ref{subsec:Derjaguin-limit}), i.e. consideration of each field mode with frequency $\omega$ separately in the Fourier space $A_{0}(\omega;\bs{x})$, we first construct the EM free energy for each mode and then sum up over all the modes. Once again, the system we are dealing with is neutral and there are no free charges, so $J^{\mu} \equiv 0$. In order to describe non-retarded effects it is sufficient to consider a vanishing magnetic field, what in the Coulomb gauge is equivalent to the choice $A_i=0$. Since description of the system in equilibrium at finite temperature reduces to the Wick rotated $t=-\i t_\ins{E}$ formulation, the modes are the Fourier transforms over the Euclidean time:
\ba
\label{mode:Matsubara}
A_0(\zeta_n;\bs{x})=\int_0^{\beta}\d t_\ins{E}\,A_0(t_\ins{E},\bs{x})e^{-\i\zeta_n{t_\ins{E}}}.
\ea
Finite temperature $T$ is accounted for by the requirement that the system is periodic in the Euclidean time $t_\ins{E}=x^0$ with the period $\beta$. Then, the classical electromagnetic action is a sum of contributions of all Matsubara modes\footnote{An interested reader may consult Appendix \ref{apx:classical-action} for the derivation in the case of non-dispersive media}
\ba\label{Free2}
\mathcal{S}_\ins{E}=\frac{1}{\beta}\sum_{n=-\infty}^{\infty} \mathcal{S}_n, \ \text{with} \ \mathcal{S}_n=-\frac{1}{2}\int \d^3x \,\varepsilon(\i\zeta_n;\bs{x})\,\partial_k A_0(\zeta_n;\bs{x})\,\partial^k A_0^{*}(\zeta_n;\bs{x}),
\ea
which is negative as it should be for dielectrics \cite{Landau:1984}; note that the free energy $\mathcal{F}$ is related to the action via $\mathcal{S}_\ins{E}=\beta \mathcal{F}$, which is sometimes used for calculation of stresses \cite{Lifshitz:1980}. Actually, this form is a more convenient physical description of dispersive media rather than a time-dependent one. Then the time-dependent formulation can be derived from \eq{Free2} with the inverse of \eqref{mode:Matsubara}:
\ba\label{FS:A0}
A_0(t_\ins{E},\bs{x})=\frac{1}{\beta}\sum_{n=-\infty}^{\infty}A_0(\zeta_n;\bs{x})e^{\i\zeta_n{t_\ins{E}}}.
\ea

For future analysis it is useful to rewrite the components \eqref{Free2} of the action as a functional of a three-dimensional (3D) metric $g_{ij}(\bs{x})$
\ba\label{Fn}
\mathcal{S}_n=-\frac{1}{2}\int \d^3x \sqrt{g} \,\varepsilon(\i\zeta_n;\bs{x})\, g^{ij}
\nabla_i A_0(\zeta_n;\bs{x}) \nabla_j A_0^*(\zeta_n;\bs{x})
\ea
and the total action $\mathcal{S}^\ins{(tot)}$ in the following 3D-covariant form
\ba\label{Ftot}
\mathcal{S}^\ins{(tot)}=\mathcal{S}^\ins{(m)} + \mathcal{S}^\ins{(em)},
\ea
where $\mathcal{S}^\ins{(m)} = \beta \int \d^3x\,\sqrt{g}\,\mathcal{L}^\ins{(m)}(\rho)$ is the `bare' action of classical matter in the absence of external electric field, $\mathcal{L}^\ins{(m)}(\rho)$ the `bare' Lagrangian density as a function of the mass density $\rho$, $g_{ij}$ an artificial 3D metric, and $g=\det(g_{ij})$. Variation of the metric $\updelta g_{ij}$ can be interpreted as a result of deformation of a fixed mass volume element of the matter and, therefore, the dependence of $\varepsilon$ on density $\rho$ can be replaced\footnote{Consider a deformation of a small volume with fixed number of molecules in a crystal as an example.} with that on the metric $g$.

After the Wick rotation, due to the change from the Minkowski to Euclidean action (cf. Appendix \ref{apx:classical-action}), it happens that the corresponding Euclidean stress-energy tensor on static backgrounds has the spatial components
\ba\label{tensor:stress-energy:E}
T_\ins{E}{}_{ij}=-T_{ij}=\sigma_{ij}.
\ea
This trivial observation appears to be useful in application to the Lifshits theory and Matsubara approach, which deal with systems in equilibrium and quantities defined as functions of imaginary frequencies. Then for the equilibrium system in question we can derive the stress tensor $\sigma_{ij}$ by variation over time-independent 3D metric. To this end note that if the Lagrangian density $\mathcal{L}[g(t_{\ins{E}},\bs{x}),A_{0}(t_{\ins{E}},\bs{x})]$ is a function of metric components $g_{0\nu}$ and $g_{ij}$, then its functional variations with respect to the spatial part of the metric satisfy the following equivalence relation
\ba
\int_0^\beta \d t_\ins{E} \int \d^3x {\updelta\mathcal{L}[t_\ins{E},\bs{x}]\over\updelta g_{ij}(t'_\ins{E},\bs{x}')}
\Leftrightarrow \int \d^3x {\updelta\mathcal{L}[\bs{x}]\over\updelta g_{ij}(\bs{x}')};
\ea
here on the left hand-side the metric is a function of time and space, while on the right hand-side it is considered as a function of space only. Given the form \eqref{tensor:stress-energy:E}, the classical EM Hilbert stress tensor is obtained by variation of the action over the 3D metric
\ba
\sigma^{ij}={2\over \beta \sqrt{g}}{\updelta \mathcal{S}_{\ins{E}}\over\updelta g_{ij}} \hh \sigma_{ij}=g_{ik}g_{jl}\sigma^{kl} .
\ea
In the general case, the dielectric permittivity $\varepsilon(\i\zeta;\bs{x})$ can also be a function of density $\rho$, it implicitly depends on the metric as well. Under deformations of an elementary volume $\sqrt{g} \, \d^3x$, which keep the number of atoms in it fixed, the total mass in this volume is constant and hence
\ba
\updelta [\sqrt{g} \, \rho]=0=\rho \updelta\sqrt{g}+\sqrt{g}\updelta\rho={1\over 2}\rho g^{ij}\sqrt{g}\updelta g_{ij}+\sqrt{g}\,\updelta\rho.
\ea
Using this relation we get
\ba
{\updelta\rho\over\updelta g_{ij}}=-{1\over 2}\rho g^{ij}
\ea
and therefore
\ba
{\updelta\varepsilon\over\updelta g_{ij}}={\updelta\varepsilon\over\updelta\rho}{\updelta\rho\over\updelta g_{ij}}=-{1\over 2}\rho {\updelta\varepsilon\over\updelta\rho}g^{ij}.
\ea
At the end, we obtain for the EM tensor, in consistency with the expressions elsewhere
\cite{Landau:1984},
\begin{align}
\label{sigmaij}
\sigma_{ij}=\frac{1}{\beta}\sum_n \widehat{\sigma}_{ij}(\zeta_n), \ \text{where} \ \widehat{\sigma}_{ij}(\zeta)=\varepsilon \nabla_i A_0 \nabla_j A_0^*-{1\over 2}g_{ij}\Big[\varepsilon-\rho {\partial\varepsilon\over\partial\rho} \Big] \nabla_k A_0 \nabla^k A_0^*,
\end{align}
and the term proportional to the derivative of $\varepsilon$ w.r.t. the density accounts for the electrostriction effects.

Variation of the total action \eqref{Ftot} over the metric produces
\ba
\sigma^{ij}{}^\ins{(tot)}=\frac{2}{\beta \sqrt{g}}\frac{\updelta \mathcal{S}^\ins{(tot)}}{\updelta g_{ij}},
\ea
where we perform variation with respect to the metric $g_{ij}$ with lower indices. For the classical free energy \eq{Ftot} we then deduce
\ba\label{sigma-tot}
\sigma_{ij}^\ins{(tot)}=\sigma_{ij}^\ins{(m)}+\sigma_{ij},
\ea
where $\sigma_{ij}^{\ins{(m)}}=-p^{\ins{(m)}} g_{ij}$ is the mechanical (elastic) stress tensor with $p^{\ins{(m)}}(\rho) = \mathcal{L}^{\ins{(m)}} - \rho \, \partial_{\rho} \mathcal{L}^{\ins{(m)}}$ having the meaning of the `bare' pressure of the medium. After the variation one can put $g_{ij}=\delta_{ij}$ and then recover \eq{sij}. The free energy \eq{Ftot} can be formally treated as a Euclidean action \cite{Gibbons:1977}, which in our case reduces to a sum over Matsubara frequencies of 3D actions. Then QFT methods can be applied to its quantization. This action is the only effective functional we need to know as long as we are working in the framework of the Lifshitz theory -- this approach is exactly equivalent to the Lifshitz theory in the non-retarded limit.

In order to compute stresses of quantum fluctuating electromagnetic field $\langle \sigma_{ij}\rangle$ one has to replace
\ba
\nabla_i A_0 \nabla_j A_0^* \to \langle\nabla_i \mathrm{A}_0 \nabla_j \mathrm{A}_0^*\rangle,
\ea
where, as before, on the right-hand side $\mathrm{A}_0$'s are to be understood as the field operators. The quantum mean value $\langle \mathrm{A}_0(\zeta;\bs{x}) \mathrm{A}_0^*(\zeta;\bs{x'})\rangle=\widehat{G}(\zeta;\bs{x},\bs{x'})$ is the Euclidean Green's function in 3D and we have
\ba
\langle\nabla_i \mathrm{A}_0(\zeta;\bs{x}) \nabla_j \mathrm{A}_0^*(\zeta;\bs{x})\rangle=\left.\nabla_i\nabla_{j'} \widehat{G}(\zeta;\bs{x},\bs{x'})\right|_{\bs{x'}=\bs{x}}.
\ea
The stress tensor takes the form \eq{hatsigmaij} symmetrized in $\bs{x}$ and $\bs{x}'$. Since in the limit of coincident points it diverges, it should be regularized and local UV divergent terms extracted. For computation of the finite part of the stress we will use the point splitting method, a preferred choice in the case of exactly solvable models for purely technical reasons. On the other hand, the structure of UV divergent terms in media with inhomogeneous dielectric permittivity is more transparent if we use regularization that keeps track of all possible divergencies and respects symmetries of the system. Proper time cut-off regularization is the best for this particular purpose.

The Euclidean Green's function $\widehat{G}(\zeta;\bs{x},\bs{x'})$ satisfies equation \eq{eqhatG}, which can be written in the form
\ba\label{eqG}
\widehat{O}\widehat{G}(\zeta;\bs{x},\bs{x'})=\delta(\bs{x}-\bs{x'}),
\ea
where
\ba\label{op1}
\widehat{O}=\delta^{ij}\,\partial_i \,\varepsilon\, \partial_j \hh \varepsilon=\varepsilon(\i\zeta;\bs{x})
\ea
is a Hermitian operator, $\int \d \bs{x}\,\Phi_1 \widehat{O} \Phi_2 =  \int \d \bs{x}\,\Phi_2 \widehat{O} \Phi_1$, in the inner product $\int \d \bs{x}\, \Phi_1 \Phi_2$ with a unit measure; here $\Phi \equiv A_0(\zeta;\bs{x})$. Operator \eqref{op1} is of the Laplace type and has the structure
\ba\label{operator:O}
\widehat{O} = \mathrm{g}^{ij}\,\partial_i \partial_j+\eta^i \partial_j
\ea
with
\ba
\mathrm{g}^{ij}=\varepsilon\delta^{ij}, \ \eta_{i} = \frac{\partial_{i} \varepsilon}{\varepsilon}, \ \eta^{i} = \mathrm{g}^{ij}\,\eta_{j},
\ea
for which powerful methods of spectral geometry \cite{Gilkey:1975,Gilkey:1994} are available. On the physical side, since the matrix $\mathrm{g}^{ij}$ in front of the second derivatives in \eqref{operator:O} can be interpreted as an effective 3D metric felt by the electric field, the electric force lines are bent (because $\varepsilon$ is inhomogeneous) exactly as if the electric field lives in a curved space with the metric $\mathrm{g}_{ij}=\varepsilon^{-1}\delta_{ij}$. On the technical side, in the path integral approach to quantization, one has to compute functional integral over the vector potential $A_0$. This requirement unambiguously fixes the inner product of two fields $A_0{}^\ins{(1)}$ and $A_0{}^\ins{(2)}$:
\ba
(A_0{}^\ins{(1)}, A_0{}^\ins{(2)})=\int \d^3x\, A_0{}^\ins{(1)}(\bs{x}) A_0{}^\ins{(2)}(\bs{x}).
\ea
Local field $A_0$ redefinition can change the form of the corresponding operator, but it also modifies the measure in the above scalar product and thus in the path integral. In the Feynman path integral approach, a local field redefinition changes the local measure in the corresponding functional integral, which results in the diverging, so-called $\delta(0)$ terms in the effective action. These terms are of UV nature and in renormalizable theories are compensated by rescaling of coupling constants in the bare Lagrangian \cite{Apfeldorf:2001}. However, the finite terms in the first order of $\hbar$ are insensitive to a local fields redefinition.

As in this work we are interested in the contribution of UV divergent terms to stresses and, in particular, to surface tension (\S \ref{sec:ST}), we have to work with the original operator $\widehat{O}$ in order to extract properly the terms that depend on the UV cut-off. As for the cut-off independent finite contributions, one can safely use regularizations such as point-splitting, zeta-function, dimensional, Pauli-Villars, and other well established in QFT techniques. Since for the exactly solvable Scarf potential we are able to compute the Green's function analytically (\S \ref{subsec:Scarf}), the Hadamard representation and point-splitting regularization better suit the computations of regularized finite stresses as will be done in \S \ref{subsec:point-splitting}. On the other hand, UV divergencies are better analyzed via the regularization that respects symmetries of the system while keeping the track of all possible divergent terms. The Schwinger proper-time cut-off regularization is the best choice for this purpose, because it is a direction-independent regularization and does not bring any artificial anisotropy to the considered quantities in contrast to the point splitting approach.

With the help of the heat kernel calculations (cf. Appendix \ref{apx:heat-kernel}) along with the use of shorthand notations $\eta^2=\eta_k\eta^k$, $\partial\eta=\partial_k\eta^k$ and the definition of the coincident points limit introduced in \eqref{dfn:limit}, we determine the divergent contributions, which are of interest to us,
\begin{multline}\label{Gij}
G_{ij}^{\ins{(div)}}(\zeta;\bs{x}) = [\partial_i\partial_{j'}\widehat{G}_{\epsilon}(\zeta;\bs{x},\bs{x'})]=-
\frac{1}{(4\pi\varepsilon)^{3/2}}\Big[\frac{1}{3\epsilon^{3/2}}\frac{1}{\varepsilon}\delta_{ij}\\
-\frac{1}{48\epsilon^{1/2}}\delta_{ij} \Big(28\partial\eta+9\eta^2\Big)-\frac{1}{24\epsilon^{1/2}}\Big(4\partial_j\eta_i-25\eta_i\eta_j\Big)\Big],
\end{multline}
as well as its contraction
\ba
G^{k\ins{(div)}}_{k}(\zeta;\bs{x}) = \delta^{ij}[\partial_i\partial_{j'}\widehat{G}_{\epsilon}(\zeta;\bs{x},\bs{x'})]
&=-\frac{1}{(4\pi\varepsilon)^{3/2}}\Big[\frac{1}{\epsilon^{3/2}}\frac{1}{\varepsilon}-\frac{1}{48\epsilon^{1/2}}
\Big(
92\partial\eta-23\eta^2
\Big)
\Big].
\ea
Its substitution into the regularized stress tensor \eq{hatsigmaij} leads to
\ba\label{hatsigmaij4}
\widehat{\sigma}_{ij}^{\ins{(div)}}(\zeta;\bs{x})
&=\varepsilon \left[
G_{ij}^{\ins{(div)}}-\frac{1}{2}\delta_{ij}\left(1-\frac{\rho}{\varepsilon}\frac{\partial\varepsilon}{\partial\rho} \right) G^{k \ins{(div)}}_k
\right].
\ea
It is convenient to collect terms of the same order in the cut-off parameter $\epsilon^{1/2}$,
\begin{subequations}
\label{hatsigmaij5}
\begin{align}
\widehat{\sigma}_{ij}^{\ins{(div)}}(\zeta;\bs{x}) =& \widehat{\sigma}_{ij}^{\ins{(div-l)}}(\zeta;\bs{x}) + \widehat{\sigma}_{ij}^{\ins{(div-s)}}(\zeta;\bs{x}), \ \text{with} \\
\label{sigma-div:leading}
\widehat{\sigma}_{ij}^{\ins{(div-l)}}(\zeta;\bs{x})=&\frac{1}{(4\pi\varepsilon)^{3/2}}\frac{1}{\epsilon^{3/2}}\delta_{ij}\left[-\frac{1}{3}+\frac{1}{2}\left(1-\frac{\rho}{\varepsilon}\frac{\partial\varepsilon}{\partial\rho} \right)\right], \\
\label{sigma-div:subleading}
\begin{split}
\widehat{\sigma}_{ij}^{\ins{(div-s)}}(\zeta;\bs{x})=&\frac{\varepsilon}{(4\pi\varepsilon)^{3/2}}\frac{1}{48\epsilon^{1/2}}\delta_{ij}
\left[28\partial\eta+9\eta^2-\frac{1}{2}\left(1-\frac{\rho}{\varepsilon}\frac{\partial\varepsilon}{\partial\rho} \right)\left(
92\partial\eta-23\eta^2
\right)\right] \\
+&\frac{\varepsilon}{(4\pi\varepsilon)^{3/2}}\frac{1}{24\epsilon^{1/2}}
\left[
4\partial_j\eta_i-25\eta_i\eta_j
\right]+O(\epsilon^{1/2}),
\end{split}
\end{align}
\end{subequations}
and also separate isotropic and anisotropic parts, omitting the terms vanishing in the limit $\epsilon \rightarrow 0$
\ba\label{sigmadiv1}
\widehat{\sigma}_{ij}^{\ins{(div)}}(\zeta;\bs{x}) = \delta_{ij} \, \widehat{\sigma}^{\ins{(iso)}}(\zeta;\bs{x}) + \widehat{\sigma}_{ij}^{\ins{(ani)}}(\zeta;\bs{x}),
\ea
where
\begin{subequations}\label{sigma-split}
\begin{align}\label{sigma-iso}
\begin{split}
\widehat{\sigma}^{\ins{(iso)}}(\zeta;\bs{x})= &\frac{1}{(4\pi\varepsilon)^{3/2}}\frac{1}{\epsilon^{3/2}} \left[-\frac{1}{3}+\frac{1}{2}\left(1-\frac{\rho}{\varepsilon}\frac{\partial\varepsilon}{\partial\rho} \right)\right]  \\
&+\frac{\varepsilon}{(4\pi\varepsilon)^{3/2}}\frac{1}{48\epsilon^{1/2}}
\left[28\partial\eta+9\eta^2-\frac{1}{2}\left(1-\frac{\rho}{\varepsilon}\frac{\partial\varepsilon}{\partial\rho} \right)\left(
92\partial\eta-23\eta^2
\right)\right],
\end{split} \\
\label{sigma-ani}
\widehat{\sigma}_{ij}^{\ins{(ani)}}(\zeta;\bs{x})= &\frac{\varepsilon}{(4\pi\varepsilon)^{3/2}}\frac{1}{24\epsilon^{1/2}}
\left[
4\partial_j\eta_i-25\eta_i\eta_j
\right].
\end{align}
\end{subequations}
The total divergent stresses are calculated then according to
\ba\label{sigma-div}
\sigma^\ins{(div)}_{ij}(\bs{x})={1\over \beta}
\sum_{n=-\infty}^{\infty}\widehat{\sigma}^\ins{(div)}_{ij}(\zeta_n;\bs{x}).
\ea
Even though the split in isotropic and anisotropic parts in \eqref{sigma-split} is not unique, it does not lead to any physical ambiguity because both parts are always present in the sum together. We choose the simplest form for the anisotropic part which, when applied to the plane interface, reduces to the only non-vanishing  component $\widehat{\sigma}_{zz}^{\ins{(ani)}}$. It should be noted that the leading divergent terms \eqref{sigma-div:leading} including the electrostriction contribution are renormalized into the mechanical pressure as will be discussed in \S \ref{subsubsec:proper-time:interpretation}, while the subleading divergent terms \eqref{sigma-div:subleading} dependent on $\eta$ are localized in the interfacial region and vanish when $\varepsilon=\const$. The latter property of subleading stresses differentiates them from the finite ones $\widehat{\sigma}_{ij}^{\ins{(fin)}}$, which stay nontrivial outside the interface as the calculations in \S \ref{subsec:Scarf} will demonstrate. It is also notable that while the finite stresses are present in the sharp interface problem \eqref{sigmas:Lifshitz}, the subleading stresses are clearly not. The latter will prove to be instrumental in explaining the origin of surface tension (\S \ref{sec:ST}).

\subsubsection{Physical interpretation} \label{subsubsec:proper-time:interpretation}

Let us now discuss physical implications of the divergent part \eq{sigmadiv1} of the EM stress tensor $\sigma_{ij}$, which contributes to the total stress tensor
\ba\label{sij1}
&\sigma_{ij}^\ins{(tot)}=\sigma_{ij}^\ins{(m)}+\sigma_{ij}.
\ea
The Lifshitz theory treats media as continuous and hence is an effective theory that describes collective phenomena of microscopic interactions of polarizable molecules separated by the distance $r_{\ins{m}}$. The leading divergency $\sim 1/r_{\ins{m}}^{3}$ at this scale is isotropic and its contribution to a force acting on any volume element is given by the integral of a total derivative, which vanishes for any compact volume element. As mentioned earlier, the standard Lifshitz prescription is to subtract at every point from the computed quantum result the stress of a homogeneous medium with the same (constant) $\varepsilon$. This prescription exactly corresponds to inclusion of the leading divergency to the renormalization of the bare stress term $\sigma_{ij}^\ins{(m)}$. However the subleading $O(\epsilon^{-1/2})$ divergency in \eq{sigma-div:subleading} cannot be omitted as it leads to $\sim 1/(r_{\ins{m}} \ell^2)$ terms and does contribute to the surface tension (\S \ref{sec:ST}) -- to distinguish from the leading UV divergence $O(\epsilon^{-3/2})$ we call it a subleading UV divergence due to its local nature, but not an IR one despite the dependence on the system size $\ell$ (and thus BCs).

On a microscopic level the forces acting on interatomic/molecular distances are phenomenologically described by the Lennard-Jones potential $\varphi_{\mathrm{LJ}} = 4 \upsilon \left[\left(r_{\ins{m}} / r\right)^{12} - \left(r_{\ins{m}} / r\right)^{6}\right]$. The last term in the Lennard-Jones potential -- the van der Waals interaction $\varphi_{\mathrm{vdW}}(r) \sim r^{-6}$ of polarized and polarizable molecules, the nature of which is the EM interactions including the ones of QM nature as per the discussion in the Introduction. The first term in the Lennard-Jones formula $r^{-12}$ is a purely heuristic way of modeling short range $O(r_{\ins{m}})$ quantum repulsion that stops two particles from being in the same location. The power $-12$ was chosen by Lennard-Jones for convenience: in reality, the actual potential may have a mixed polynomial-exponential form \cite{Rackers:2019}. For example, in the case of helium atoms the repulsive interaction between them is caused by a depletion in electron density in the overlap region that descreens the nuclei from each other resulting in internuclear repulsion \cite{Salem:1961}. In general, while the $r^{-12}$ term is often attributed to the Pauli exclusion principle \cite{Rackers:2019} for fermions due to anti-symmetry of their wave functions resulting in an exchange interaction between identical particles (which also applies to bosons) -- this interaction is different from Coulomb electrostatic repulsion and, in fact, stronger on shorter distances and ultimately responsible for the stability of matter \cite{Dyson:1967a,Dyson:1967b,Lenard:1968} -- it could also originate from the Heisenberg principle of uncertainty \cite{Volovik:2003fe}. Regardless of the nature of this short-range interaction, on the larger scales we are interested in, one may consider this interaction as that between billiard balls and hence \textit{isotropic}. In other words, in the context of the macroscopic Lifshitz theory, i.e. when the microscopic details at the length-scales below intermolecular one $r_{\ins{m}}$ are not accounted for, repulsive molecular interactions are treated as that between points and thus lead to locally isotropic contributions to the effective stresses at the length-scales $O(r_{\ins{m}})$. Therefore, in the effective theory we can identify its role as the bare material stress tensor \cite{Dzyaloshinskii:1961}
\be
\sigma_{ij}^\ins{(m)}(\bs{x})=-\delta_{ij} p^\ins{(m)}(\bs{x}),
\ee
where the isotropic pressure $p(\bs{x})$ can depend on the point $\bs{x}$. In equilibrium, this pressure is always adjusted in such a way that the local force density vanishes
\begin{align}\label{equilibrium:general-condition}
\nabla^j\sigma_{ij}^\ins{(tot)}=0
\end{align}
everywhere in the bulk. Because we consider the system with a smooth inhomogeneous interface, which is static, the total local force density vanishes at every point inside the interface too. The terms in \eq{sigmadiv1} which are isotropic $\sim\delta_{ij}$, including the electrostriction terms\footnote{Physically, this absorption of electrostriction to the renormalized isotropic pressure follows from the chemical potential \cite{Dzyaloshinskii:1961,Landau:1984}, which must be constant for media in equilibrium. One may think of the electrostriction stress as analogous to the gravity in the ocean compensated by the mechanical stresses in water: if it is balanced in the $r$-direction, then due to isotropy it must be balanced in the $\theta$-direction as well. Notably, the $\theta$-dependence of the electrostriction stress leads to non-uniform compression of the matter which is stronger near the interface. However, due to isotropy, electrostriction does not contribute to surface tension (\S \ref{sec:ST}).}, should be combined with the bare $\sigma_{ij}^\ins{(m)}$ to produce the renormalized isotropic stress tensor
\be\label{sij5}
\sigma_{ij}^\ins{(ren)}(\bs{x})=-\delta_{ij} p^\ins{(ren)}(\bs{x}).
\ee
This procedure exactly corresponds to renormalization in QFT, where $p^{(m)}$ is formally defined by the microscopic ``bare'' Lagrangian of a classical theory of non-EM interactions. In condensed matter, however, not every system is renormalizable, because the structure of the classical theory is not always the same as the structure of quantum divergencies. This is exactly our case, because the term $\sim\eta_i\eta_j$ in \eq{sigma-div:subleading} is not isotropic and together with finite terms proves to contribute to the surface tension (\S \ref{sec:ST}).

In the light of the above discussion, we rearrange all the contributions to the stress tensor in the following way
\ba\label{sij4}
\sigma_{ij}^\ins{(tot)}=\sigma_{ij}^\ins{(m)}+\sigma_{ij}
=\sigma_{ij}^\ins{(m)}+\sigma_{ij}^\ins{(div)}+\sigma_{ij}^\ins{(fin)}
=\delta_{ij}\left[-p^\ins{(m)} + \sigma^{\ins{(iso)}}\right]+\sigma_{ij}^\ins{(ani)}+\sigma_{ij}^\ins{(fin)},
\ea
where $\sigma_{ij}^\ins{(fin)}$ is defined by \eqref{stress:split:divfin} and will be calculated in \S \ref{subsec:Scarf}, while
\ba
\sigma^{\ins{(iso)}}(\bs{x}) =\frac{1}{\beta} \sum_{n}{\widehat{\sigma}^{\ins{(iso)}}(\zeta_{n};\bs{x})} \ \text{and} \ \sigma_{ij}^{\ins{(ani)}}(\bs{x}) =\frac{1}{\beta} \sum_{n}{\widehat{\sigma}_{ij}^{\ins{(ani)}}(\zeta_{n};\bs{x})}.
\ea
For our system $\varepsilon(\i\zeta;\bs{x})$ is a function of the coordinate $z$ only. Then with the notation
\be\label{dfn:eta}
\eta_i=(0,0,\eta)\hh \eta={\partial_z\varepsilon(\i\zeta;z)\over \varepsilon(\i\zeta;z)}.
\ee
we can write the $zz$-component of the stress tensor as
\ba\label{zz1}
\sigma_{zz}^\ins{(tot)}=\left[-p^\ins{(m)}(z)+\sigma^{\ins{(iso)}}(z)\right] + \sigma_{zz}^{\ins{(ani)}} +\sigma_{zz}^\ins{(fin)}.
\ea
In the $x-y$ plane the stress is diagonal, so we need to know only the $xx$-component
\ba\label{xx1}
\sigma_{xx}^\ins{(tot)}=\left[-p^\ins{(m)}(z)+\sigma^{\ins{(iso)}}(z)\right] +\sigma_{xx}^\ins{(fin)}.
\ea
The quantities in square brackets of (\ref{zz1},\ref{xx1}) correspond to isotropic contributions to the stress tensor. Because all quantities depend only on the $z$-coordinate, the equilibrium condition \eqref{equilibrium:general-condition}
reduces to
\be\label{condition:equilibrium}
\partial_z\sigma_{zz}^\ins{(tot)}=0,
\ee
that is
\be
\label{sij3}
\sigma_{zz}^\ins{(tot)}=\lim_{z \to \pm\infty}\left[-p^\ins{(m)}(z)
+\frac{1}{\beta}\sum_n\frac{1}{(4\pi\varepsilon)^{3/2}\epsilon^{3/2}}\left\{- \frac{1}{3}+\frac{1}{2}\left(1-{\rho(z)\over \varepsilon(\i\zeta_n;z)}{\partial\varepsilon(\i\zeta_n;z)\over\partial\rho(z)}\right)\right\}\right] \equiv - p_0^{\ins{(ren)}},
\ee
where the constant $p_0^{\ins{(ren)}}$ is the value of the renormalized (observable) pressure in both asymptotics $z \to \pm\infty$ and corresponds to the sum of the first two terms in square brackets of (\ref{zz1},\ref{xx1}) in the limit $z \to \pm\infty$. As for the physical interpretation of \eqref{sij3}, note that in liquids and solids $p^\ins{(m)}>0$ since it originates from the repulsion between molecules, while the added EM terms are generally positive as they originate from attraction forces; altogether, we get $p_0^{\ins{(ren)}}<0$ since liquids and solids are a condensed matter (as opposed to gases), i.e. due to the cohesive nature of liquids and solids their pressure  $p_0^{\ins{(ren)}}$ must be negative. Note that $p(z)$ is a function of the $z$-coordinate only, but this dependence always compensates the $z$-dependence of $\sigma_{zz}^\ins{(div)}+\sigma_{zz}^\ins{(fin)}$. Thus
we obtain
\ba\label{sigmaxx-tot}
\sigma_{xx}^\ins{(tot)}=-p_0^{\ins{(ren)}}- \sigma_{zz}^{\ins{(ani)}} +\sigma_{xx}^\ins{(fin)} - \sigma_{zz}^\ins{(fin)}.
\ea
This tangent to the interface pressure does depend on $z$, but in the asymptotic regions ${z \to \pm\infty}$ it approaches the same constant
\be\label{sij3xx}
\sigma_{xx}^\ins{(tot)}=\sigma_{yy}^\ins{(tot)} \to - p_0^{\ins{(ren)}}.
\ee

The nature of the divergent part $\sigma_{ij}^\ins{(div)}(z,z')$ when $z' \rightarrow z$ requires an additional clarification. In QFT it is usually omitted or combined with the bare stress $\sigma_{ij}^\ins{(m)}(z,z')$ of the classical theory of non-EM intermolecular interactions to produce a renormalized stress $\sigma_{ij}^\ins{(ren)}(z,z')$. The remaining part, which cannot be absorbed into the bare stresses, is then interpreted as the observable stress. This renormalization procedure works well for renormalizable theories, when all quantum UV divergencies can be absorbed in the redefinition of parameters of a ``bare'' classical theory. Following this approach the first choice is to combine all $\epsilon^{1/2}$-dependent, including subleading, terms into the renormalized stress tensor
\ba\label{sij2}
\sigma_{ij}^\ins{(ren)}(z,z')=\sigma^\ins{(m)}_{ij}(z,z')+\sigma^\ins{(div)}_{ij}(z,z').
\ea
Although condensed matter systems may not be renormalizable, it is logical to consider all microstructure dependent terms as a renormalized mechanical stress tensor \eqref{sij5}. As one can see from \eq{sij4} and \eq{hatsigmaij5}, divergent terms of quantum EM fluctuations consist of a leading universal part $\sigma_{ij}^{\ins{(div-l)}}$ (discarded by Lifshitz \cite{Lifshitz:1956} due to being independent of the distance $\ell$ between the bodies) and a subleading part $\sigma_{ij}^{\ins{(div-s)}}$, which depends on the EM properties of the media. The other choice  would be to combine with the bare mechanical stresses only the leading divergent terms to produce the renormalized mechanical stress tensor. In both approaches, far away from the interface the renormalized stress tensor is taken in the form \eqref{sij5}. In the second approach the subleading divergent part depends on both the UV cut-off $\epsilon^{1/2}$ and the gradient of $\varepsilon$ thus leading to observable effects such as surface tension (\S \ref{sec:ST}). The presented here scheme of regularization goes in parallel with the standard in QFT renormalization and can be formally elevated to the renormalization of a bare (material) Lagrangian $\mathcal{L}^{\ins{(m)}}$, which in our case cannot be written explicitly, by adding the Lagrangian with counterterms $\mathcal{L}^{\ins{(CT)}}$
\begin{align}
\mathcal{L}^{\ins{(m)}} = \mathcal{L}^{\ins{(ren)}} + \mathcal{L}^{\ins{(CT)}},
\end{align}
as was done in some model problems such as for scalar fields and $\delta$-potentials \cite{Graham:2002fw,Graham:2003ib}. In spirit, the exploited renormalization is along the lines of the original Wilsonian approach \cite{Wilson:1971a,Wilson:1971b,Wilson:1983}, which is always defined with a physical cut-off, so that there is no fundamental difference between renormalizable and non-renormalizable theories. The non-renormalizable scales in our problem appear to be molecular $\epsilon^{1/2} \sim r_{\ins{m}}$, which are not sources of any trouble if we are interested in the physics at scales above the cut-off as in the Lifshitz theory.

It should be mentioned that observations similar to ours were made by Ravndal \cite{Ravndal:2000kn} in the context of calculations of the Casimir energy between parallel plates without taking into account a dispersion. Namely, if the Casimir energy of a scalar field between the plates separated by a distance $\ell$ with pure Dirichlet boundary conditions is considered, the regularized energy density consists of three terms: (i) the leading UV divergent term $3\Lambda^4/(2\pi^2)$, where $\Lambda$ is the cut-off wavenumber beyond which the theory is inapplicable; (ii) the subleading UV divergent term $\Lambda^3/(4\pi \ell)$; (iii) and the pure Casimir term $-\pi^2/(1440 \ell^4)$. For the Neumann boundary conditions the leading UV divergent and Casimir terms are the same, while the subleading term appears with the opposite sign. In the case of the EM field between perfectly conducting parallel plates, the transverse electric multipoles correspond to the Dirichlet modes while the transverse magnetic multipoles correspond to the Neumann modes \cite{Candelas:1982,Brevik:2000hk,Lutken:1984qr}. It happens that for perfectly conducting plane plates the EM stress-energy tensor divergencies cancel and only finite Casimir effect survives \cite{Candelas:1982}. However, this cancellation is not universal, but applicable only to Casimir effect with retardation as an artifact of idealized BCs in nondispersive media. In a physically realistic system of dielectrics with dispersion the role of magnetic field is negligible because magnetic permeability of dielectrics is almost the same as that of the vacuum, i.e. an interface is not felt by a magnetic field. Therefore, there is no compensation of subleading UV divergent terms coming from quantum fluctuations of interacting electric dipoles which depend on an inhomogeneity of the system. As we will demonstrate, this effect has observable physical consequences and together with the finite van der Waals contribution eventually leads to the surface tension of the interface. Our goal is to calculate this contribution and explain the origin of surface tension in the framework of the Lifshitz theory (\S \ref{sec:ST}).

\subsection{Point splitting regularization} \label{subsec:point-splitting}

As in the previous section, we assume that the permittivity of a dielectric depends only on the $z$-coordinate and approaches asymptotically a constant greater than one at $z \to -\infty$ and one at $z \to +\infty$. We will use the point-splitting regularization in the $z$-direction, so that in the Green's function \eq{FT:Green-thermal} we put $t_\ins{E}=0$, $x=x'$, and $y=y'$. The resulting  Green's function $\widetilde{G}(\zeta,q;z,z')$ defined via \eq{hatG} for every mode depends only on $q$ and satisfies the equation\footnote{Because from now on we will work mostly with the Fourier transforms of all quantities, we implicitly assume that they depend on the Matsubara frequencies $\zeta_n$ and $q$, and, for simplicity, we omit these arguments in functions. Thus, in these notations we have $\varepsilon(\i\zeta;\bs{x})\equiv\varepsilon(\bs{x})$, $\widehat{G}(\zeta;\bs{x},\bs{x'})\equiv\widehat{G}(\bs{x},\bs{x'})$, etc.}
\ba
\label{eq:Schrodinger}
\big[\partial_{z}\,\varepsilon(\i\zeta;z)\partial_{z}-q^2\varepsilon(\i\zeta;z)\big]
\widetilde{G}(\zeta,q;z,z')
=\delta(z-z')\hh q=\sqrt{q_x^2+q_y^2}.
\ea
Let $\varepsilon(\i\zeta;z)$ change considerably only in the layer of width ${w}$ near $z=0$. Then we can use this length-scale to introduce dimensionless quantities $\overline{z}$, $k$, $\overline{\mathcal{D}}$, $\widehat{\tau}_{ij}$ via\footnote{Because $\varepsilon$ is already dimensionless, we use the same letter for the functions $\varepsilon=\varepsilon(\overline{z})=\varepsilon(\i\zeta;{w}\overline{z})$. It will not lead to any confusions in computations.}
\ba\label{scaling:width}
z={w}\overline{z}
\hh
q={k\over {w}}
\hh
\ea
\ba\label{calD}
\overline{\mathcal{D}}(\overline{z},\overline{z}')={1\over {w}}\widetilde{G}(\zeta,{w}k;{w}\overline{z},{w}\overline{z}'),
\ea
\ba\label{sigmatau}
\widehat{\sigma}^\ins{(fin)}_{ij}(\zeta_n)={1 \over {w}^3} \, \widehat{\tau}^\ins{(fin)}_{ij}(\zeta_n)
\hh
\widehat{\sigma}^\ins{(div)}_{ij}(\zeta_n)={1 \over {w}^3} \, \widehat{\tau}^\ins{(div)}_{ij}(\zeta_n).
\ea
To remove the first-order derivative in \eq{eq:Schrodinger}, let us rescale $\overline{\mathcal{D}}$ according to
\ba\label{calD}
\overline{\mathcal{D}}(\overline{z},\overline{z}')
={1\over\sqrt{{\varepsilon}{\varepsilon}' }}\mathcal{D}(\overline{z},\overline{z}') .
\ea
Then the Green's function $\mathcal{D}(\overline{z},\overline{z}')$ obeys the simpler equation
\be\label{eqG1}
[\partial^2_{\overline{z}}-k^2-V(\overline{z})]\,\mathcal{D}(\overline{z},\overline{z}')=\delta(\overline{z}-\overline{z}'),
\ee
where $k$ plays the role of an effective mass of the mode and the dimensionless potential $V(\overline{z})$ is given by the expression
\be\label{V}
V={1\over 2}\partial_{\overline{z}}\overline{\eta}+{1\over 4}\overline{\eta}^2\hh
\overline{\eta}={\partial_{\overline{z}}\varepsilon \over\varepsilon }=w\,\eta.
\ee
The Hadamard expansion of the Green's function can be derived using the heat
kernel representation
\be\label{GK}
\mathcal{D}(\overline{z},\overline{z}')=\int_0^{\infty}\d s\, K(s|\overline{z},\overline{z}'),
\ee
where the heat kernel $K(s|\overline{z},\overline{z}')$ corresponding to the one-dimensional operator
\be
\widehat{O}=\partial^2_{\overline{z}}-k^2-V(\overline{z})
\ee
satisfies
\be\label{hatO}
[\partial_s - \widehat{O}] K(s|\overline{z},\overline{z}')=-\delta(s)\delta(\overline{z}-\overline{z}').
\ee
Its formal solution
$
K(s|\overline{z},\overline{z}')=e^{s\widehat{O}}\delta(\overline{z}-\overline{z}')
$
can be represented as a power series in the proper time parameter $s$:
\be
K(s|\overline{z},\overline{z}')=-{1\over(4\pi s)^{1/2}}e^{-k^2 s-{(\overline{z}-\overline{z}')^2\over 4s}}
[a_0(\overline{z},\overline{z}')+a_1(\overline{z},\overline{z}')s+a_2(\overline{z},\overline{z}')s^2+\dots],
\ee
where $a_n(\overline{z},\overline{z}')$ are called the Seeley-DeWitt (aka HAMIDEW or Gilkey-Seeley) coefficients. For the calculation of the divergent parts of the stress tensor we need to know only the first two coefficients $a_0(\overline{z},\overline{z}')$ and
$a_1(\overline{z},\overline{z}')$. For the operator defined by \eq{op1} the zero-order coefficient
\be\label{a0}
a_0(\overline{z},\overline{z}')=1,
\ee
while the first-order coefficient $a_1$ we need to know only in the limit $\overline{z}\to \overline{z}'$:
\be\label{a1}
a_1(\overline{z},\overline{z})=-V(\overline{z}).
\ee
All higher order coefficients $a_n(\overline{z},\overline{z}')$  for $n\ge 2$ do not contribute to the divergent parts of the stress tensor $\sigma_{ij}(\bs{x},\bs{x'})$ in 3D because in the non-retarded regime the EM field reduces to the scalar potential $A_0$, which satisfies equation \eq{eqhatG} with the 3D operator. Our procedure guarantees regularity of the contribution of each mode to the renormalized stresses and energy density. Of course, we have to find a sum over all modes with the Matsubara frequencies $\zeta_{n}$, which may diverge in principle. However, for all realistic materials\footnote{This formula is usually deduced phenomenologically \cite{Jackson:1998}, but has a QM justification \cite{Adler:1962}.}
\begin{align}\label{formula:Debye}
\varepsilon(\omega;\bs{x}) = 1 + \frac{\const_{1}}{1-\i \, \omega/\omega_{\ins{rot}}} + \frac{\const_{2}}{1-\omega^{2}/\omega_{\ins{e}}^{2}},
\end{align}
with $\omega_{\ins{rot}}$ being the rotational relaxation frequency and $\omega_{\ins{e}}$ the main electronic absorption frequency, the summation over Matsubara frequencies $\zeta_n$ converges meaning that all the stresses become regular as soon as we take care of 3D spatial divergencies, because dielectric permittivity at high frequencies approaches to that of the vacuum fast enough. In view of dispersion, the corresponding operator of the EM filed effectively becomes integral-differential when expressed in terms of real time, rather than frequency. Therefore, the UV behavior of the system is governed by the properties of the 3D operator only. This is why the problems related to the coefficient $a_2$ \cite{Bordag:2002}, which describes logarithmic divergencies in 4D QFT, do not arise in the Lifshitz approach.

It is important to keep in mind that the UV divergencies of the stress tensor are stronger than those of the Green's function. Therefore, in the Hadamard expansion of the Green's function one has to keep all leading divergent parts and the first subleading term. Even though this subleading term vanishes at the coincident points limit $\overline{z}=\overline{z}'$, its derivatives do not. So, here we have to include in the divergent part of the Green's function $G^\ins{(div)}$ all the terms that after integration over the momenta $k$ lead to the divergent parts of the stress tensor in 3D \cite{Moretti:1998rs}. Retaining only divergent terms in the integral over $s$ in \eq{GK} and substituting (\ref{a0},\ref{a1},\ref{V}), we get the following structures
\begin{subequations}
\begin{align}
\label{G:div}
\mathcal{D}^\ins{(div)}(\overline{z},\overline{z}')
&=-{1\over 2 k}e^{-k|\overline{z}-\overline{z}'|}\left[a_0(\overline{z},\overline{z}')+a_1(\overline{z},\overline{z}')
{1+k|\overline{z}-\overline{z}'|\over 2k^2}\right],
\\
\partial_{\overline{z}} \mathcal{D}^\ins{(div)}(\overline{z},\overline{z}')
&=-{1\over 2}e^{-k|\overline{z}-\overline{z}'|}\left[-a_0(\overline{z},\overline{z}')-a_1(\overline{z},\overline{z}'){|\overline{z}-\overline{z}'|\over 2k}\right],
\\
\partial_{\overline{z}'} \mathcal{D}^\ins{(div)}(\overline{z},\overline{z}')
&=-{1\over 2}e^{-k|\overline{z}-\overline{z}'|}\left[a_0(\overline{z},\overline{z}')+a_1(\overline{z},\overline{z}'){|\overline{z}-\overline{z}'|\over 2k}\right],
\\
\partial_{\overline{z}}\partial_{\overline{z}'} \mathcal{D}^\ins{(div)}(\overline{z},\overline{z}')
&=-{k\over 2}e^{-k|\overline{z}-\overline{z}'|}\left[-a_0(\overline{z},\overline{z}')+a_1(\overline{z},\overline{z}'){1-k|\overline{z}-\overline{z}'|\over 2k^2}\right].
\end{align}
\end{subequations}
Integration of these expressions over the momenta $k$ is formally divergent, which reflects divergencies of the corresponding quantities in the coordinate space in the limit of coincident points of the divergent part of the Green's function $\widehat{G}(\bs{x},\bs{x}')$ defined in \eqref{eqG}, i.e. the part which leads to divergent stresses:
\begin{subequations}
\begin{align}
\big[\partial_x\partial_{x'}\widehat{G}^\ins{(div)}(\bs{x},\bs{x'})\big]\Big|_{\stackrel{x=x'}{\inds{y=y'}}}
&=\big[\partial_y\partial_{y'}\widehat{G}^\ins{(div)}(\bs{x},\bs{x'})\big]
\Big|_{\stackrel{x=x'}{\inds{y=y'}}}
={1\over 2{w}^3}\int\limits_{0}^{\infty} {\d k\,k \over 2\pi}\,k^2 \mathcal{D}^\ins{(div)}(\overline{z},\overline{z}'),
\\
\big[\partial_z\partial_{z'}\widehat{G}^\ins{(div)}(\bs{x},\bs{x'})\big]
\Big|_{\stackrel{x=x'}{\inds{y=y'}}}&={1\over {w}^3}\int\limits_{0}^{\infty} {\d k\,k \over 2\pi}\,
\partial_{\overline{z}}\partial_{\overline{z}'} \mathcal{D}^\ins{(div)}(\overline{z},\overline{z}').
\end{align}
\end{subequations}
These formulas provide us with a prescription how to extract divergencies of the stress-energy tensor \eq{sigmatau} using the mode representation \eqref{FT:Green-thermal}, which is consistent with the point-splitting approach in the coordinate space.

For computation of the divergent stresses we need to know the divergent part of the unscaled Green's function $\overline{\mathcal{D}}$ \eq{calD}, which comes from \eqref{G:div}, its derivatives $\partial_{\overline{z}}\partial_{\overline{z}'}\overline{\mathcal{D}}^\ins{(div)}$
and $k^2\overline{\mathcal{D}}^\ins{(div)}$. Their explicit form is
\begin{subequations}
\begin{align}
\sqrt{\varepsilon \varepsilon' }\,
\partial_{\overline{z}}\partial_{\overline{z}'}\overline{\mathcal{D}}^\ins{(div)}(\overline{z},\overline{z}')
&=-{k\over 2}e^{-k|\overline{z}-\overline{z}'|}
\left[-1+{(1+k|\overline{z}-\overline{z}'|)\over 2k^2}
\Big(-{1\over 2}\partial_{\overline{z}}\overline{\eta}+{1\over 4}\overline{\eta}^2\Big)\right],\\
\sqrt{\varepsilon \varepsilon' }\,k^2\, \overline{\mathcal{D}}^\ins{(div)}(\overline{z},\overline{z}')
&=-{k\over 2 }e^{-k|\overline{z}-\overline{z}'|}
\left[1-{1+k|\overline{z}-\overline{z}'|\over 2k^2}
\Big({1\over 2}\partial_{\overline{z}}\overline{\eta}+{1\over 4}\overline{\eta}^2\Big)\right],
\end{align}
\end{subequations}
which together produce
\ba
\sqrt{\varepsilon \varepsilon' }\big[\partial_{\overline{z}}\partial_{\overline{z}'}\overline{\mathcal{D}}^\ins{(div)}(\overline{z},\overline{z}')+k^2\, \overline{\mathcal{D}}^\ins{(div)}(\overline{z},\overline{z}')\big]
&={k\over 2}e^{-k|\overline{z}-\overline{z}'|}
\left[-{(1+k|\overline{z}-\overline{z}'|)\over 2k^2} \partial_{\overline{z}}\eta\right].
\ea
Then for the mode contributions \eq{sigmatau} to the diverging part of the stress tensor we have
\begin{subequations}
\begin{align}
\label{tauxx}
\widehat{\tau}^\ins{(div)}_{xx}(\zeta_n)&=-{1\over 16\pi}\int\limits_{0}^{\infty}\d k\,e^{-k|\overline{z}-\overline{z}'|}
\left[k^2+{1\over 2}\left({1\over 2}\partial_{\overline{z}}\overline{\eta}-{1\over 4}\overline{\eta}^2
-{\rho\over\varepsilon}{\partial\varepsilon\over\partial \rho}\partial_{\overline{z}}\overline{\eta}
\right)
(1+k|\overline{z}-\overline{z}'|)
\right],
\\
\label{tauzz}
\widehat{\tau}^\ins{(div)}_{zz}(\zeta_n)&=-{1\over 16\pi}\int\limits_{0}^{\infty}\d k\,e^{-k|\overline{z}-\overline{z}'|}
\left[-2k^2+\left({1\over 4}\overline{\eta}^2
-{1\over 2}{\rho\over\varepsilon}{\partial\varepsilon\over\partial \rho}\partial_{\overline{z}}\overline{\eta}\right)
(1+k|\overline{z}-\overline{z}'|)
\right],
\end{align}
\end{subequations}
where the dependence of $\widehat{\tau}^\ins{(div)}_{ij}$ on the Matsubara frequencies $\zeta_n$ comes through $\varepsilon(\i\zeta_n;\overline{z})$, while the finite stresses for every frequency mode $\zeta_n$ are computed from
\ba\label{taufin}
\widehat{\tau}^\ins{(fin)}_{ij}(\zeta_n)&=\widehat{\tau}_{ij}(\zeta_n)-\widehat{\tau}^\ins{(div)}_{ij}(\zeta_n),
\ea
where
\ba
\widehat{\tau}_{ij}(\zeta_n)=\int_0^{\infty}\frac{\d k k}{4\pi}\left\{\varepsilon \partial_i\partial_j\overline{\mathcal{D}}
-\frac{\varepsilon}{2}\delta_{ij}\Big[
1-\frac{\rho}{\varepsilon}\frac{\partial\varepsilon}{\partial\rho}
\Big]\partial^k\partial_k\overline{\mathcal{D}}\right\}
\ea
and the components of $\widehat{\tau}^\ins{(div)}_{ij}$ are given by \eq{tauxx}-\eq{tauzz}. When written explicitly, the components of the total EM stresses $\widehat{\tau}_{ij}$ are
\begin{subequations}
\label{stresses:tau-all}
\begin{align}
&\widehat{\tau}_{zz}(\zeta_n)=\int_0^{\infty}\frac{\d k k}{4\pi}\left\{\varepsilon \partial_{\overline{z}}\partial_{\overline{z}'}\overline{\mathcal{D}}
-\frac{\varepsilon}{2}\Big[
1-\frac{\rho}{\varepsilon}\frac{\partial\varepsilon}{\partial\rho}
\Big]\big(\partial_{\overline{z}}\partial_{\overline{z}'}\overline{\mathcal{D}}+k^2\overline{\mathcal{D}}\big)\right\},
\\
&\widehat{\tau}_{xx}(\zeta_n)=\widehat{\tau}_{yy}=\int_0^{\infty}\frac{\d k k}{4\pi}\left\{\frac{1}{2}\varepsilon k^2 \overline{\mathcal{D}}
-\frac{\varepsilon}{2}\Big[
1-\frac{\rho}{\varepsilon}\frac{\partial\varepsilon}{\partial\rho}
\Big]\big(\partial_{\overline{z}}\partial_{\overline{z}'}\overline{\mathcal{D}}+k^2\overline{\mathcal{D}}\big)\right\},
\\
&\widehat{\tau}_k^k(\zeta_n)=-\int_0^{\infty}\frac{\d k k}{4\pi}\left\{\frac{\varepsilon}{2}\Big[
1-3\frac{\rho}{\varepsilon}\frac{\partial\varepsilon}{\partial\rho}
\Big]\big(\partial_{\overline{z}}\partial_{\overline{z}'}\overline{\mathcal{D}}+k^2\overline{\mathcal{D}}\big)\right\}.
\end{align}
\end{subequations}
We also have to sum over all Matsubara frequencies in order to get the complete finite stress tensor for a particular dielectric:
\ba\label{sigmatau1}
\sigma^\ins{(fin)}_{ij}={1\over \beta}
\sum_{n=-\infty}^{\infty}\widehat{\sigma}^\ins{(fin)}_{ij}(\zeta_n) = {1 \over \beta \, {w}^3} \, \sum_{n=-\infty}^{\infty}{\widehat{\tau}^\ins{(fin)}_{ij}(\zeta_n)},
\ea
while the diverging part is found from an analogous expression \eqref{sigma-div}. Recall that the finite stress $\sigma^\ins{(fin)}_{ij}$ does not depend on the regularization scheme: both proper time cut-off and the point splitting Hadamard regularization lead to the same finite stresses, while UV divergent terms in different regularization schemes assume different forms. Also, note that the total stresses \eqref{stresses:tau-all} contain the electrostriction contributions, which are isotropic and therefore do not contribute to the local force density.

In order to compute the finite part of the EM stress tensor $\sigma^\ins{(fin)}_{ij}$, in the next section we consider a particular model of a smoothed out interface that can be solved exactly. Among the exactly solvable models it is the one reducing to the Schrodinger equation with the Scarf potential \cite{Scarf:1958,Derezinski:2011}, which satisfies all the properties necessary to account for a smoothed out step-like transition between a dielectric medium and the vacuum or between two dielectric media.

\subsection{Exact solution: the Scarf potential} \label{subsec:Scarf}

\begin{figure}[!htb]
\centering \setlength{\labelsep}{0.0mm}
\begin{tabular}{l l l l}
\hspace{-0.05in}\imagetop{(a)} & \hspace{-0.065in}\imagetop{\includegraphics[width=2.5in]{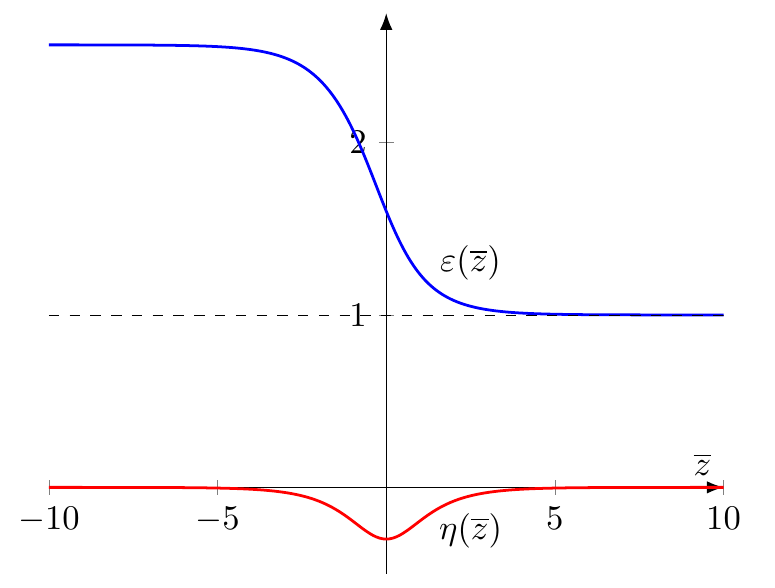}} &
\hspace{0.05in}\imagetop{(b)} & \hspace{-0.065in}\imagetop{\includegraphics[width=2.5in]{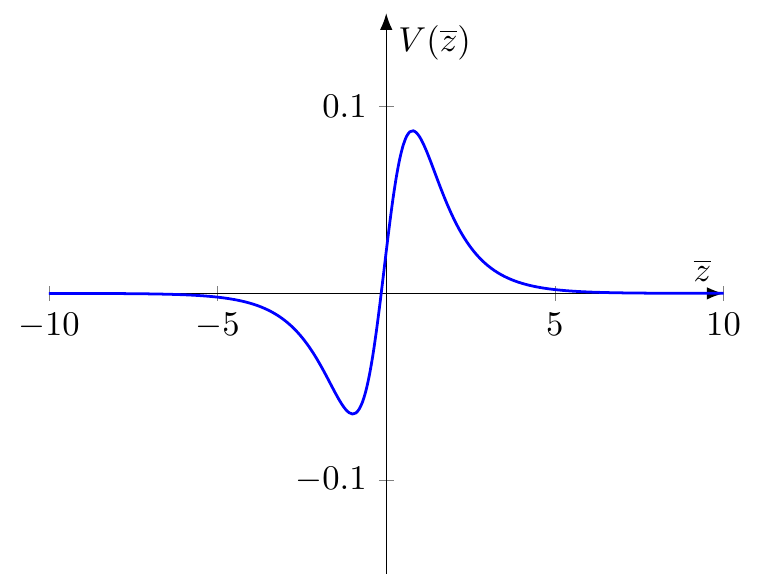}}
\end{tabular}
\caption{(a) The dielectric permittivity $\varepsilon(\overline{z})$ (blue) and $\eta(\overline{z})$ (red) for
    $c=0.15$ ($\varepsilon_{\infty}=2.57$). (b) The Scarf potential $V(\overline{z})$ for
    $c=0.15$.}\label{fig:scarf}
\end{figure}

Let us consider a particular model of a smoothed out interface (cf. figure~\ref{fig:scarf}a):
\be\label{fcn:shape}
\varepsilon(\overline{z};\zeta)=e^{4c(\zeta)\arctan\big(e^{-\overline{z}}\big)},
\ee
where $c>0$ is a dimensionless constant. Since inside the dielectric and away from the interface $\overline{z}\to-\infty$ we have
\ba
\varepsilon(-\infty;\zeta)=\varepsilon_{\infty}(\zeta),
\ea
the constant $c$ is given by the asymptotic value of the dielectric constant
\ba\label{c}
c(\zeta)=\frac{1}{2\pi}\ln \varepsilon_{\infty}(\zeta),
\ea
which depends on the Matsubara frequency $\zeta$. Then, the non-dimensionalized $\overline{\eta}$ introduced in \eqref{V} and the potential $V(\overline{z})$ become
\be\label{V:Scarf}
\overline{\eta}(\overline{z})=-2c{1\over\cosh \overline{z}}\hh V(\overline{z})=c^2{1\over\cosh^2 \overline{z}}+c{\sinh \overline{z}\over\cosh^2 \overline{z}},
\ee
respectively; here we suppressed the dependence on $\zeta$ for brevity. This form of the potential $V$ is known as the Scarf potential \cite{Scarf:1958,Derezinski:2011}, cf. figure~\ref{fig:scarf}b.

We need to find the Green's functions $\overline{\mathcal{D}}(\overline{z},\overline{z}')$ \eq{calD} or, equivalently,  $\mathcal{D}(\overline{z},\overline{z}')$ which satisfies equation \eq{eqG1} and should also vanish at $\overline{z}\to \pm \infty$. To this end, we first have to find modes obeying the homogeneous part of equation \eq{eqG1}
\ba\label{ddu}
\big[\partial_{\overline{z}}^2 -k^2-V(\overline{z})\big]u(\overline{z})=0,
\ea
which has the form of the Schrodinger equation with the Scarf potential and can be solved in terms of hypergeometric functions. Two independent solutions to \eq{ddu} are
\begin{subequations}\label{u1u2}
\begin{align}
u_1&=e^{\pi c\over 2}Z^{-\i{c\over 2}}(1-Z)^{\i{c\over 2}}
\,{}_\ins{2}F_\ins{1}\Big(k,-k;\varsigma;Z\Big)
\\
u_2&=e^{\i{\pi\over 4}}\,Z^{{1\over 2}+\i{c\over 2}}(1-Z)^{\i{c\over 2}}
\,{}_\ins{2}F_\ins{1}\Big(1+k-\varsigma,1-k-\varsigma;2-\varsigma;Z\Big),
\end{align}
\end{subequations}
where $Z$ is a complex function of $\overline{z}$
\be
Z={1\over 2}(1-\i\sinh \overline{z})
\ee
and
\be
\varsigma={1\over 2}-\i c
\ee
is a complex number.

For computational purposes it is convenient to use another pair of solutions expressed in terms of the hypergeometric functions of an inverse argument \cite{Abramowitz:1965}:
\begin{subequations}\label{w1w2}
\begin{align}
&w_1=Z^{-\i{c\over 2}}(1-Z)^{\i{c\over 2}}Z^{-k} \,{}_\ins{2}F_\ins{1}\Big(k,1+k-\varsigma;1+2k;{1\over Z}\Big),\\
&w_2=Z^{-\i{c\over 2}}(1-Z)^{\i{c\over 2}}Z^{k}\,{}_\ins{2}F_\ins{1}\Big(-k,1-k-\varsigma;1-2k;{1\over Z}\Big).
\end{align}
\end{subequations}
Solutions \eq{u1u2} are related to \eq{w1w2} as follows, depending on the sign of $\overline{z}$, i.e. for $\overline{z}>0$:
\begin{align}
\begin{split}
e^{- \frac{\pi c}{2}} u_1&=e^{-\i\pi k}\,{\Gamma(-2k)\Gamma(\varsigma)\over
\Gamma(-k)\Gamma(\varsigma-k)}w_1
-e^{\i\pi k} \,{\Gamma(-2k)\Gamma(1+2k)\Gamma(\varsigma)\over
\Gamma(k)\Gamma(1-2k)\Gamma(k+\varsigma)}w_2,\\
e^{\i{\pi\over 4}} e^{- \pi c} u_2&=e^{-\i\pi k}\,{\Gamma(-2k)\Gamma(2-\varsigma)\over
\Gamma(1-k)\Gamma(1-k-\varsigma)}w_1-e^{\i\pi k} \,{\Gamma(-2k)\Gamma(1+2k)\Gamma(2-\varsigma)\over
\Gamma(1+k)\Gamma(1-2k)\Gamma(1+k-\varsigma)}w_2,
\end{split}
\end{align}
while for $\overline{z}<0$:
\begin{align}
\begin{split}
e^{-\frac{\pi c}{2}}u_1&=e^{\i\pi k}\,{\Gamma(-2k)\Gamma(\varsigma)\over
\Gamma(-k)\Gamma(\varsigma-k)}w_1-e^{-\i\pi k}
\,{\Gamma(-2k)\Gamma(1+2k)\Gamma(\varsigma)\over
\Gamma(k)\Gamma(1-2k)\Gamma(k+\varsigma)}w_2,\\
e^{-\i{3\pi\over 4}}e^{\pi c}u_2&=e^{\i\pi k}\,{\Gamma(-2k)\Gamma(2-\varsigma)\over
\Gamma(1-k)\Gamma(1-k-\varsigma)}w_1
-e^{-\i\pi k}\,{\Gamma(-2k)\Gamma(1+2k)\Gamma(2-\varsigma)\over
\Gamma(1+k)\Gamma(1-2k)\Gamma(1+k-\varsigma)}w_2.
\end{split}
\end{align}
These different representations for positive and negative $\overline{z}$ appear because the hypergeometric functions in $w_1,w_2$ have a branch cut for $1/Z\in[1,\infty)$ on the real axis.

The Green's function satisfying \eq{eqG1} we are looking for has to be real, so the complexity of the modes $u_1$, $u_2$ is an obstacle. Fortunately, there exist a pair of independent real solutions $v_1$,$v_2$, which are linear combinations of $u_1$ and $u_2$:
\ba
\label{slns-v1v2}
v_1:=A u_1-B u_2  \hh~~~~~~~~~~ v_2:=C u_1-E u_2,
\ea
where the coefficients are complex
\begin{subequations}
\begin{align}
A&=e^{\i\pi k\over 2}\,{1\over \Gamma(1+k-\varsigma)\Gamma(\varsigma)}, &
B&=e^{{\i\pi\over 2}(1+k-\varsigma)}\,{k\over \Gamma(k+\varsigma)\Gamma(2-\varsigma)}, \\
C&=e^{-2\i\pi \varsigma}e^{-\i\pi(1+k)}A, &
E&=e^{-\i\pi(1+k)}B.
\end{align}
\end{subequations}
The solution $v_1$ vanishes at $z\to +\infty$ and grows at $z\to -\infty$, while the solution $v_2$ grows at $z\to +\infty$ and vanishes at $z\to -\infty$.

\begin{figure}
\centering \setlength{\labelsep}{-4.0mm}
\begin{tabular}{l l l l}
\hspace{-0.05in}\imagetop{(a)} & \hspace{-0.265in}\imagetop{\includegraphics[height=1.965in]{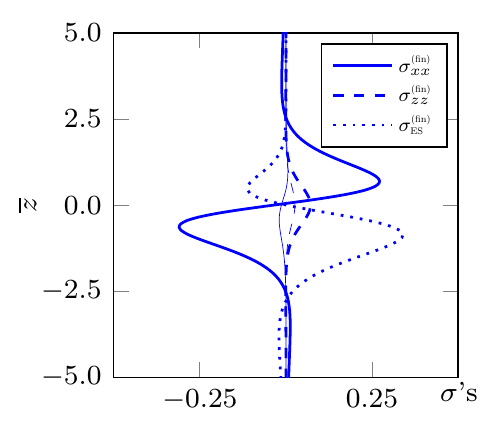}} &
\hspace{0.05in}\imagetop{(b)} & \hspace{-0.265in}\imagetop{\includegraphics[height=1.965in]{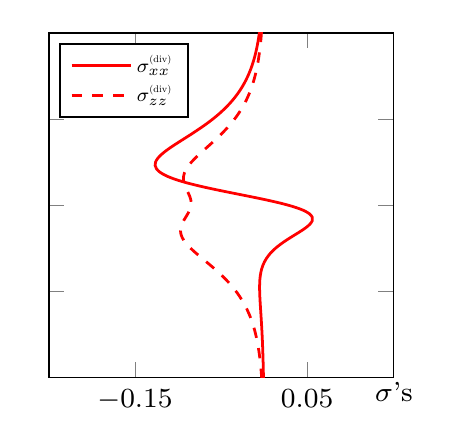}} \\
\hspace{-0.05in}\imagetop{(c)} & \hspace{-0.265in}\imagetop{\includegraphics[height=1.965in]{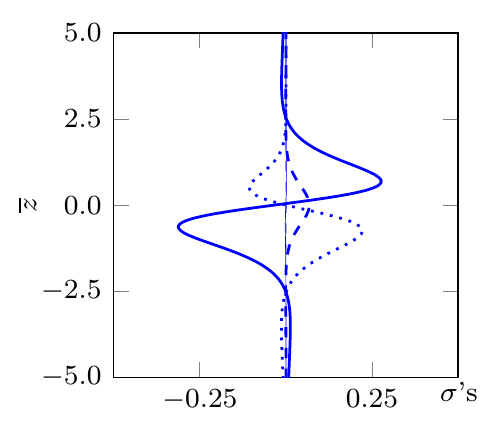}} &
\hspace{0.05in}\imagetop{(d)} & \hspace{-0.265in}\imagetop{\includegraphics[height=1.965in]{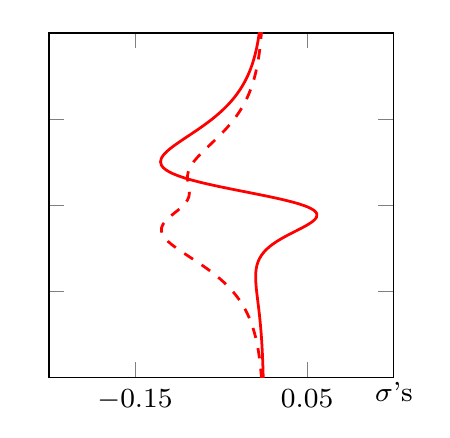}}
\end{tabular}
\caption{Stress distributions across the interfacial region for (a-b) water and (c-d) benzene. (a,c) Finite contributions \eqref{sigmatau1} without the prefactor $\beta^{-1} w^{-3}$ along with the zero Matsubara frequency contribution only (thin lines); the finite electrostriction contribution $\sigma^{(\ins{fin})}_{\ins{ES}}$ is also shown separately (the electrostriction terms in \eqref{stresses:tau-all} are singled out and renormalized with the corresponding counterparts in the divergent stresses according to \eqref{taufin}). (b,d) Subleading contributions $\sigma^{(\ins{div})}_{xx}=\sigma^{(\ins{iso})}$ and $\sigma^{(\ins{div})}_{zz}=\sigma^{(\ins{iso})} + \sigma^{(\ins{ani})}$ calculated from \eqref{sigma-iso} and \eqref{sigma-ani} without the prefactor $\beta^{-1} \epsilon^{-1/2} w^{-2}$ and with both $\eta$ and $z$ scaled with respect to the interface width $w$.}\label{fig:stress-distributions}
\end{figure}
The real solutions \eq{slns-v1v2} can be also written in the form depending upon the sign of $\overline{z}$, i.e. for $\overline{z}>0$:
\begin{subequations}
\begin{align}
v_1&={e^{-\i{\pi \over 2}k}e^{{\pi c\over 2}}\cosh(\pi c)\over 2^{2k}\pi^{1\over 2}\,\Gamma\big(k+{1\over 2}\big)}\,w_1, \\
v_2&=-{e^{-\i{\pi \over 2}k}e^{-{\pi c\over 2}}\sinh(2\pi c)\,\Gamma\big(-k+{1\over 2}\big)\over 2^{2k+1}\pi^{3\over 2}\,}\,w_1
+{e^{\i{\pi \over 2}k}e^{-{\pi c\over 2}}\cosh(\pi c)2^{2k} \,\Gamma\big(k+{1\over 2}\big)\over \pi^{1\over 2}\,\Gamma\big(k+{1\over 2}-\i c\big)\Gamma\big(k+{1\over 2}+\i c\big)}\,w_2,
\end{align}
\end{subequations}
while for $\overline{z}<0$:
\begin{subequations}
\begin{align}
v_1&={e^{\i{\pi \over 2}k}e^{-{\pi c\over 2}}\sinh(2\pi c)\over 2^{2k+1}\pi^{1\over 2}\cos(\pi k)\,\Gamma\big(k+{1\over 2}\big)}~w_1
+{e^{-\i{\pi \over 2}k}e^{-{\pi c\over 2}}\cosh(\pi c)2^{2k} \,\Gamma\big(k+{1\over 2}\big)\over \pi^{1\over 2}\,\Gamma\big(k+{1\over 2}-\i c\big)\Gamma\big(k+{1\over 2}+\i c\big)}~w_2, \\
v_2&={e^{\i{\pi \over 2}k}e^{-{3\pi c\over 2}}\cosh(\pi c)\over 2^{2k}\pi^{1\over 2}\,\Gamma\big(k+{1\over 2}\big)}~w_1 .
\end{align}
\end{subequations}
The Wronskian of these two solutions can be computed analytically
\be\label{W}
W[v_1,v_2]=e^{\pi c}\big(1+e^{-2\pi c}\big)^2
\,{k\over 2\pi\, \Gamma(k+\varsigma)\Gamma(1+k-\varsigma)},
\ee
which is, evidently, also real despite the complex $\varsigma$ entering the expression. Now we have everything at hand to construct the Green's function $\overline{\mathcal{D}}(k;\overline{z},\overline{z}')$ in \eqref{calD} we are looking for, where $\mathcal{D}(k;\overline{z},\overline{z}')$ is given by
\ba\label{GvvW}
\mathcal{D}(k;\overline{z},\overline{z}')=-{v_1(\overline{z}_\ins{max})v_2(\overline{z}_\ins{min})\over W[v_1,v_2]},
\ea
and $\overline{z}_\ins{max}=\max(\overline{z},\overline{z}')$, $\overline{z}_\ins{min}=\min(\overline{z},\overline{z}')$.

Substituting the constructed Green's function (\ref{GvvW},\ref{calD}) in equation \eqref{stresses:tau-all} and subtracting $\widehat{\tau}^\ins{(div)}_{ij}$, as per \eqref{taufin} we find the finite stresses $\widehat{\tau}^\ins{(fin)}_{ij}$ at each frequency $\zeta_n$. The result of this procedure for two types of dielectrics -- water and benzene chosen here as examples of polar and non-polar liquids, respectively -- is shown in figure~\ref{fig:stress-distributions}.

\section{The quantum nature of surface tension} \label{sec:ST}

\subsection{Calculation of surface tension} \label{subsec:ST-calculation}

Given the calculated total stresses \eqref{sigma-tot}, we can compute the force density
\ba\label{fi}
f_i=\nabla^{k}\sigma_{ik}^{\ins{(tot)}}=f^\ins{(m)}_i + E_i\nabla_{k}(\varepsilon E^k)
-{1\over 2}\bs{E}^2 \nabla_{i}\varepsilon
+{1\over 2} \nabla_{i}\Big( \bs{E}^2 \rho{\partial\varepsilon\over\partial\rho}\Big),
\ea
where $\bs{E}^2=E_k E^k$. The first term on the right hand side is a mechanical force $f^\ins{(m)}_i=\nabla^{k}\sigma_{ik}^\ins{(m)}=-\nabla_{i} p^\ins{(m)}$ due to the gradients of pressure. In the absence of free charges the second term $\nabla_{k}(\varepsilon E^k)$ vanishes. The third term $-{1\over 2}\bs{E}^2 \nabla_{i}\varepsilon$ is the electrostatic force density for any inhomogeneous dielectric in an external electric field $\bs{E}$. The last term is due to the electrostriction effect.

Since we study the system which is in a state of mechanical equilibrium, i.e. all forces \eq{fi} in the bulk compensate each other at each point, when \eqref{fi} is seen in the QM sense the equilibrium condition leads to
\ba\label{fi1}
f^\ins{(m)}_i - {1\over 2\beta}\sum_{\zeta_n}\left[\langle \mathrm{E}_k \mathrm{E}^k\rangle \nabla_{i}\varepsilon- \nabla_{i}\Big(\langle \mathrm{E}_k \mathrm{E}^k\rangle \rho{\partial\varepsilon\over\partial\rho}\Big)\right]=0,
\ea
where the operators $\mathrm{E}_k$ and $\varepsilon$ are functions of the Matsubara frequencies $\zeta_n$ (cf. \S \ref{subsec:Derjaguin-limit}) and point $\bs{x}$.
The correlator $\langle \mathrm{E}_k \mathrm{E}^k\rangle$ can be expressed in terms of the Green's function \eq{GG} as follows
\be
\langle \mathrm{E}_k \mathrm{E}^k\rangle=\lim_{\bs{x}\to\bs{x'}}[g^{kk'}\, \nabla_{k}\nabla_{k'}\widehat{G}(\zeta;\bs{x},\bs{x'})].
\ee

\begin{figure}
\centering \setlength{\labelsep}{-4.0mm}
\begin{tabular}{l l l l l l}
\hspace{-0.05in}\imagetop{(a)} & \hspace{-0.385in}\imagetop{\includegraphics[height=1.965in]{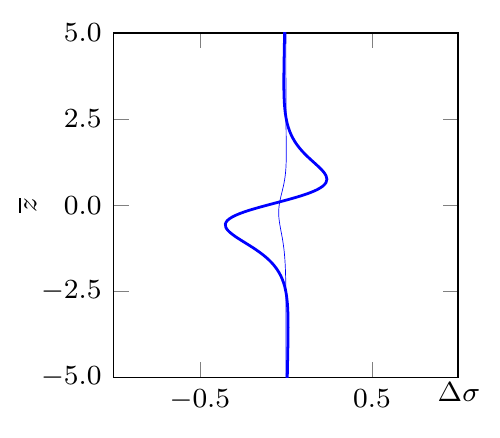}} &
\hspace{-0.15in}\imagetop{(b)} & \hspace{-0.325in}\imagetop{\includegraphics[height=1.965in]{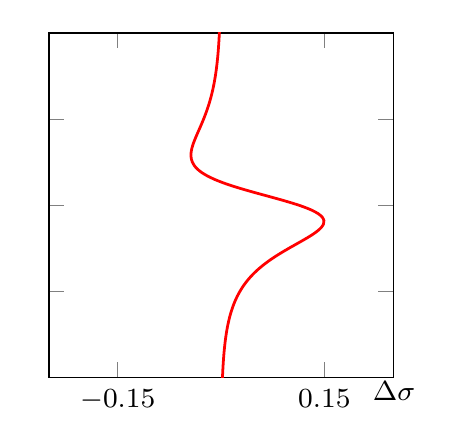}} &
\hspace{-0.15in}\imagetop{(c)} & \hspace{-0.325in}\imagetop{\includegraphics[height=1.965in]{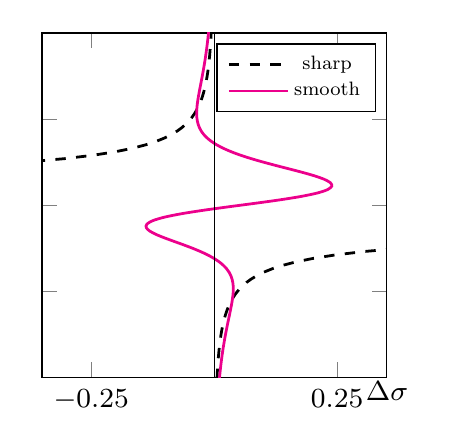}} \\
\hspace{-0.05in}\imagetop{(d)} & \hspace{-0.385in}\imagetop{\includegraphics[height=1.965in]{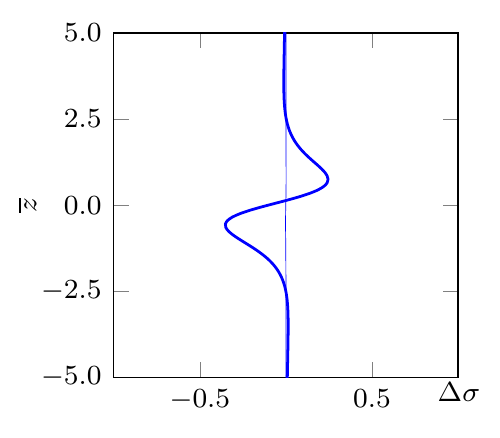}} &
\hspace{-0.15in}\imagetop{(e)} & \hspace{-0.325in}\imagetop{\includegraphics[height=1.965in]{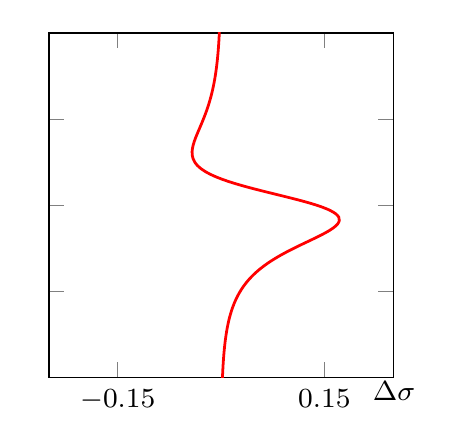}} &
\hspace{-0.15in}\imagetop{(f)} & \hspace{-0.325in}\imagetop{\includegraphics[height=1.965in]{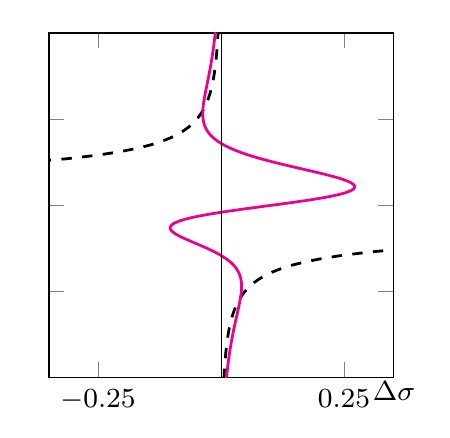}}
\end{tabular}
\caption{The difference of stress distributions $\Delta \sigma=\sigma_{xx}-\sigma_{zz}$ across the interfacial region for (a-c) water and (d-f) benzene. (a,d) Finite contributions $\sigma^{(\ins{fin})}_{xx}-\sigma^{(\ins{fin})}_{zz}$ as per equations \eqref{sigmatau1} without the prefactor $\beta^{-1} w^{-3}$ along with the zero Matsubara frequency contribution only (thin lines). (b,e) Subleading contributions $\sigma^{(\ins{div})}_{xx}-\sigma^{(\ins{div})}_{zz}=-\sigma^{(\ins{ani})}$ calculated from \eqref{sigma-iso} and \eqref{sigma-ani} without the prefactor $\beta^{-1} \epsilon^{-1/2} w^{-2}$ and with both $\eta$ and $z$ scaled with respect to the interface width $w$. (c,f) Finite contributions $\sigma^{(\ins{fin})}_{xx}-\sigma^{(\ins{fin})}_{zz}$ in the case of a sharp interface (dashed), as per equation \eqref{Delta-sigma:sharp} without the prefactor $\beta^{-1} w^{-3}$, are compared with the total $\Delta \sigma=\sigma_{xx}-\sigma_{zz}$ in the case of a smoothed out interface, i.e. sum of finite (a,d) and subleading (b,e) contributions: $\Delta \sigma^{(\ins{fin})} + (w/\epsilon^{1/2}) \Delta \sigma^{(\ins{div})}$. Here we took $\epsilon^{1/2}=r_{\ins{m}}$ and the measured values $w=4.49 \, \angstrom$, $r_{\ins{m}} = 3.1 \, \angstrom$ for water and $w=7.12 \, \angstrom$, $r_{\ins{m}} = 4.75 \, \angstrom$ for benzene \cite{Douillard:2009}.}\label{fig:stress-diff-distributions}
\end{figure}

Consider now a system with a single smoothed out interface, which is assumed to be flat and, hence, $\varepsilon$ is a function of the coordinate $z$ only. In order to define the surface tension we consider a \textit{strip} of infinite length in the $z$-direction and of unit length in the $y$-direction, with the normal $\bs{n}$ in the $x$-direction correspondingly. In the asymptotic regions $z\to \pm\infty$ the total stress tensor of a fluid is isotropic (\ref{sij3},\ref{sij3xx}), i.e. given by $\sigma_{ij}^\ins{(tot)}= - \delta_{ij} p_0^{\ins{(ren)}}$. Note that because our system with a free interface is in a mechanical equilibrium, there is no net force in the $z$-direction acting on every volume element of the system. Therefore $\sigma_{zz}^\ins{(tot)}=-p_0^{\ins{(ren)}}=\const$\footnote{Recall that the pressure is the force per unit area of the wall should it be bounding the medium, while normal stress is the force per unit area acting on the medium itself.} for all $z$ as per \eqref{sij3}: below, above, and inside the smoothed out interface. Recall that the total stress tensor consists of a sum of the mechanical and EM contributions \eqref{sij1}. The leading order UV-divergent EM stresses in the $x,y$-directions are compensated by the bare mechanical (elastic) stresses producing an isotropic renormalized pressure \eqref{sij5}, which does not depend on the $z$-coordinate. The remaining subleading and finite parts of stresses in the $x,y$-directions do depend on $z$. At the same time, all the quantities in the bulk do not depend on $x$ and $y$ due to translational invariance and the equilibrium condition \eq{fi1} in these directions is satisfied automatically for all $z$. In the absence of the interface, i.e. when the entire space is filled with a homogeneous fluid, the force acting on the considered \textit{strip} would be just $-\int_{-\infty}^{\infty} \d z\, p_0^{\ins{(ren)}} > 0$ since $p_0^{\ins{(ren)}}<0$. Thus, the surface tension of the interface, in hydrodynamics \cite{Lifshitz:2013} and theoretical physics \cite{Schwinger:1978} traditionally defined as a positive quantity because physically it corresponds to tensile stresses tending to compress the medium, is computed as the difference of the integral of $\int_{-\infty}^{\infty}\d z \, \sigma_{xx}^\ins{(tot)}(z)$ and the latter force:
\ba\label{gamma}
\gamma=\int_{-\infty}^{\infty}\d z \big[\sigma_{xx}^\ins{(tot)}(z)+p_0^{\ins{(ren)}}\big]=\int_{-\infty}^{\infty}\d z \big[\sigma_{xx}^\ins{(tot)}(z)-\sigma_{zz}^\ins{(tot)}\big].
\ea
The formula \eqref{gamma} is notable: while both EM (namely, the calculated here London) and non-EM (such as the ones responsible for the repulsive core in the Lennard-Jones potential, cf. \S \ref{subsubsec:proper-time:interpretation}) interactions contribute to the total energy of the interface, the equilibrium condition \eqref{condition:equilibrium} and its consequence \eqref{sigmaxx-tot} enable us to express the total surface tension solely in terms of EM stresses. While above we provided a mechanistic derivation of \eqref{gamma}, it is also in line with an energy-based derivation of \eqref{gamma} by Fisher \cite{Fisher:1964}: the total energy per unit area of the interface is determined by the work of the surface tension forces, which is why the energy density per unit area of the interface exactly equals to the surface tension $\gamma$. Due to the structure of \eqref{gamma}, it is important to emphasise that any isotropic contributions, should they originate from electrostriction or other leading diverging terms in \eqref{sigma-div:leading}, cannot contribute to surface tension according to \eqref{gamma}; these isotropic terms can only affect the (renormalized) mechanical pressure (\S~\ref{subsubsec:proper-time:interpretation}). Because of the asymptotic behavior \eqref{sij3xx}, only the region in the vicinity of the interface contributes to this integral. Remarkably, even if the interface is between a dielectric and the vacuum, polarization of the vacuum also contributes to surface tension -- this is not an artifact of the theory as the measurable Casimir effect \cite{Casimir:1948b} is also due to the vacuum polarization. Finally, it is interesting to note that because of formula \eqref{gamma}, the surface of tension defined by the first moment of $\sigma_{xx}^\ins{(tot)}-\sigma_{zz}^\ins{(tot)}$ is, in general, different from the equimolar dividing surface \cite{Walton:1983}, which is the Gibbs definition of a position of an interface. Given that surface tension is dictated by the distribution of $\sigma_{xx}^\ins{(tot)}-\sigma_{zz}^\ins{(tot)}$ across the interface, this behavior is shown in figure~\ref{fig:stress-diff-distributions} with separated finite (a,d) and subleading (b,e) contributions as well as the total stress difference (c,f) compared to that obtained from the sharp interface problem \eqref{sigmas:Lifshitz}, which in the limit $\ell \rightarrow \infty$ leads to
\begin{align}\label{Delta-sigma:sharp}
\sigma_{xx}^\ins{(fin)}-\sigma_{zz}^\ins{(fin)} = - \frac{1}{32 \, \pi \, \beta \, w^{3} \, \overline{z}^{3}} \, \sum_{n}{\frac{\varepsilon_{1}-\varepsilon_{3}}{\varepsilon_{1}+\varepsilon_{3}}}.
\end{align}
Note that \eqref{Delta-sigma:sharp} is an antisymmetric function of $\overline{z}$ and hence it would be impossible to get a non-zero surface tension or contribution to it in the framework of the sharp interface formulation (\S \ref{sec:Lifshitz})! It must be also mentioned that the electrostriction contributions, which would have to be added to \eqref{sigmas:Lifshitz} in the dielectric phase, cancel out in \eqref{Delta-sigma:sharp} due to the isotropic nature of electrostriction, cf. \eqref{sigma-iso}.

Let us next analyse the structure of $\sigma_{xx}^\ins{(tot)}(z)$. As per \S \ref{subsec:proper-time-regularization}, the EM contribution consists of the divergent and regular parts $\sigma_{ij}=\sigma^\ins{(div)}_{ij}+\sigma^\ins{(fin)}_{ij}$. The UV divergent part  appears because of our treatment of media as continuous, which is why this model is expected to work only down to the length-scales on the order of intermolecular distance $r_{\ins{m}}$. Thus, the UV divergent part is in fact a large, but finite quantity that depends on $r_{\ins{m}}$ or a cut-off parameter $\epsilon^{1/2}$. The finite part $\sigma^\ins{(fin)}_{ij}$ describes the $\epsilon^{1/2}$-independent contribution of the vacuum polarization effects in inhomogeneous media. The equilibrium condition \eqref{fi1} tells us that all mechanical and EM forces counterbalance each other; in particular, in the $z$-direction we get \eqref{sij3}. As per \eq{gamma} and \eq{sigmaxx-tot}, the EM tangent pressure $\sigma^\ins{(fin)}_{xx}$ depends on $z$ and contributes to the finite EM part of surface tension
\ba\label{ST-fin}
\gamma^\ins{(fin)}=\int\limits_{-\infty}^{\infty}\d z \big[\sigma_{xx}^\ins{(fin)}-\sigma_{zz}^\ins{(fin)}\big]\equiv\int\limits_{-\infty}^{\infty}\d z \, \Delta\sigma^\ins{(fin)}.
\ea
This integral is finite because $\sigma_{xx}^\ins{(fin)}\to 0$ and $\sigma_{zz}^\ins{(fin)}\to 0$ fast enough at $z\to\pm\infty$. It does not depend on the UV cut-off $\epsilon^{1/2}$ and is proportional to ${w}^{-2}$; this dependence on a characteristic width of the interface ${w}$ can be deduced from dimensional considerations because the finite part depends only on $w$. One can compute the coefficient of proportionality for a particular interface shape function \eqref{fcn:shape}, which allows an exact representation for the quantum fluctuating modes (\S \ref{subsec:Scarf}).

Summing up the finite \eqref{ST-fin} and divergent contributions, the surface tension \eq{gamma} assumes the form
\ba\label{gamma1}
\hspace{-0.3cm}\gamma=\gamma^\ins{(fin)}+\gamma^\ins{(div)}, \
\ea
with
\begin{align}
\gamma^\ins{(div)}= - \int_{-\infty}^{\infty}{\d z \, \sigma_{zz}^{\ins{(ani)}}}\equiv\int\limits_{-\infty}^{\infty}\d z \, \Delta\sigma^\ins{(div)}={23\over 24 (4\pi)^{3/2} \beta\epsilon^{1/2}}\sum_n \int\limits_{-\infty}^{\infty}\d z \,\frac{ \eta^2}{\varepsilon^{1/2}},
\end{align}
and $\eta$ is given by \eqref{dfn:eta}; in the case of the Scarf model, $\gamma^\ins{(div)}$ can be calculated explicitly:
\ba\label{gammaScarf}
\gamma^\ins{(div)}={23\over 6 (4\pi)^{3/2}  \beta \epsilon^{1/2}w}\sum_n \frac{(\ln\varepsilon_{\infty})^2}{(\ln\varepsilon_{\infty})^2+4\pi^2}\frac{1+\sqrt{\varepsilon_{\infty}}}{\sqrt{\varepsilon_{\infty}}}.
\ea
It should be emphasized that this expression depends only on the EM properties $\varepsilon$ of the media. The integrals in \eqref{gamma1} converge because $\sigma_{ij}^\ins{(fin)}$ and $\eta$ vanish fast enough far away from the interface. Note that expression \eqref{gamma1} for $\gamma^\ins{(div)}$ is general and can be easily evaluated for any profile $\varepsilon(\i\zeta;z)$ of a smoothed out interface. Computation of $\gamma^\ins{(fin)}$ in \eqref{ST-fin} is much more tricky because, as in the case of the Casimir effect, the stress tensor $\sigma_{ij}^\ins{(fin)}$ does not vanish outside the localized interface. Based on the results of \S \ref{subsec:point-splitting}, one can compute $\sigma_{ij}^\ins{(fin)}$ and show that $\gamma^\ins{(fin)}$ is non-trivial and proportional to $\sim w^{-2}$, where $w>\epsilon^{1/2}$ is a characteristic width of the interface and $\epsilon^{1/2}$ is the cut-off length-scale. This contribution is smaller than $\gamma^\ins{(div)}$, which is of the order of $\sim 1/(\epsilon^{1/2} w)$. Also, while both $\gamma^\ins{(div)}$ and $\gamma^\ins{(fin)}$ in \eqref{gamma1} seem to be proportional to temperature $T=\beta^{-1}$, as discussed in \S \ref{subsec:Derjaguin-limit} these quantities do not vanish in the limit $T \rightarrow 0$.

\begin{figure}[ht!]
\centering
\centering \setlength{\labelsep}{0.0mm}
\begin{tabular}{l l l l}
\hspace{-0.05in}\imagetop{(a)} & \hspace{-0.165in}\imagetop{\includegraphics[height=2.25in]{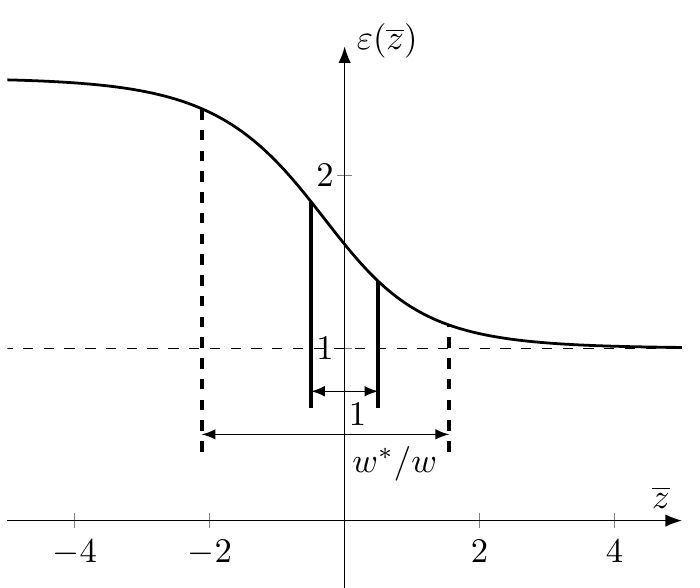}} &
\hspace{-0.15in}\imagetop{(b)} & \hspace{-0.165in}\imagetop{\includegraphics[height=2.25in]{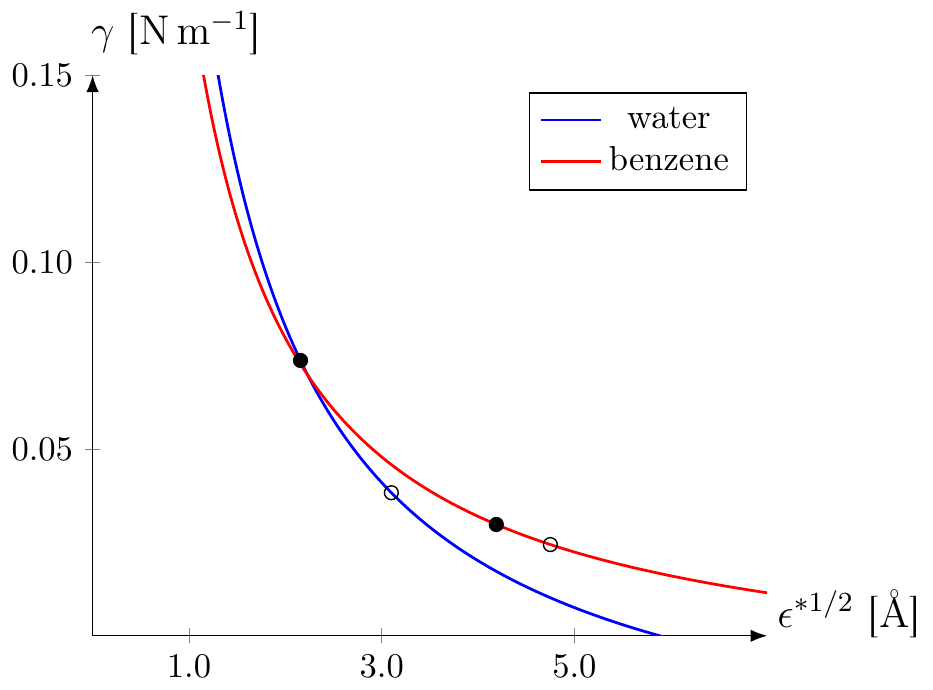}}
\end{tabular}
\caption{(a) On the interface width definition: $w$ is the width scale introduced in \eqref{scaling:width}, set in this figure to be $1$; $w^{*}$ is the experimentally measured width based on the length over which the density $\rho(z)$ changes from $90$ to $10 \%$ of the bulk value \cite{Penanen:2000}, which is assumed here to correlate with the profile of $\varepsilon(z)-1$. (b) Dependence of surface tension $\gamma$ on the cut-off parameter $\epsilon^{*1/2}$ for water at $287.5 \, \mathrm{K}$ and benzene at $286 \, \mathrm{K}$ according to formula \eqref{eqn:ST-total}; the data for the interface width $w$ are taken from figure~\ref{fig:stress-diff-distributions}. The cut-off parameter required to obtain the corresponding correct surface tension (marked with a solid circle) at the reported temperature \cite{Douillard:2009}: $\epsilon^{*1/2} = 2.16 \, \angstrom$ for $0.0737 \, \mathrm{N \, m^{-1}}$ (water) and $\epsilon^{*1/2} = 4.19 \, \angstrom$ for $0.0298 \, \mathrm{N \, m^{-1}}$ (benzene). For comparison, we also show the values of surface tension predicted by taking $\epsilon^{*1/2}$ to be equal to the intermolecular distance (marked with empty circles) producing $0.0383 \, \mathrm{N \, m^{-1}}$ for $\epsilon^{*1/2} = 3.1 \, \angstrom$ (water) and $0.0244 \, \mathrm{N \, m^{-1}}$ for $\epsilon^{*1/2} = 4.75 \, \angstrom$ (benzene).}\label{fig:ST-empirical}
\end{figure}
To demonstrate the dependence of surface tension on the cut-off parameter $\epsilon^{1/2}$ and identify the value of the latter required to fit to the experimentally measured surface tension, we used the calculations from figure~\ref{fig:stress-diff-distributions} along with equation \eqref{gamma1} to produce in SI units
\begin{align}\label{eqn:ST-total}
\gamma = \frac{k_{\ins{B}} T}{w^{*2}} \left[\int\limits_{-\infty}^{\infty}\d z \, \Delta \sigma^{(\ins{fin})} + \frac{w^{*}}{\epsilon^{*1/2}} \int\limits_{-\infty}^{\infty}\d z \, \Delta \sigma^{(\ins{div})}\right],
\end{align}
where we took into account that scaling \eqref{scaling:width} with respect to $w$, which sets the length-scale in the problem, is different from the procedure used in experiments \cite{Penanen:2000} to determine the interfacial width $w^{*}$. This fact is illustrated in figure~\ref{fig:ST-empirical}a for the chosen interfacial profile \eqref{fcn:shape}; as a result, the cut-off parameter $\epsilon^{*1/2}$ must also be rescaled to the same factor $w^{*}/w$. Calculations for \eqref{fcn:shape} show that as $\varepsilon$ drops from its value of $80$ at zero frequency to $1$ at high frequencies, the factor $w^{*}/w$ varies in the range $3.29-3.69$. Using $w^{*}/w=3.67$ typical for the most values of $\varepsilon$ over the entire frequency domain, we plot \eqref{eqn:ST-total} in figure~\ref{fig:ST-empirical}b. It is notable that for the chosen substances the finite stresses contribution in \eqref{eqn:ST-total} is actually negative.

The reported in figure~\ref{fig:ST-empirical}b values of $\gamma$ evaluated at $\epsilon^{*1/2}=r_{\ins{m}}$ are indicative of the validity of the theory, but the value of the cut-off parameter cannot be predicted within the framework of the Lifshitz theory as the physics behind $\epsilon^{1/2}$ lies beyond its scope. While a more detailed comparison between theory and experiment would also require a more precise reconciliation with how the interface width $w$ is measured in experiments \cite{Douillard:2009,Penanen:2000}, figure~\ref{fig:ST-empirical}b demonstrates that all these differences can be absorbed into the cut-off parameter $\epsilon^{1/2}$. The departure in the prediction is the largest in the case of water, i.e. about three times, which is because it is a polar liquid with strong hydrogen bonds. First, this leads to the Keesom effect \cite{Keesom:1915} contributing to van der Waals forces due to attraction of permanent dipoles with moments $\mathrm{p}$ -- in the case of water the ratio of the Keesom to London energies of interaction $\varphi_{K}/\varphi_{L} = 4 \mathrm{p}^{4}/(9 h_{\ins{B}} T \alpha_{w}^{2} \omega_{a}) \approx 6.32$, where $\alpha_{w}$ is the water molecular polarizability; most importantly, the hydrogen bond energy $3.82 \cdot 10^{-20} \, \mathrm{J}$ per bond is comparable to that of the latent heat of evaporation per molecule $6.78 \cdot 10^{-20} \, \mathrm{J}$ and much larger than $k_{\ins{B}} T \approx 4.14 \cdot 10^{-21} \, \mathrm{J}$ thus indeed being capable of affecting the value of surface tension. Second, the water molecule polarity may give rise to a local structure in the bulk and potentially at the interface thus, in particular, invalidating assumptions behind the isotropy of the dielectric permittivity upon which the Lifshitz theory is based -- if one thinks of water molecules as permanent dipoles, they may orient themselves along the interface thus increasing surface tension as opposed to, say, surfactants, which, when added to water, stick their polar heads in water and thus orient themselves perpendicularly to the interface, thus decreasing surface tension due to repulsion between themselves.

While for water and benzene the determined values of the cut-off parameter $\epsilon^{*1/2}$ prove to be reasonably close to the intermolecular distance $r_{\ins{m}}$ for a given substance, the true physics behind the cut-off parameter $\epsilon^{1/2}$, responsible for the dependence on the alluded ``trans-Planckian'' physics (\S \ref{subsec:proper-time-regularization}), remains an open question. Since the Lifshitz theory is based upon the macroscopic dielectric permittivity $\varepsilon(\omega)$, which does not know about the origin of the nature of quantum EM fluctuations, one can equally assert that the cut-off parameter $\epsilon^{1/2}$ depends on the intermolecular distance $r_{\ins{m}}$ as on the charge separation distance $\lambda_{\mathrm{d}}$ of the fluctuating dipoles responsible for the EM stresses and hence surface tension. Since the latter can be estimated independently as the bond energy $\sim k_{\ins{B}}$ divided by $r_{\ins{m}}^{2}$, and the energy of fluctuating dipoles having moments $\mathrm{p} = \lambda_{\mathrm{d}} \, e$ due to polarization of molecules is $\mathrm{p}^{2}/(4 \pi \varepsilon_{0} r_{\ins{m}}^{2})$, we find, for example, for water
\begin{align}
\lambda_{\mathrm{d}} \sim \sqrt{4\pi\varepsilon_{0} k_{\ins{B}} T r_{\ins{m}}^{3}/e^{2}} \simeq 0.25 \, \angstrom \ll r_{\ins{m}} = 3.1 \, \angstrom,
\end{align}
where $e$ is the elementary (electron) charge. In particular, originally Dzyaloshinskii et al. \cite{Dzyaloshinskii:1961} suggested that $\epsilon^{1/2}$ might be considerably smaller than the interatomic spacing $r_{\ins{m}}$. Regardless of the actual value, the dependence of surface tension on the cut-off parameter as per \eqref{eqn:ST-total} may appear astonishing. Nevertheless, such kind of behavior is typical for macroscopic models, which are \textit{not closed}, e.g. the Navier-Stokes equations governing motion of a fluid or the Navier–Cauchy equations governing deformation of solids at macroscopic length-scales ``remember'' their microscopic origin through their respective dependence on viscosity or Lam\'{e} parameters, which can be accounted for by a microscopic theory only. This is a general property of macroscopic systems in which microscopic effects propagate to macroscopic scales. Our cut-off parameter $\epsilon^{1/2}$ plays such a role in the Lifshitz theory -- as a ``macroscopic'' theory it remembers its origin at the microscopic scales, at which fluctuations originate.

The cut-off dependence of surface tension was envisioned by Schwinger et al. \cite{Schwinger:1978} (see also \cite{Hoye:2017}). These authors considered a sharp interface and found that the renormalized energies, and hence contribution to surface tension, which appears to be non-zero if stresses are taken into account only on one side of the interface, and latent heat of liquid helium, are ``cut-off dependent'' -- the idea that ``remains provocative yet unresolved'' \cite{Milton:1998ku}. Leaving aside the fact that stresses on both sides of the interface must be taken into account \eqref{eqn:ST-total}, the leading order contribution to surface tension, i.e. due to non-retarded stresses, in the sharp interface formulation considered by the authors of \cite{Schwinger:1978} should vanish due to antisymmetry of the contributing stresses \eqref{Delta-sigma:sharp}; also UV divergencies at a sharp interface are not well defined and in some cases may even cancel each ether locally \cite{Candelas:1982} or in the integral sense \cite{Milton:2011}. Notably, while Schwinger et al. \cite{Schwinger:1978} reported the calculated magnitude of surface tension three times higher than the actual one $\gamma=0.00037 \, \mathrm{N/m}$, our calculations produce $0.00048 \, \mathrm{N/m}$ based on the known interface thickness $6.5 \, \angstrom$ \cite{Penanen:2000} and interatomic distance $3.75 \, \angstrom$. Moreover, they have not taken into account energies of non-EM nature: as we demonstrated, smoothing out the interface and invoking the equilibrium condition (\ref{condition:equilibrium},\ref{sigmaxx-tot}) make the problem of computing surface tension well defined. More recently \cite{Volovik:2003fe}, in the context of quantum liquids, for which a bare Lagrangian exists, it was also noted that surface tension phenomena are the lowest-order corrections to the vacuum energy dependent on the intermolecular distance $r_{\ins{m}}$ -- in our study the latter are the leading and the former are the subleading UV divergent terms in \eqref{hatsigmaij5}, which depend on the cut-off $\epsilon^{1/2}$. This cut-off dependence survives despite that in the UV limit the matter becomes transparent as $\varepsilon \rightarrow 1$.

\begin{figure}[t!]
\centering
\includegraphics[height=3.75in]{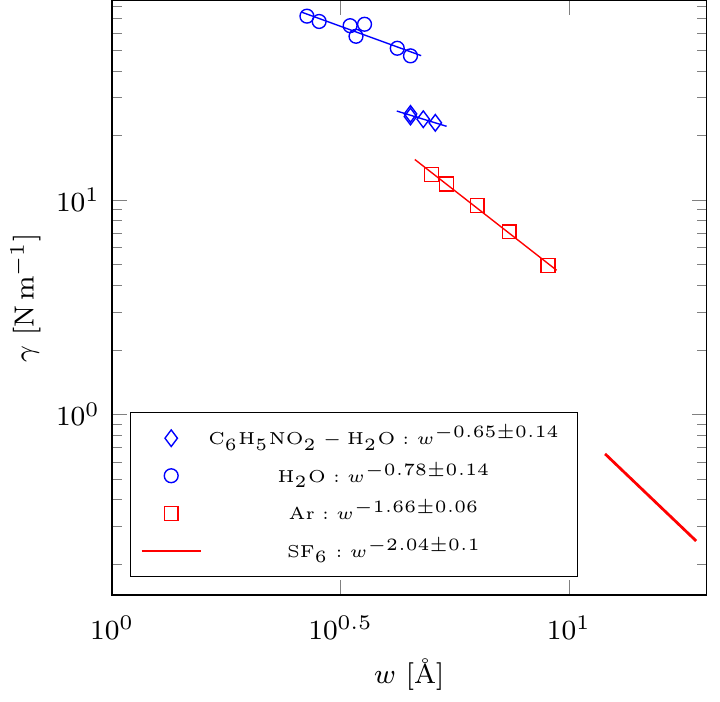}
\caption{Scaling of interfacial tension $\gamma$ with interfacial thickness $w$ far from critical point (based on experimental data for Nitrobenzene-water \cite{Luo:2006} in the temperature range $T=298-328 \, \mathrm{K}$ with $T_{\ins{c}}=513 \, \mathrm{K}$ and molecular dynamics simulations for water \cite{Taylor:1996} in the temperature range $T=268-373 \, \mathrm{K}$ with $T_{\ins{c}}=647 \, \mathrm{K}$) and near critical point (based on experimental data for liquid Argon \cite{Beaglehole:1980} in the temperature range $T=85-120 \, \mathrm{K}$ with $T_{\ins{c}}=150.9 \, \mathrm{K}$ and Sulfur-Hexafluoride \cite{Wu:1972} in the temperature range $T=305.952-318.573 \, \mathrm{K}$ with $T_{\ins{c}}=318.7 \, \mathrm{K}$). All experiments used ellipsometry to study departures of reflection of polarized light from properties predicted by Fresnel's equations. The power law scalings reported in the legend of the figure are obtained using least square fit to the data taking into account the errors reported in the above referenced papers.}\label{fig:ST-scaling}
\end{figure}
Lastly, we would like to point out that our theory of surface tension involving both subleading $\gamma^\ins{(div)} \sim \epsilon^{-1/2} w^{-1}$ and finite $\gamma^\ins{(fin)} \sim w^{-2}$ contributions does not contradict the experimental \cite{Luo:2006,Beaglehole:1980,Wu:1972} and molecular simulations \cite{Taylor:1996} data available in the literature, cf. figure \ref{fig:ST-scaling}. Since the developed theory is applicable not only at zero temperature but also at finite temperatures (see the discussion in \S \ref{subsec:Derjaguin-limit}), formula \eqref{eqn:ST-total} can be put in the context of known scalings below and near a critical point for corresponding substances. As one can gauge from figure~\ref{fig:ST-scaling} for Nitrobenzene-water and water-vapor interfaces, far below the critical point the scaling $\gamma \sim w^{-m}$ is closer to $m=1$ rather than $m=2$. For temperatures approaching critical one the exponent $m$ increases, e.g. for Argon becomes $1.66$, and for temperatures really close to the critical point in the case of Sulfur-Hexafluoride it reaches the value $m=2$, which is in accord to the critical phenomena scaling (capillary wave) theory \cite{Buff:1965} widely confirmed by experimental measurements \cite{Pressing:1973,Beysens:1987}. In general, the interfacial thickness $w$ and tension $\gamma$ near a critical point scale as \cite{Pressing:1973}
\begin{align}
w \sim \left(T_{c}-T\right)^{-\nu} \ \text{and} \ \gamma \sim \left(T_{c}-T\right)^{\mu},
\end{align}
which from the classical theory of van der Waals \cite{Waals:1894} and Cahn-Hilliard \cite{Cahn:1958} amount to the values $\nu=1/2$ and $\mu=3/2$. However, experimental measurements, cf. \cite{Pressing:1973} and references therein, deviate appreciably yielding values closer to $\nu=2/3$ and $\mu=4/3$, e.g. for Sulfur-Hexafluoride, which is commonly used in electrical power industry for high-voltage circuit breakers and gas-insulated transmitters and therefore well-studied, one measures $\nu=0.658 \pm 0.012$ and $\mu=1.34 \pm 0.06$. Altogether, this yields $\gamma \sim w^{-2.04}$, which is close to the scaling of $\gamma^\ins{(fin)} \sim w^{-2}$ predicted in our work. Having said that, the difference between the subleading and finite contributions to surface tension in most cases is elusive, however. Indeed, far from a critical point $w$ is just a few intermolecular distances $r_{\ins{m}}$, so that both contributions are comparable, which is why surface tension can be reasonably estimated \cite{Weisskopf:1985} from the enthalpy of evaporation divided by $r_{\ins{m}}^{2}$ producing the right order of magnitudes for a wide range of substances \cite{Lautrup:2011}.

\subsection{Further interpretation}

First, let us discuss the mathematical aspects of how surface tension phenomena compare between sharp and smoothed out interface formulations. In the sharp (plane) interface case we can compute $\sigma_{xx}^{(\ins{fin})}-\sigma_{zz}^{(\ins{fin})}$ if we take the limit $\ell=\infty$ in the case of the classical Lifshitz problem, cf. figure~\ref{fig:Lifshitz} and \S \ref{sec:Lifshitz}. From the resulting equation \eqref{Delta-sigma:sharp} it is easy to see that $\sigma_{xx}^{(\ins{fin})}-\sigma_{zz}^{(\ins{fin})}$ grows without bound as the interface is approached (figure~\ref{fig:stress-diff-distributions}), but represents an antisymmetric function of $z$. Therefore, integration over $z$ in the Cauchy principal value sense yields zero, which implies a vanishing contribution of renormalized (finite) terms to the surface tension of a sharp interface. This implies that the problem of calculating surface tension represents \textit{a distinguished limit}: the result depends on whether one starts with a sharp interface formulation or first computes surface tension for a smoothed out interface and then takes a limit to the sharp one. The presence of the distinguished limit is the reflection of the internal structure of the interface, which is lost in the sharp interface formulation: while the BCs \eqref{BCs:sharp} are recovered in the limit from smoothed out to sharp interface (cf. Appendix \ref{apx:Schwartz}), the stresses \eqref{hatsigmaij} involve derivatives of higher order of the Green's function than in \eqref{BCs:sharp}, which leads to the distinguished limit in the behavior of stresses.

Also, we can consider a step-like interface of width $w$ with an intermediate dielectric constant $\varepsilon_2<\varepsilon_3<\varepsilon_1$, cf. Appendix \ref{apx:ST-slab}. Analogous integration of $\sigma_{xx}^{(\ins{fin})}-\sigma_{zz}^{(\ins{fin})}$ over all $z$ leads to a non-zero $\gamma^\ins{(fin)} \sim w^{-2}$ diverging as $w \to 0$, rather than vanishing as in the single sharp interface case. This contradistinction is a consequence of inaccurate dealing with UV divergencies in the Lifshitz' and subsequent works discussed in the Introduction. Namely, if some divergent terms are omitted in the calculation of stresses such as the subleading ones \eqref{sigma-div:subleading}, then the resulting surface tension depends on the order of performing integration w.r.t. $z$ and taking a limit to the sharp interface. Only proper regularization scheme, such as the one used here based on the heat kernel method, allows extraction of all terms remaining after subtraction of the leading divergencies: this enables one to extract both finite contributions that do not depend on the regularizer and UV divergent ones that depend on it.

\begin{figure}
\centering %\setlength{\labelsep}{-0.0mm}
\begin{tabular}{l l l l}
\hspace{-0.05in}\imagetop{(a)} & \hspace{-0.065in}\imagetop{\includegraphics[height=2.0in]{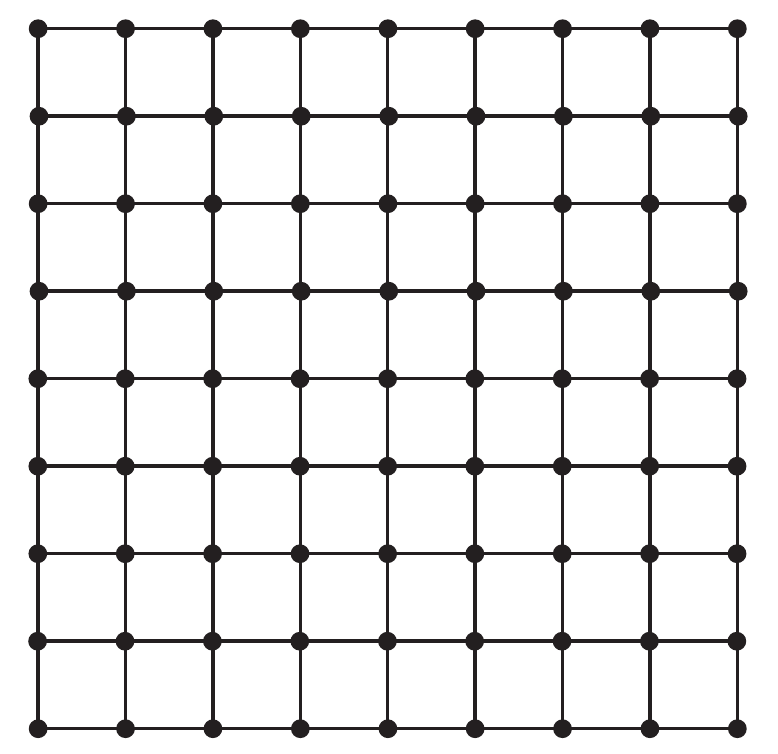}} &
\hspace{0.05in}\imagetop{(b)} & \hspace{-0.065in}\imagetop{\includegraphics[height=2.0in]{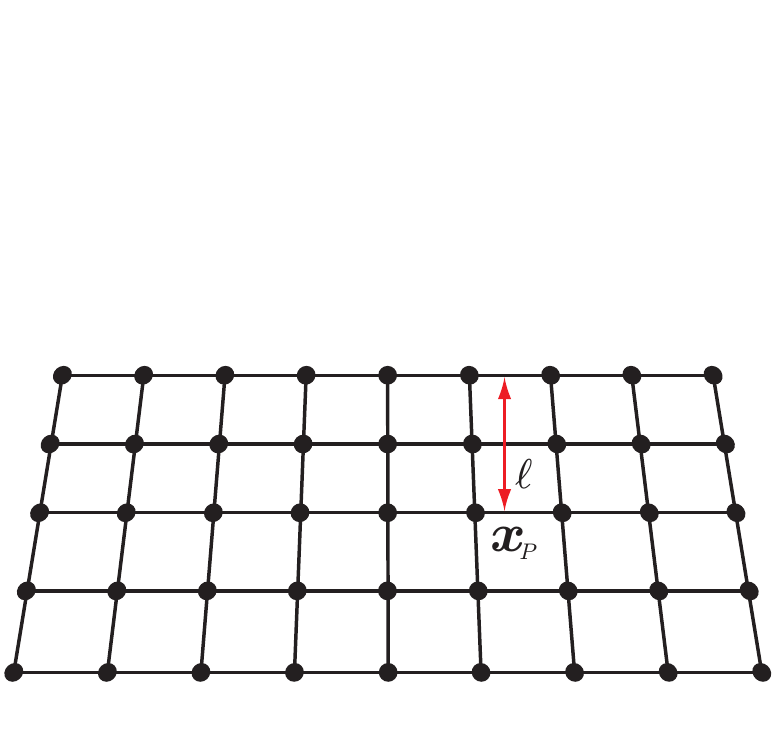}}
\end{tabular}
%\subfigure[Bulk]{\includegraphics[height=1.75in]{bulk.pdf}} \qquad
%\subfigure[Interface]{\includegraphics[height=1.75in]{interface.pdf}}
\caption{On the qualitative change in the intermolecular distance (and hence density profile) between the configurations in the bulk (a) and at the interface (b) in a crystal.}\label{fig:intermolecular-distance}
\end{figure}
Next, to highlight the physics of surface tension, let us first consider a crystal surface. In the near-the-interface region (figure~\ref{fig:intermolecular-distance}b) the molecules are expected to be at a different equilibrium intermolecular distance compared to that in the bulk (figure~\ref{fig:intermolecular-distance}a) because of the anisotropy, i.e. molecules are not experiencing the same interaction with the vacuum above the interface (cf. figure~\ref{fig:intermolecular-distance}b) as with the substance (cf. figure~\ref{fig:intermolecular-distance}a) -- this effect is due to van der Waals forces $\varphi_{\mathrm{vdW}} \sim r^{-6}$, which are long range compared to the intermolecular distance. This effect was anticipated theoretically by Lennard-Jones and Dent \cite{Lennard-Jones:1928} along with earlier experiential indications discussed in the aforementioned reference. While in the bulk the situation is isotropic, i.e. for the two orthogonal orientations of molecules in figure~\ref{fig:intermolecular-distance}a, the van der Waals forces acting on the molecules are the same, in the case of the interface in figure~\ref{fig:intermolecular-distance}b the vertically oriented pair no longer experiences the attraction from the upper space, while the horizontally oriented pair of molecules is now pulled apart by a force about half of that compared to the bulk configuration -- altogether this leads to anisotropy in the pressure, i.e. it becomes a tensor as we also saw from the direct calculations (\ref{sigmas:Lifshitz},\ref{hatsigmaij5}). As a result, the matter is compressed near the interface, though anisotropically, and therefore the near-the-interface region acts like a skin. In liquids, of course, the behavior of the intermolecular distance and hence the density profile is more complicated compared to that in crystals as a liquid phase must transition to its vapor state in a less abrupt fashion. The fact that the normal $p_{N} \equiv - \sigma_{zz}$ and tangent $p_{T} \equiv - \sigma_{xx} = - \sigma_{yy}$ components of renormalized pressure are different near the interface indicates that a liquid in its vicinity starts behaving more like a solid. This anisotropy is an inherent property of inhomogeneous media, e.g. atmosphere, ocean, and rocks stratified under gravity field, though would require exotic conditions under which the length-scale of inhomogeneity is comparable to the range of Derjaguin or Casimir effects.

Despite being convincing, the classical considerations of the effect of van der Waals forces above are based on pairwise molecular interaction $\phi_{\mathrm{vdW}}^{ij} = - 4 \, \upsilon \, (r_{\ins{m}}/r_{ij})^{6}$, i.e. require molecules $i$ and $j$ to be present, and therefore fail to account for the vacuum polarization. Moreover, as we observed in the previous section \S \ref{subsec:ST-calculation}, the Lifshitz theory with a sharp interface gives null surface tension due to antisymmetry of the stress \eqref{Delta-sigma:sharp}. It is only the physically justified smoothed out interface formulation with smoothly varying background $\varepsilon(z)$ is capable to properly account for the vacuum polarization and hence the surface tension phenomena. Therefore, all theories based on an effective Lennard-Jones type intermolecular potential, should it be combined with statistical approach (molecular distribution functions) \cite{Jaffe:1942,Kirkwood:1949,Fisher:1964,Shoemaker:1970,Berry:1972,Hansen:2006} or direct molecular simulations \cite{Walton:1983,Laghaei:2005,Yamaguchi:2018,Jabbarzadeh:1997,Trokhymchuk:1999}, cannot truly capture the quantum-mechanical nature of the underlying local stresses and hence surface tension. Therefore, it is not surprising that surface tension calculated with these approaches often results in values substantially different from experimentally measured. In the light of the presented theory this is not surprising as (i) the involved van der Waals forces cannot be accounted for by addition of the individual intermolecular potentials due to screening effects -- the fact realized by Lifshitz \cite{Lifshitz:1956} -- and (ii) the quantum mechanical effects prove to play a significant role in settling the value of surface tension. Based on the gained understanding, one can also expect that surface tension value may vary along the interface if the latter is not planar, though this variation would become noticeable only if the interfacial radius of curvature is comparable to the range of van der Waals forces contributing to the integrals in \eqref{gamma1}.

Lastly, the gained understanding of the stress distributions in figure~\ref{fig:stress-distributions} being concentrated in the neighborhood of the interface brings up yet another physical interpretation of surface tension phenomena: if one considers molecules as dielectric dipoles in the fluctuating EM field, then, after averaging the intensity over fluctuations, they (molecules) are dragged in the direction of the gradient of the EM fluctuations intensity, i.e. to the interfacial region. As also envisioned above from the (pairwise) van der Waals intermolecular potential considerations, this effect naturally results in a higher molecular density near the interface, i.e. it acts as optical tweezers, which have an analogous underlying physical mechanism.

\section{Conclusions}

First, let us recap the basic elements of the presented here story in simple terms. The key effect addressed in our study -- divergence of the EM stresses acting along the interface -- can be gleaned even from the pairwise summation \cite{Kirkwood:1949} of intermolecular van der Waals potentials $\phi_{ij}=-4 \, \upsilon \, r_{\ins{m}}^{6} \, r_{ij}^{-6}$, which for the stress $\sigma_{xx}(\bs{x}_{P})$ along the interface at a point $\bs{x}_{P}=(x_{P},y_{P},z_{P})$ located at the distance $\ell$ from the interface, cf. figure~\ref{fig:intermolecular-distance}b, leads to
\begin{align}
\sigma_{xx}(\bs{x}_{P}) = - 12 \, \upsilon \, r_{\ins{m}}^{6} \int_{-\infty}^{0}\d z\iint_{\-\infty}^{+\infty}{\d x \d y \, \frac{(x_{P}-x)^{2}}{|\bs{x}_{P}-\bs{x}|^{8}}} = - \frac{\pi}{3} \frac{\upsilon \, r_{\ins{m}}^{6}}{\ell^{3}},
\end{align}
and obviously diverges as one approaches the interface $\ell \rightarrow 0$, qualitatively being analogous to the calculation from the Lifshitz theory \eqref{Delta-sigma:sharp}. Based on a similar calculation one can compute the disjoining pressure \eqref{pressure:Derjaguin}, as it was originally done by Derjaguin \cite{Derjaguin:1936}, but the Hamaker constant $A_{H}$ would be incorrect because non-additive effects are neglected in this approach. While the Lifshitz approach adopted here accounts for non-additivity, the divergence remains. The nature of the latter comes from the nonphysical nature of the sharp interface associated with the jump in $\varepsilon$ and the high wavenumber (momenta) $\bs{q}$ in the along-the-interface direction of the virtual photons contribution to the stress spectral representation leading to the divergent integrand $\sim q^{2}$ in \eqref{sigmas:Lifshitz}. In reality, however, there must be a cut-off \cite{Dzyaloshinskii:1961} at some $q_{\mathrm{max}}$, which is considerably smaller than the reciprocal of the interatomic spacing $r_{\ins{m}}^{-1}$. The equivalence of the wavenumber $q$-divergence to that due to sharp interface condition is easy to see from the corresponding equation \eqref{eq:Schrodinger} for the Fourier components of the Green's function when interface is smeared on the length-scale $w$:
\begin{align}
\left\{\partial_{z} \left[\varepsilon(z/w) \, \partial_{z}\right] - q^{2}\right\} \widetilde{G} = \delta(z-z^{\prime});
\end{align}
clearly, the problem is scale-free under the transformations $z \rightarrow w \widetilde{z}$ and $q^{2} \rightarrow w^{2} q^{2}$, thus indicating that the limit of a sharp interface $w \rightarrow 0$ is equivalent to the limit $q \rightarrow \infty$.

In summary, we demonstrated that in the sharp interface formulation it is impossible to obtain surface tension due to asymmetry of the stresses \eqref{Delta-sigma:sharp}. Hence, in order to reveal the origin of surface tension, one must consider a smoothed out interface, the internal structure of which (cf. figure~\ref{fig:stress-diff-distributions}) explains the surface tension phenomena. Next, without proper renormalization, i.e. if one disregards all diverging contributions, surface tension could be even negative. Hence, subleading divergent terms must be kept to produce correct surface tension, which is also consistent with the a priori fact that the Lifshitz theory cannot be fully renormalizable as it is an effective and not a closed one (cf. discussion in \S~\ref{subsec:ST-calculation}). Another way to look at this is to realize that the Lifshitz theory can be seen as an outer solution of a general problem involving microscopic scales as well. Based on Kaplun's extension theorem \cite{Lagerstrom:1988}, it must have a region of validity overlapping at the scales $O(r_{\ins{m}})$ with the inner solution corresponding to a microscopic theory operating at the length-scales $\le r_{\ins{m}}$. This implies that the Lifshitz theory can be applied at the intermolecular scales $O(r_{\ins{m}})$. While the cut-off parameter $\epsilon^{1/2}$ can be determined theoretically by matching these inner and outer solutions, should a microscopic theory exist, it can also be determined empirically as done here (\S~\ref{subsec:ST-calculation}). The theoretical basis for such a matching comes from the employed here proper time regularization scheme (\S~\ref{subsec:proper-time-regularization}), which also makes the Coulomb potential regularized at small distances provided the cut-off parameter $\epsilon$ is finite. Hence, when applied to dipole-dipole interaction, this regularization makes the effective potential between dipoles to be attractive at large distances and repulsive at the distances on the order of $\epsilon^{1/2}$, the behavior qualitatively similar to the Lenard-Jones potential. This observation also explains why the properly regularized Lifshitz theory with a finite cut-off parameter can match a microscopic theory and remain quite accurate at the intermolecular distances. As demonstrated in the present work (\S~\ref{sec:ST}), applicability of the Lifshitz theory at the length-scales $O(r_{\ins{m}})$ also explains its ability to capture the surface tension phenomena accurately. It is one of those occasions when ``very beautiful quantum field theories can arise in condensed matter physics as effective theories. In addition to their beauty, effective field theories are also very effective in answering certain questions that the more microscopic versions cannot'' \cite{Shankar:1999}.

At the technical level, in this paper we studied the vacuum polarization in inhomogeneous media with an embedded smoothed out interface. Because the problem of UV divergent terms appears already for a single sharp interface, we consider a model of a single smoothed out interface. We used the heat kernel representation, proper time cut-off regularization, and the Hadamard expansion for the Green's function to determine the divergent part. The heat kernel technique was also used by Bordag et al. \cite{Bordag:1998} to study UV-divergencies in the EM effective action in the case of inhomogeneous dielectric media without dispersion. Here we considered dispersive media in the non-retarded (Deryaguin) limit, in which the operator containing only spatial derivatives appears, though the fluctuating fields are time-dependent. These fluctuations are quantum fluctuations of the polarizable atoms dipole moments. Given the equivalence of the Green's function at finite temperatures $G^{\beta}$ and that at zero temperature $G^{F}$ after Wick's rotation, we expand all time-dependent quantities in the Fourier series over the Matsubara modes. Then the quantization scheme reduces to the spectral theory of the 3D operator $\widehat{O} = \delta^{ij} \partial_{i} \varepsilon \partial_{j}$. We found that while there is no Casimir force acting normally to the interface, the vacuum polarization still plays an important role in the origin of surface tension. The value of total energy density, as well as of total pressure in the liquid, cannot be determined solely by the effective theory -- the microscopic physics must be evoked for that \cite{Volovik:2003fe}. Vacuum contributions to the stress-tensor of an inhomogeneous system depend on $\epsilon^{1/2}$ and the characteristic scale $w$ of inhomogeneity, i.e. the width of the interface, and can be classified as follows. (i) The leading UV contribution to the pressure is of the order $\epsilon^{-3/2}$, does not depend on the geometry of the system, is homogeneous and structurally the same as for the vacuum. For this reason it does not lead to any observable effects. The standard Lifshitz procedure is to truncate these terms. (ii) The subleading UV divergent term, which is on the order of $\epsilon^{-1/2} w^{-2}$, is related to surface tension of the interface and can be measured in experiments. (iii) The finite contribution $\sim w^{-3}$ is due to van der Waals interactiosn, does not depend on the UV cut-off scale $\epsilon^{1/2}$, but also contributes to surface tension. In the limit of a very sharp interface $w\sim \epsilon^{1/2}$, the subleading and finite contributions can be comparable. (iv) If one takes into account retardation effects, then $w^{-4}$ terms would appear and correspond to the Casimir-type effects. This classification is in accordance with the one proposed by Volovik \cite{Volovik:2003fe} for a description of quantum effects in quantum liquids. It is important to emphasise that while our renormalization technique is in essence the same as that used by Casimir and Lifshitz -- namely, subtraction of the divergent free space contribution (with the value of dielectric permittivity $\varepsilon(\bs{x})$ corresponding to a given point $\bs{x}$) -- the presence of a smoothed out interface leads not only to finite stresses \eqref{sigmatau1} (present in the sharp interface case), but also subleading divergent stresses \eqref{sigma-div:subleading} dependent on the cut-off parameter, both of which in turn give rise to surface tension (\ref{gamma1},\ref{eqn:ST-total}) favorably compared here with the available experimental data. The proposed approach for computation of surface tension can also be used for calculating its values in solids, which are not only difficult to measure in practice, but also differ from surface energies and may vary with direction along the solid surface \cite{Gibbs:1876,Orowan:1970}.

Extension of the theory laid out here and applicable in the Derjaguin (non-retarded) limit to the general case including retarded effects as well as massive fields would be a straightforward, though technically involved, generalization of the present work. In this context it should be mentioned that while a general expression for the force density \eqref{fi} is still a subject of extensive debates \cite{Brevik:1979,Shevchenko:2011,Mansuripur:2012,Cho:2012,Mansuripur:2013,Webb:2016} since the times of Minkovski, Helmholtz, Einstein \& Laub, and Abraham, which was left mute in our case as we deal with the electrostatic limit only, it does not prevent one to include the magnetic field (retarded effects) for the same reason as to why the Lifshitz theory of calculating local stresses and hence force density was made possible \cite{Lifshitz:1956,Dzyaloshinskii:1959,Dzyaloshinskii:1961} -- it requires one to deal with the fluctuating EM field only, which is in a thermodynamic equilibrium with the medium. Calculations of stresses and surface tension can also be improved by using more accurate spectral data for dielectric constants \cite{Bergstrom:1997} instead of the simplified formula \eqref{formula:Debye} and accounting for finite temperature corrections.

\appendix

\section{The limit of sharp interface} \label{apx:Schwartz}

Consider the equation for the electrostatic potential $\widehat{\phi}$, transformed in the Fourier space in the $x$ and $y$ directions, when the dielectric permittivity is a function in the remaining $z$-direction:
\begin{align}
\label{problem:electrostatic}
\frac{\d}{\d z}\left(\varepsilon(z)\frac{\d\widehat{\phi}}{\d z}\right) - \varepsilon(z) q^{2} \widehat{\phi} = 0,
\end{align}
which would be subject to the usual boundary conditions in the limit of sharp interface:
\begin{align}
\label{BCs:electrostatics}
\left[\widehat{\phi}\right]_{1}^{2}(0) = 0, \ \left[\varepsilon \, \widehat{\phi}_{z}\right]_{1}^{2}(0) = 0,
\end{align}
where $\left[\cdot\right]_{1}^{2}(0)$ stands for the jump across the interface between two media $1$ and $2$ at $z=0$; that is, the potential $\widehat{\phi}$ is continuous (which is obvious since in the physical space the tangential component of the electrostatic field $\mathbf{t} \cdot \nabla \phi = E_{\mathbf{t}}$ is continuous), while $\widehat{\phi}_{z}$ is not. In this limit the problem \eqref{problem:electrostatic} becomes that of a linear ODE with discontinuous and singular coefficients, since $\varepsilon$ experiences a jump, and $\widehat{\phi} \in C$ since its first derivative is discontinuous at $z=0$. Hence, it can be treated from the point of view of linear distributions; namely, multiplying the first term in \eqref{problem:electrostatic} with a test function $f$ in the Schwartz space $f \in \mathcal{D}(\Bbb{R})$, applying the differentiation rules of distributions and integrating by parts \cite{Vladimirov:1971} we find:
\begin{align}
\left(\left(\varepsilon \, \widehat{\phi}'\right)',f\right) &= - \left(\varepsilon \, \widehat{\phi}',f'\right) = - \left(\widehat{\phi}', \varepsilon \, f'\right) = \left(\widehat{\phi},\left(\varepsilon \, f'\right)'\right)= \left(\varepsilon \, f'\right)(0) \left[\widehat{\phi}\right]_{1}^{2}(0)  - \int{\varepsilon \, f' \widehat{\phi}' \, \d z} \nonumber \\
&= \left(\varepsilon \, f'\right)(0) \left[\widehat{\phi}\right]_{1}^{2}(0) - \left[\varepsilon \, \widehat{\phi}'\right]_{1}^{2}(0) f(0) + \int{\left(\varepsilon \, \widehat{\phi}'\right)' \, f \, \d z},
\end{align}
where the integrand in the last term must cancel with $- \int{\varepsilon(z) q^{2} \widehat{\phi} \, f \, \d z}$ due to \eqref{problem:electrostatic} and arbitrariness of $f$, while in the first two terms we must have
\begin{align}
\left[\widehat{\phi}\right]_{1}^{2}(0) = \left[\varepsilon \, \widehat{\phi}'\right]_{1}^{2}(0) = 0
\end{align}
in order for $\widehat{\phi}$ to be a generalized solution in the sense of linear distributions. Hence, we naturally arrive at the BCs \eqref{BCs:electrostatics} dictated by the physics of electrostatic field (Nature somehow knows about Sobolev and Schwartz distributions).

Now, if we make the substitution $\phi(z) = \varepsilon^{-1/2}(z) \, u(z)$, we get the Schr\"{o}dinger type equation $u'' - \left(V-E\right) u = 0$ with the potential $V$ involving $\left(\varepsilon^{\prime}\right)^{2}$ and $\varepsilon^{\prime\prime}$, i.e. it may appear that we leave the space of linear distributions due to the $\delta^{2}$ term. However, this change of dependent variables makes $u(z)$ more singular than $\phi$ and the $\sim \delta^{2}$ term identically disappears after the transformation to the original variable $\phi$. Also, while it is known that $\delta^{2}$ does not make a unique sense as different delta-sequences produce different results (also depending upon the choice of test functions), it was shown by Mikusinki \cite{Mikusinski:1966} that $\delta^{2}$ does make sense in combination with another distribution as it happens here in the formula for the potential $V(z)$. Also, one can say that specifying a particular choice of $\varepsilon(z/w)$ produces a unique delta sequence and hence the unique limit of $\delta^{2}$ when $w \rightarrow 0$.

\section{Classical action for non-dispersive media} \label{apx:classical-action}

We start with the classical action $\mathcal{S}$ of the electromagnetic field in an isotropic dielectric without dispersion, i.e. when the real permittivity $\varepsilon$ does not depend on the frequency
\ba
\mathcal{S}=-\frac{1}{4}\int \d t\d^3x\,\sqrt{|g|}\varepsilon(\bs{x}) F_{\mu\nu} F^{\mu\nu}+\int \d t\d^3x\,\sqrt{|g|} A_{\mu} J^{\mu}.
\ea
Description of the system in equilibrium at finite temperature reduces to the Wick rotated $t=-\i t_\ins{E}$ version of electrodynamics with the corresponding Euclidean action \cite{Schwinger:1959}
\ba
\mathcal{S}_\ins{E}=\frac{1}{4}\int\d t_\ins{E}\d^3x\,\sqrt{g_\ins{E}}\varepsilon(\bs{x}) F_{\mu\nu} F^{\mu\nu}-\int \d t_\ins{E}\d^3x\,\sqrt{g_\ins{E}} A_{\mu} J^{\mu},
\ea
which leads to the change in signs in \eqref{tensor:stress-energy:E}. Since the system we are dealing with is neutral, i.e. there are no free charges, so $J^{\mu} \equiv 0$. In order to describe non-retarded effects it is sufficient to consider a vanishing magnetic field, which in the Coulomb gauge is equivalent to the choice $A_i=0$. Then the Euclidean action takes the form
\ba\label{action:Euclidean}
\mathcal{S}_\ins{E}=-\frac{1}{2}\int_0^{\beta} \d t_\ins{E} \int \d^3x \,\varepsilon(\bs{x})\, \partial_k A_0(t_\ins{E},\bs{x})\,\partial^k A_0(t_\ins{E},\bs{x}),
\ea
where the sign is chosen such that the Boltzmann factor makes the Euclidean path integral convergent. Finite temperature $T$ is accounted for by the requirement that the system is periodic in the Euclidean time $t_\ins{E}=x^0$ with the period $\beta$.
Now let us expand the potential $A_0$ in Fourier series \eqref{FS:A0} over the Matsubara frequencies $\zeta_n={2\pi n / \beta}$, the inverse of which is \eqref{mode:Matsubara}. Note that dimensionality of the Fourier component $A_0(\zeta_n;\bs{x})$ differs from that of the potential $A_0(t_\ins{E},\bs{x})$. As a result, \eqref{action:Euclidean} can be rewritten as
\ba
\mathcal{S}_\ins{E}&=-\frac{1}{2\beta^2}\int_0^{\beta} \d t_\ins{E} \int \d^3x \,\varepsilon(\bs{x})\, \sum_{n=-\infty}^{\infty}\partial_k A_0(\zeta_n;\bs{x})e^{\i\zeta_n{t_\ins{E}}}\,\sum_{n'=-\infty}^{\infty} \partial^k A_0(\zeta_n';\bs{x})e^{\i\zeta_n'{t_\ins{E}}}.
\ea
Using the representation
\ba
{1\over\beta}\int_0^{\beta} \d t_\ins{E}e^{\i\zeta_n{t_\ins{E}}}e^{\i\zeta_n'{t_\ins{E}}}={1\over\beta}\int_0^{\beta} \d t_\ins{E}e^{\i{2\pi(n+n')\over\beta}{t_\ins{E}}}=\int_0^{1} \d \tau e^{2\pi\i(n+n')\tau}=\delta_{n+n',0},
\ea
where
\ba
\delta_{n+n',0}=\begin{cases}
1& \mbox{if}~~~ n+n'=0,\\
0& \mbox{if}~~~n+n'\neq 0,
\end{cases}
\ea
we obtain
\ba
\mathcal{S}_\ins{E}&=-\frac{1}{2\beta}\int \d^3x \,\varepsilon(\bs{x})\, \sum_{n=-\infty}^{\infty}\partial_k A_0(\zeta_n;\bs{x})\,\partial^k A_0(\zeta_{-n};\bs{x}),
\ea
where $\partial^k A_0(\zeta_{-n};\bs{x}) = \partial^k A_0^{*}(\zeta_{n};\bs{x})$. Generalization of the above considerations to dispersive media leads to \eqref{Free2}.

\section{Heat kernel calculations} \label{apx:heat-kernel}

The coefficient in front of the second derivatives in the operator \eqref{op1} defines the effective metric
\ba
\mathrm{g}^{ij}=\varepsilon\delta^{ij}  \hh \mathrm{g}_{ij}=\frac{1}{\varepsilon}\delta_{ij}\hh \sqrt{\mathrm{g}}=\varepsilon^{-3/2},
\ea
so that the operator \eqref{op1} can be represented as
\ba
\widehat{O} = \varepsilon\delta^{ij}\,\partial_i \partial_j+\delta^{ij}\,(\partial_i \varepsilon) \partial_j = \mathrm{g}^{ij}\,\partial_i \partial_j+\eta^i \partial_j,
\ea
where
\be
\eta_i=\frac{\partial_i\varepsilon}{\varepsilon}\hh \eta^i=\mathrm{g}^{ij}\,\eta_j .
\ee
Let us define the new field
\ba
\phi=\mathrm{g}^{-1/4}\Phi=\varepsilon^{3/4}\Phi
\ea
and the corresponding Green's function
\ba\label{Green-new}
\mathbb{G}(\zeta;\bs{x},\bs{x'})=\langle\upphi(\bs{x})\upphi(\bs{x'})\rangle=\mathrm{g}^{-1/4}(\bs{x})\,\mathrm{g}^{-1/4}(\bs{x}')\,\widehat{G}(\zeta;\bs{x},\bs{x'}),
\ea
where $\upphi(\bs{x})$ is the operator corresponding to the field $\phi(\bs{x})$. Then the inner product for this new field takes the covariant form
\ba
\int \d \bs{x}\, \Phi_1 \Phi_2=\int \d \bs{x}\,\varepsilon^{-3/2}\, \phi_1 \phi_2=\int \d \bs{x}\sqrt{\mathrm{g}}\, \phi_1 \phi_2.
\ea
Equation \eq{eqG} transforms to
\ba\label{eqG1-apx}
\mathrm{g}^{1/4}(\bs{x}')\,\widehat{O}\,\mathrm{g}^{1/4}(\bs{x})\,\mathbb{G}(\zeta;\bs{x},\bs{x'})=\delta(\bs{x}-\bs{x'})
\ea
or
\ba\label{eqn:Green-new}
\mathbb{O} \, \mathbb{G}(\zeta;\bs{x},\bs{x'})=\frac{\delta(\bs{x}-\bs{x'})}{\sqrt{\mathrm{g}}}, \ \text{where} \ \mathbb{O}=\mathrm{g}^{-1/4}(\bs{x})\,\widehat{O}\,\mathrm{g}^{1/4}(\bs{x}).
\ea
Taking into account that
\ba
\partial_j \varepsilon^{-3/4}&=\varepsilon^{-3/4}\Big(\partial_j-\frac{3}{4}\eta_j \Big)\\
\partial_i\partial_j \varepsilon^{-3/4}&=\varepsilon^{-3/4}
\Big(\partial_i-\frac{3}{4}\eta_i \Big)\Big(\partial_j-\frac{3}{4}\eta_j \Big)\\
&=\varepsilon^{-3/4}
\Big(\partial_i\partial_j-\frac{3}{4}\eta_i\partial_j -\frac{3}{4}\eta_j\partial_i-\frac{3}{4}(\partial_i\eta_j)+\frac{9}{16}\eta_i\eta_j \Big),
\ea
the new operator $\mathbb{O}$ becomes
\ba
\mathbb{O}&=\mathrm{g}^{ij}\,\Big(\partial_i\partial_j-\frac{3}{4}\eta_i\partial_j -\frac{3}{4}\eta_j\partial_i-\frac{3}{4}(\partial_i\eta_j)+\frac{9}{16}\eta_i\eta_j \Big)+
            \mathrm{g}^{ij}\Big( \eta_i \partial_j-\frac{3}{4}\eta_i \eta_j
            \Big)\\
            &=\mathrm{g}^{ij}\,\Big(\partial_i\partial_j-\frac{1}{2}\eta_i\partial_j -\frac{3}{4}(\partial_i\eta_j)-\frac{3}{16}\eta_i\eta_j \Big)
            \\
            &=\mathrm{g}^{ij}\,\Big(\partial_i\partial_j-\frac{1}{2}\eta_i\partial_j\Big) -\frac{3}{4}\mathrm{g}^{ij}\,\Big(\partial_i\eta_j+\frac{1}{4}\eta_i\eta_j \Big)
            \\
            &=\varepsilon\delta^{ij}\,\Big(\partial_i\partial_j-\frac{1}{2}\eta_i\partial_j\Big) -\frac{3}{4}\varepsilon\delta^{ij}\,\Big(\partial_i\eta_j+\frac{1}{4}\eta_i\eta_j \Big).
\ea
Note that
\ba
\nabla_k\eta^k=\varepsilon\delta^{ij}\Big(\partial_i\eta_j-\frac{1}{2}\eta_i\eta_j\Big)=\mathrm{g}^{ij}\Big(\partial_i\eta_j-\frac{1}{2}\eta_i\eta_j\Big)
\ea
and the covariant Laplacian in metric $\mathrm{g}_{ij}$ reads
\ba
\lap &=\mathrm{g}^{ij}\nabla_i\nabla_j=\frac{1}{\sqrt{\mathrm{g}}}\partial_i\Big(\mathrm{g}^{ij}\sqrt{\mathrm{g}}\,\partial_j\Big)
=\varepsilon^{3/2}\partial_i\Big(\delta^{ij}\varepsilon\, \varepsilon^{-3/2}\,\partial_j\Big)
=\varepsilon \delta^{ij}\,\Big(\partial_i\partial_j-\frac{1}{2}\eta_i\partial_j \Big)\\
&=\mathrm{g}^{ij}\,\Big(\partial_i\partial_j-\frac{1}{2}\eta_i\partial_j \Big).
\ea
Therefore, we obtain the covariant form for the operator in \eqref{eqn:Green-new}
\ba\label{OV1}
\mathbb{O}=\lap -\mathbb{V}, \ \text{where} \ \mathbb{V}=\frac{3}{4}\Big(\nabla^k\eta_k+\frac{3}{4}\eta^k\eta_k\Big)
=\frac{3}{4}\varepsilon\delta^{ij}\,\Big(\partial_i\eta_j+\frac{1}{4}\eta_i\eta_j \Big),
\ea
which is Hermitian with the covariant measure $\sqrt{\mathrm{g}}$. The potential $\mathbb{V}$ can also be obtained following the computations along the lines of Gilkey \cite{Gilkey:1975,Gilkey:1994} and Bordag et al. \cite{Bordag:1998}, which require the knowledge of the Christoffel symbols
\begin{subequations}
\begin{align}
\begin{split}
\Gamma^i_{jk}&=\frac{1}{2}\mathrm{g}^{im}\big[\mathrm{g}_{mj,k}+\mathrm{g}_{mk,j} -\mathrm{g}_{jk,m} \big]\\
&=-\frac{1}{2}\delta^{im}\big[\delta_{mj}\eta_k+\delta_{mk}\eta_j -\delta_{jk}\eta_m \big]\\
&=-\frac{1}{2}\big[\delta_{ij}\eta_k+\delta_{ik}\eta_j -\delta_{jk}\eta_i \big],
\end{split} \\
\mathrm{g}^{jk}\Gamma^i_{jk}&=\frac{1}{2} \varepsilon\,\eta_i,
\end{align}
\end{subequations}
and the connection
\ba
\omega_i=\frac{1}{2}\big[\eta_i+\mathrm{g}_{ik}\mathrm{g}^{mn}\Gamma^k_{mn}\big]=\frac{3}{4}\eta_i.
\ea
Then, the endomorphism  $\mathbb{E}=-\mathbb{V}$ gives the potential
\ba\label{OV2}
\mathbb{V}=\mathrm{g}^{ij}\big[\partial_i\omega_i+\omega_i\omega_i -\omega_k\Gamma^k_{ij}\big]
=\frac{3}{4}\varepsilon\,\delta^{ij}\Big(\partial_i\eta_j+\frac{1}{4}\eta_i\eta_i\Big),
\ea
which coincides with \eq{OV1}.

Heat kernel corresponding to the operator $\widehat{O}$ defined in \eqref{op1} is
\ba
\widehat{K}(s|\bs{x},\bs{x'})=\mathrm{g}^{1/4}(\bs{x})\mathrm{g}^{1/4}(\bs{x}')\mathbb{K}(s|\bs{x},\bs{x'}),
\ea
where
\ba
\mathbb{K}(s|\bs{x},\bs{x}')=-e^{s\mathbb{O}}=-\frac{\Delta^{1/2}(\bs{x},\bs{x}')}{(4\pi s)^{3/2}}e^{-\frac{\sigma(\bs{x},\bs{x}')}{2s}}\Big(a_0(\bs{x},\bs{x}')+s\,a_1(\bs{x},\bs{x}')+\dots\Big);
\ea
here $\sigma(\bs{x},\bs{x}')$ is a world function defined as one half of the square of geodesic distance between points in the metric $\mathrm{g}_{ij}$ and $\Delta^{1/2}(\bs{x},\bs{x}') \equiv \mathrm{g}^{-1/2}(\bs{x}) \frak{D}(\bs{x},\bs{x}') \mathrm{g}^{-1/2}(\bs{x}')$, where $\frak{D}(\bs{x},\bs{x}') = -\det\left(-\sigma_{;\mu\nu'}\right)$ is the VanVleck-Morette determinant. For the computation of the heat kernel of this operator we need to know $\Delta^{1/2}(\bs{x},\bs{x}')$, $a_0(\bs{x},\bs{x}')$, and $a_1(\bs{x},\bs{x}')$ and their derivatives. In particular they include the curvature terms:
\begin{subequations}
\begin{align}
R&=2\Big(\eta^k{}_{;k}+\frac{1}{4}\eta^k\eta_k\Big) = 2\varepsilon\,\delta^{km}\Big(\partial_k\eta_m-\frac{1}{4}\eta_k\eta_m\Big), \\
R_{ij}&= \frac{1}{2}\eta_{i;j}-\frac{1}{4}\eta_i \eta_j +\frac{1}{2} \mathrm{g}_{ij}\eta^k{}_{;k}+\mathrm{g}_{ij}\frac{1}{4}\eta^k\eta_k \nonumber \\
&=\frac{1}{2}\partial_j\eta_i+\frac{1}{4}\eta_i \eta_j
+\frac{1}{4} \delta_{ij}\delta^{km}\big(2\partial_k\eta_m - \eta_k\eta_m\big);
\end{align}
\end{subequations}
here we used the relations
\begin{subequations}
\begin{align}
\eta_{i;j}&=\partial_j\eta_i+\eta_i\eta_j-\frac{1}{2}\mathrm{g}_{ij}\eta^{k}\eta_k=\partial_j\eta_i+\eta_i\eta_j-\frac{1}{2}\delta_{ij}\delta^{km}\eta_{k}\eta_m, \\
\eta^k{}_{;k}&=\varepsilon\delta^{ij} \big(\partial_j\eta_i-\frac{1}{2}\eta_{i}\eta_j), \\
\partial_i (\mathrm{g}^{1/4})&=-\frac{3}{4}\varepsilon^{-3/4}\eta_i.
\end{align}
\end{subequations}

The regularized Green's function \eqref{Green-new} assumes the form
\ba
\mathbb{G}_{\epsilon}(\zeta;\bs{x},\bs{x'})=\int_{\epsilon}^{\infty} \d s \, \mathbb{K}(s|\bs{x},\bs{x}')
\ea
or, in the original notation,
\ba
\widehat{G}_{\epsilon}(\zeta;\bs{x},\bs{x'})=\int_{\epsilon}^{\infty} \d s\, \widehat{K}(s|\bs{x},\bs{x}')
=\mathrm{g}^{1/4}(\bs{x})\mathrm{g}^{1/4}(\bs{x}')\int_{\epsilon}^{\infty} \d s\,\mathbb{K}(s|\bs{x},\bs{x'}),
\ea
which can be expanded as
\ba
\widehat{G}_{\epsilon}(\zeta;\bs{x},\bs{x'})=-\int_{\epsilon}^{\infty} \d s\,
\frac{\frak{D}^{1/2}(\bs{x},\bs{x}')}{(4\pi s)^{3/2}}e^{-\frac{\sigma(\bs{x},\bs{x}')}{2s}}\Big(a_0(\bs{x},\bs{x}')+s\,a_1(\bs{x},\bs{x}')+\dots\Big)
\ea
Derivation of the regularized  stress tensor then reduces to the following steps:

i) Compute $\Delta^{1/2}(\bs{x},\bs{x}')$, $a_0(\bs{x},\bs{x}')$, and $a_1(\bs{x},\bs{x}')$;

ii) Compute the second partial derivative of the Green's function $\partial_{\bs{x}}\partial_{\bs{x}'}\widehat{G}_{\epsilon}(\zeta;\bs{x},\bs{x'})$;

iii) Take the limit $\bs{x}=\bs{x'}$;

iv) Integrate over the Schwinger proper time parameter $s$ and then expand the result in the small cut-off parameter $\epsilon^{1/2}$;

v) Compute $\widehat{\sigma}_{ij}(\zeta;\bs{x})$ by substituting the result to \eqref{hatsigmaij}.

\noindent For the quantities in the limit of coincident points we introduce notation $[\dots]\equiv (\dots)|_{\bs{x}=\bs{x'}}$. The necessary expressions that we need to compute divergent parts of the Green's function $\widehat{G}_{\epsilon}(\zeta;\bs{x},\bs{x'})$ are \cite{Christensen:1976}
\begin{subequations}
\begin{align}
[\sigma]&=0,  \ [\sigma_{;i}]=0, \ [\sigma_{;i'}]=0, \ [\sigma_{;ij}]=\mathrm{g}_{ij}, [\sigma_{;ij'}]=-\mathrm{g}_{ij}, \\
[a_0]&=1, \ [a_0{}_{;i}]=0, \ [a_0{}_{;ij'}]=0, \ [a_1]=\frac{1}{6}R-\mathbb{V}, \  [a_1{}_{;i}]=\frac{1}{2}\big(\frac{1}{6}R_{;i}-\mathbb{V}_{;i}\big),\\
[\Delta^{1/2}]&=1, \ [(\Delta^{1/2})_{;i}]=0, \ [(\Delta^{1/2})_{;ij}]=\frac{1}{6}R_{ij}, \ [(\Delta^{1/2})_{;ij'}]=-\frac{1}{6}R_{ij},
\end{align}
\end{subequations}
and
\ba
&[\partial_i\partial_{j'}\big(\mathrm{g}^{1/4}(\bs{x})\mathrm{g}^{1/4}(\bs{x'})\Delta^{1/2}\big)]
=\varepsilon^{-3/2}\left(\frac{9}{16}\eta_i\eta_j-\frac{1}{6}R_{ij}\right),
\ea
which lead to \eqref{Gij}.

\section{Surface tension of a slab} \label{apx:ST-slab}

The surface tension $\gamma^\ins{(fin)}$ of a slab of width $w$, cf. figure~\ref{fig:Lifshitz}, is
\be
\gamma^\ins{(fin)}=\int_{-\infty}^{\infty}dz\, \left[\sigma_{xx}(z)-\sigma_{zz}(z)\right],
\ee
where take into account that in regions 1 and 2 $\sigma_{zz}(z)=0$ and in region 3 the stress $\sigma_{zz}(z)$ is given by \eqref{sigmazz}.
We see that the integral of $-\sigma_{zz}(z)$ is twice more than the integral of $+\sigma_{xx}(z)$. Altogether, we obtain
\ba
\gamma^\ins{(fin)}=-\frac{3w}{4\pi c^2\beta}\sum_{n=-\infty}^{\infty}
\int_0^{\infty} dq\,q^2\,\left[{1\over
{\frac{(\varepsilon_1+\varepsilon_3)(\varepsilon_2+\varepsilon_3)}{({\varepsilon_1}-{\varepsilon_3})
({\varepsilon_2}-{\varepsilon_3})}}e^{2qw}
-1}\right]
\ea
or
\ba
\gamma^\ins{(fin)}=-{3\over 4\pi c^2\beta w^2}\sum_{n=-\infty}^{\infty}
\int_0^{\infty} d\bar{q}\displaystyle{\bar{q}^2\over
M\,e^{2\bar{q}}
-1} .
\ea
where
\ba
M={(\varepsilon_1+\varepsilon_3)(\varepsilon_2+\varepsilon_3)\over ({\varepsilon_1}-{\varepsilon_3})
({\varepsilon_2}-{\varepsilon_3})}  .
\ea
If medium 3 has intermediate dielectric constant $\varepsilon_2<\varepsilon_3<\varepsilon_1$, then $M<0$ and hence $\gamma^\ins{(fin)} >0$:
\ba
\gamma^\ins{(fin)}=-{1\over 32\pi c^2\beta w^2}\sum_{n=-\infty}^{\infty}
\Big[6\polylog(3,M)+\pi^2\ln(-M)+\ln(-M)^3\Big] >0.
\ea
For $M<0$ and $|M|\gg1$ we get an asymptotic approximation:
\ba
\gamma^\ins{(fin)}={3\over 16\pi c^2\beta w^2}\sum_{n=-\infty}^{\infty}
\left[\frac{1}{|M|}+\frac{1}{8|M|^2}+O(M^{-3})\right],
\ea
or, if we keep only the first term,
\ba
\gamma^\ins{(fin)}={3\over 16\pi c^2\beta w^2}\sum_{n=-\infty}^{\infty}\left|
{(\varepsilon_1-\varepsilon_3)(\varepsilon_2-\varepsilon_3)\over ({\varepsilon_1}+{\varepsilon_3})
({\varepsilon_2}+{\varepsilon_3})}\right|.
\ea
Note that $1/3$ of this value comes from of $+\sigma_{xx}$ and $2/3$ of it from of $-\sigma_{zz}$.

\section*{Acknowledgments}

This work was partially supported by the Natural Sciences and Engineering Research Council of Canada (NSERC). A.Z. also acknowledges financial support by the Killam Trust.

\bibliographystyle{JHEP}

\providecommand{\href}[2]{#2}\begingroup\raggedright\endgroup

\end{document}